\documentclass[twocolumn]{aastex631}

\makeatletter
\let\frontmatter@title@above=\relax
\makeatother
\usepackage{lipsum}
\usepackage{array,url,kantlipsum}
\usepackage{booktabs}
\usepackage{verbatim} 
\usepackage{natbib}
\usepackage{rotating}

\usepackage{graphicx}

\shorttitle{AGE-PRO II: Ophiuchus Disks}
\shortauthors{Ruiz-Rodriguez et al.}

\begin{document}

\title{The ALMA Survey of Gas Evolution of PROtoplanetary Disks (AGE-PRO): \\
II. Dust and Gas Disk Properties in the Ophiuchus Star-forming Region}

\author[0000-0003-3573-8163]{Dary A. Ruiz-Rodriguez}
\affiliation{National Radio Astronomy Observatory; 520 Edgemont Rd., Charlottesville, VA 22903, USA}
\affiliation{Joint ALMA Observatory, Avenida Alonso de Córdova 3107, Vitacura, Santiago, Chile}

\author[0000-0003-4907-189X]{Camilo Gonz\'alez-Ruilova}
\affiliation{Instituto de Estudios Astrofísicos, Universidad Diego Portales, Av. Ejercito 441, Santiago, Chile}
\affiliation{Millennium Nucleus on Young Exoplanets and their Moons (YEMS), Chile and Center for Interdisciplinary Research in Astrophysics}
\affiliation{Space Exploration (CIRAS), Universidad de Santiago, Chile}

\author[0000-0002-2828-1153]{Lucas A. Cieza}
\affiliation{Instituto de Estudios Astrofísicos, Universidad Diego Portales, Av. Ejercito 441, Santiago, Chile}

\author[0000-0002-0661-7517]{Ke Zhang}
\affiliation{Department of Astronomy, University of Wisconsin-Madison, 475 N Charter St, Madison, WI 53706, USA}

\author[0000-0002-8623-9703]{Leon Trapman}
\affiliation{Department of Astronomy, University of Wisconsin-Madison, 475 N Charter St, Madison, WI 53706, USA}

\author[0000-0002-5991-8073]{Anibal Sierra}
\affiliation{Departamento de Astronom\'ia, Universidad de Chile, Camino El Observatorio 1515, Las Condes, Santiago, Chile}
\affiliation{Mullard Space Science Laboratory, University College London, Holmbury St Mary, Dorking, Surrey RH5 6NT, UK}

\author[0000-0001-8764-1780]{Paola Pinilla}
\affiliation{Mullard Space Science Laboratory, University College London, Holmbury St Mary, Dorking, Surrey RH5 6NT, UK}

\author[0000-0001-7962-1683]{Ilaria Pascucci}
\affiliation{Lunar and Planetary Laboratory, the University of Arizona, Tucson, AZ 85721, USA}

\author[0000-0002-1199-9564]{Laura M. P\'erez}
\affiliation{Departamento de Astronom\'ia, Universidad de Chile, Camino El Observatorio 1515, Las Condes, Santiago, Chile}

\author[0000-0003-0777-7392]{Dingshan Deng}
\affiliation{Lunar and Planetary Laboratory, the University of Arizona, Tucson, AZ 85721, USA}

\author[0000-0002-7238-2306]{Carolina Agurto-Gangas}
\affiliation{Departamento de Astronom\'ia, Universidad de Chile, Camino El Observatorio 1515, Las Condes, Santiago, Chile}

\author[0000-0003-2251-0602]{John Carpenter}
\affiliation{Joint ALMA Observatory, Avenida Alonso de Córdova 3107, Vitacura, Santiago, Chile}

\author[0000-0002-1103-3225]{Benoît Tabone}
\affiliation{Institut d'Astrophysique Spatiale, Université Paris-Saclay, CNRS, Bâtiment 121, 91405, Orsay Cedex, France}

\author[0000-0003-4853-5736]{Giovanni P. Rosotti}
\affiliation{Dipartimento di Fisica, Università degli Studi di Milano, Via Celoria 16, I-20133 Milano, Italy}

\author[0009-0004-8091-5055]{Rossella Anania}
\affiliation{Dipartimento di Fisica, Università degli Studi di Milano, Via Celoria 16, I-20133 Milano, Italy}

\author[0000-0002-1575-680X]{James Miley}
\affiliation{Departamento de Física, Universidad de Santiago de Chile, Av. Victor Jara 3659, Santiago, Chile}
\affiliation{Millennium Nucleus on Young Exoplanets and their Moons (YEMS), Chile and Center for Interdisciplinary Research in Astrophysics}
\affiliation{Space Exploration (CIRAS), Universidad de Santiago, Chile}

\author[0000-0002-6429-9457]{Kamber Schwarz}
\affiliation{Max-Planck-Institut fur Astronomie (MPIA), Konigstuhl 17, 69117 Heidelberg, Germany}

\author[0000-0002-6946-6787]{Aleksandra Kuznetsova}
\affiliation{Center for Computational Astrophysics, Flatiron Institute, 162 Fifth Ave., New York, New York, 10025}

\author[0000-0002-4147-3846]{Miguel Vioque}
\affiliation{European Southern Observatory, Karl-Schwarzschild-Str. 2, 85748 Garching bei München, Germany}
\affiliation{Joint ALMA Observatory, Alonso de Córdova 3107, Vitacura, Santiago 763-0355, Chile}

\author[0000-0002-2358-4796]{Nicolas Kurtovic}
\affiliation{Max-Planck-Institut fur Astronomie (MPIA), Konigstuhl 17, 69117 Heidelberg, Germany}

\begin{abstract}

The ALMA survey of Gas Evolution in PROtoplanetary disks (AGE-PRO) Large Program aims to trace the evolution of gas disk mass and size throughout the lifetime of protoplanetary disks. This paper presents Band-6 ALMA observations of 10 embedded (Class I and Flat Spectrum) sources in the Ophiuchus molecular cloud, with spectral types ranging from M3 to K6 stars, which serve as the evolutionary starting point in the AGE-PRO sample. While we find 4 nearly edge-on disks ($\geq$70 deg.), and 3 highly inclined disks ($\geq$60 deg.) in our sample, we show that,  as a population, embedded disks in Ophiuchus are not significantly contaminated by more-evolved, but highly inclined sources. We derived dust disk masses from the Band-6 continuum and estimated gas disk masses from the C$^{18}$O $J$=2$-$1 and C$^{17}$O $J$=2$-$1 lines. The mass estimates from the  C$^{17}$O line are slightly higher, suggesting C$^{18}$O emission might be partially optically thick. While the  $^{12}$CO  and $^{13}$CO  lines are severely contaminated by extended emission and self-absorption, the C$^{18}$O  and C$^{17}$O lines allowed to trace the radial extent of the gaseous disks.   From these measurements, we found that the C$^{18}$O $J$=2$-$1 and C$^{17}$O $J$=2$-$1 fluxes correlate well with each other and with the continuum fluxes. Furthermore, the C$^{18}$O and C$^{17}$O lines present a larger radial extension than disk dust sizes by factors ranging from  $\sim$1.5 to $\sim$2.5, as it is found for Class II disks using the radial extension of the $^{12}$CO. In addition, we have detected outflows in three disks from $^{12}$CO observations.

\end{abstract}


\section{Introduction} \label{sec:intro}

In recent years, a large diversity and a high incidence of planetary systems have been discovered, challenging our understanding of planet formation \citep[e.g.][]{Zhu2021}. The variety of planetary architectures observed is impressive,  ranging from very compact systems, like TRAPPIST-1,  with seven rocky planets within 0.07 au \citep{Gillon2017}, to very extended configurations, as in the case of HR 8799, with four gas giants planets at tens of au separations \citep{Marois2010}. However, this heterogeneity is not too surprising considering that circumstellar disk populations in nearby molecular clouds are also widely diverse in masses and sizes  \citep[e.g.][]{Ansdell2016, Barenfeld2016, Pascucci2016, Cieza2019}. On the other hand, the high incidence of exoplanets is more difficult to explain given the many known obstacles that need to be overcome to form a planet in the standard core-accretion model \citep{Drazkowska2023}, including the drift barrier \citep{Brauer2008}, the fragmentation barrier \citep{Dullemond2005}, and the bouncing barrier \citep{Dominik2024}. 

ALMA high-resolution observations have shown that gaps compatible with dynamical clearing by planetary-mass objects are ubiquitous in protoplanetary disks \citep[e.g.,][]{Andrews2020,Bae2023_PPVII}.  Such structures are seen in some very young ($\lesssim$1 Myr) circumstellar disks still embedded in their natal envelopes \citep{ALMAPartnership2015, Yen2016,  Segura2020, Cieza2021, Ohashi2023}. The origin of these gaps is not well established. Still, if protoplanets cause such structures, as is often interpreted \citep{Zhang2018}, this would imply that planet formation must proceed faster than previously thought,  even at the large radii($>$ 10 au). In particular, current planet formation models are unable to reproduce the birth of planets within disk lifetimes of a few Myr at the large radial distances (many tens of au) due to the long dynamical timescales and the relatively low surface densities involved \citep{Morbidelli2020}. Explaining the formation of objects massive enough to form gaps in the embedded phase is even more challenging. However, the direct detection of protoplanets at tens of au in the PDS 70 system \citep{Keppler2018, Haffert2019} demonstrates that giant planets can indeed form at those distances in a few Myr and underscores the current tension between planet formation theory and observations. Alleviating this tension will likely require a better characterization of disk properties as a function of age and significant theoretical work.  

Given the fundamental importance of gas and dust in the formation of different exoplanet types and planetary architectures, studying disk masses and sizes throughout the lifetime of protoplanetary disks is essential for linking disk properties, from the earliest stages, to the final outcome: mature planetary systems.

The ALMA survey of Gas Evolution in PROtoplanetary disks (AGE-PRO) program \citep{Zhang2024} is designed to measure and trace how disk components, especially the gas, change over time using a well-defined sample of 30 disks between $\lesssim$ 0.5 Myr and 6 Myr old located in three different star-forming regions within 160 pc. AGE-PRO selected objects with previous detections of mm continuum and $^{12}$CO line emission and with stellar masses ranging between 0.3 and 0.8 M$_{\odot}$ from the Ophiuchus (embedded sources, $\lesssim$ 1 Myr), Lupus ($\sim$1-3 Myr) and Upper Sco ($\sim$2-6 Myr) star-forming regions. Overall, AGE-PRO aims to provide disk masses and sizes using dust continuum, CO isotopologues, and N$_{2}$H$^{+}$ molecular lines to constrain global disk evolution mechanisms such as viscous spread, photoevaporation,  disk winds, gain growth, and radial drift.

Here, we report ALMA Band-6 observations and results of the 10 disks from the Ophiuchus cloud complex located in the Gould Belt at an average distance of $\sim$138.4$\pm$2.4 pc from the Sun \citep{Ortiz2018}. This nearby region of low-mass star formation has $\sim$300  Young Stellar Objects (YSOs) initially identified by NASA's \emph{Spitzer} Infrared Space Telescope as part of its Legacy Project ``From Molecular Cores to Planet-Forming Disks" \citep{Evans2009}. 

\begin{figure*}
\includegraphics[width=\linewidth]{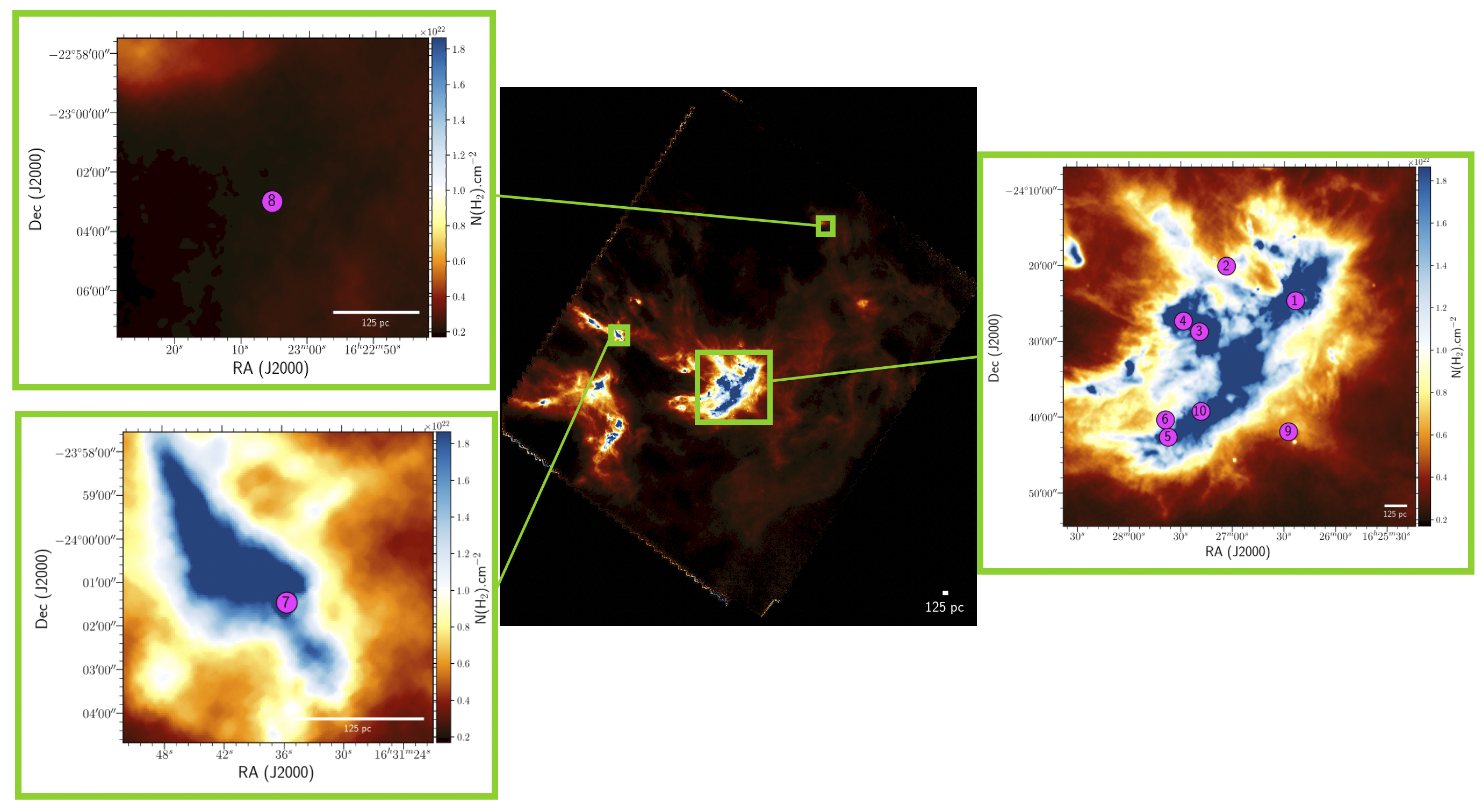}
\figcaption{Column density map of the Ophiuchus star-forming region derived from Herschel data \citep{Ladjelate2020}. The effective  Half Power Beamwidth (HPBW) resolution is 18.2 arcsec. The zoomed-in rectangles show the location of the Ophiuchus targets studied in this paper. Magenta stars represent our 10 selected objects.  Most objects are highly embedded into the molecular cloud, except for Oph 8 with a relatively low visual extinction, see Table \ref{Table:Sample}.
\label{Fig:Density}
}
\end{figure*}

The Ophiuchus cloud complex extends over 10 deg$^{2}$ on the sky and contains objects with a wide range of ages ($\lesssim$ 1-6 Myr) and evolutionary stages, from deeply embedded to diskless stars \citep{Cieza2007, Padgett2008, Esplin2020}. Specifically, the embedded stars (Class I and Flat Spectrum sources) represent $\sim$20$\%$ of the YSO population in the cloud, see Figure \ref{Fig:Density}, and are mostly located in the L1688 cluster \citep{Padgett2008, Ladjelate2020}.

Previous ALMA studies of embedded sources in Ophiuchus have provided interesting results on the dust continuum \citep{Cieza2019,Cieza2021}.  Namely, embedded objects have 
a distribution of dust disk sizes that is indistinguishable from that of Class II sources \citep{Dasgupta2025}, but have mm fluxes that are, on average, a factor of $\sim$3 higher \citep{Williams2019}. They also show gaps that seem narrower and shallower than those seen in their more evolved counterparts \citep{Segura2020, Cieza2021}. However, much less is known about the gas content of these embedded objects. By focusing on molecular line data of the youngest sources in Ophiuchus, this work aims to constrain the initial conditions of embedded disks around stars in the 0.3-0.8 M$_{\odot}$ mass range and to serve as the starting point in the AGE-PRO sample to study disk evolution throughout the lifetime of protoplanetary disks. The details of the sample selection and its limitations are described in Section \ref{sec:sample}. In Sections \ref{Sec:Obs} and \ref{sec:Reduction}, we describe the ALMA Band-6 observations and the data reduction process. In Section \ref{sec:Results}, we present our results regarding disk sizes and masses, which are then discussed in Section \ref{Sec:Discussion}. In Section \ref{Sec:Summary}, we provide a summary of our main conclusions.

\section{Selected Sample in Ophiuchus} \label{sec:sample}

The AGE-PRO program was designed to measure gas disk masses and sizes at three evolutionary phases: the embedded disk phase, the middle age, and the end of the gas-rich phase \citep{Zhang2024}. The Ophiuchus disk sample was selected to represent the embedded disk phase. We employed the following selection criteria for the Ophiuchus sample.

(1) We first chose the IR spectral energy distribution (SED) classes corresponding to protostars still embedded in their natal envelopes (Class I and FS sources) to ensure that our sample is younger than $\sim$1 Myr (\citealt{Evans2009}). Our selection adopts the following definitions from \citet{WilliamsCieza2011}, using spectral slope $\alpha_{IR}$ measured between $\sim$2 and 20 $\mu$m \citep{lada1987, Evans2009}.

\begin{itemize}
    \item Class I~: $\alpha_{IR}$ $>$ +0.3
    \item Flat Spectrum (FS): +0.3  $>\alpha_{IR} >$ -0.3 
    \item Class II: -0.3  $>\alpha_{IR} >$ -1.6 
    \item Class III: -1.6  $>\alpha_{IR}$
\end{itemize}

However, there is no one-to-one correlation between the observational SED classifications and the theoretical ``evolutionary stage". In particular, foreground extinction and disk inclination can affect the SED shape of a YSO, rendering the SED classification an imperfect indicator of the evolutionary stage. Therefore, a protoplanetary disk without a significant envelope (evolutionary stage Class II) might be classified as Class I or FS based on its SED if its foreground extinction and/or inclination are particularly high. The SEDs of the 10 objects in the Ophiuchus sample (from the optical to the mm) are shown in Fig.~\ref{Fig:SED}. The SEDs include data from Gaia, Pan-STARRS \citep{chambers2016}, 2MASS \citep{cutri2003}, WISE  (Cutri et al 2012), \emph{Spitzer} \citep{Evans2009}, \emph{Herschel} \citep{rebollido2015} and ALMA \citep{Cieza2019}. The SEDs show both the observed and extinction-corrected fluxes using the \citet{Fitzpatrick1999} parameterization by applying the extinction law ${R}_v$ = 5.0 \citep{Mathis1990} and with the ${A}_v$ values adopted by AGE-PRO (see Table \ref{Table:Sample} and discussion below). We performed a conservative modeling of the photosphere of each object by fitting only the BT-Settl models \citep{Allard2012} without excess emission from a disk, and scaled assuming an average distance of $\sim$138.4$\pm$2.4 pc \citep{Ortiz2018} for all our selected sample since there are no Gaia distances for the AGE-PRO Ophiuchus targets.

(2) We selected sources with known stellar spectral types between M3-K6, \citep[][Ruiz-Rodriguez et al. In prep]{Furlan2009, McClure2009, McClure2010, Ricci2010, Manara2015}.  As these objects are highly embedded, we chose spectral types derived from optical and infrared spectra rather than photometry data. Unfortunately, the spectral type classification for highly embedded sources is sensitive to the extinction correction applied to the observed spectra which typically are compared to a standard template \citep[e.g.][]{Furlan2009, Manara2015}.  As a result, the resulting spectral types are usually more uncertain than those of Class II sources. Hence, for consistency, we adopt the visual extinction values (${A}_v$) applied as part of the spectral classification\footnote{Since extinction can significantly impact the stellar classification of embedded objects, Appendix \ref{App:Extinction} presents the large range of extinction values -- and stellar luminosities -- that can result from following different approaches.}. Having determined SpT and ${A}_v$ in our sample, and since the SpT is a good proxy for effective temperature ($T_{\rm eff}$), the SpTs were transformed to $T_{\rm eff}$ using the conversion for pre-main-sequence (PMS) stars from \citet{Pecaut2013}. Given the difficulty of determining reliable extinction values (and hence stellar luminosities), we are unable to derive stellar masses and ages from the comparison to theoretical evolutionary tracks. Instead, we adopt the following approach: we assume an age of 1 Myr for all the embedded sources and then use $T_{\rm eff}$ to estimate stellar masses (${M_\ast}$) and luminosities (${L_\ast}$) according to theoretical evolutionary models of \citet{Baraffe2015}. Table \ref{Table:Sample} summarizes the stellar properties adopted by AGE-PRO for the Ophiuchus sample, including ${A}_v$, SpT, ${M_\ast}$, and ${L_\ast}$, together with the SED classifications.

For uniformity and completeness, we also estimated photospheric luminosities (${L_\ast}$) by applying a bolometric correction to the extinction-corrected K-band magnitudes from 2MASS. We opted to use the K-band magnitude as a compromise between the level of extinction (much higher at shorter wavelengths) and IR excess (higher at longer wavelengths). The extinction in K-band (${A}_K$) is calculated by adopting the relation ${A}_K$/ ${A}_v$ = 0.09 \citep{Cieza2005}. The bolometric corrections in the K-band (BCK, see Table \ref{Table:Extinction}) for the corresponding  spectral type are computed from the BT-Settl theoretical evolutionary models of \citet{Baraffe2015}. The resulting ${L_\ast}$ values are listed in Table \ref{Table:Extinction} in the Appendix \ref{App:Extinction}. We emphasize that the luminosity values calculated this way are unreliable because they require a large (and uncertain) extinction correction and do not account for the likely presence of K-band excesses. 

In summary, AGE-PRO adopted the stellar parameters listed in Table \ref{Table:Sample}, which rely on the strong but reasonable assumption that the embedded objects are very young, with statistical ages taken from \citet{Evans2009}.

(3) Sources were further selected from previous detections of mm continuum and CO line emission \citep{Cieza2019}. We excluded sources with known companions or that are in wide-separation binaries ($>$600\,AU), as close binaries may evolve differently due to tidal interactions \citep[e.g.,][]{Cuello2023}. Ten disks were selected to cover the spread of continuum luminosities in the region, including weak to bright sources from \citet{Cieza2019}. 

\begin{figure}
\centering
\includegraphics[width=0.95\linewidth]{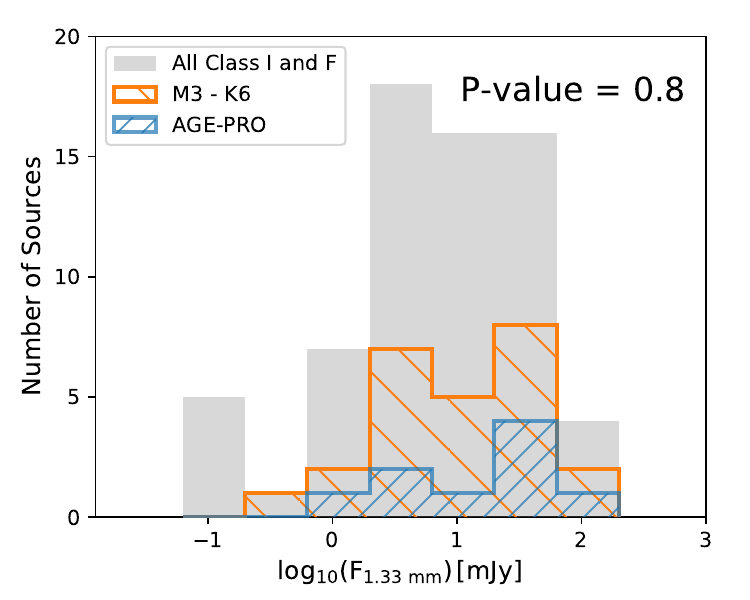}
\figcaption{In gray, histogram of the distribution of flux values at 1.33 mm for the Class I and FS population in the Ophiuchus star-forming region \citep{Cieza2019}. The orange histogram shows all objects ranging from M3 and K6 spectral types, while the AGE-PRO sample is shown in the blue histogram. After performing a two-sample K-S test, we find a P-value of 0.8, indicating that the AGE-PRO Ophiuchus sample is indeed representative of the Class I and FS disks around $\sim$ M3-K6 stars in the Ophiuchus star-forming region. \label{Fig:Hist}
}
\end{figure}

\begin{figure*}
\centering
\includegraphics[width=0.95\textwidth]{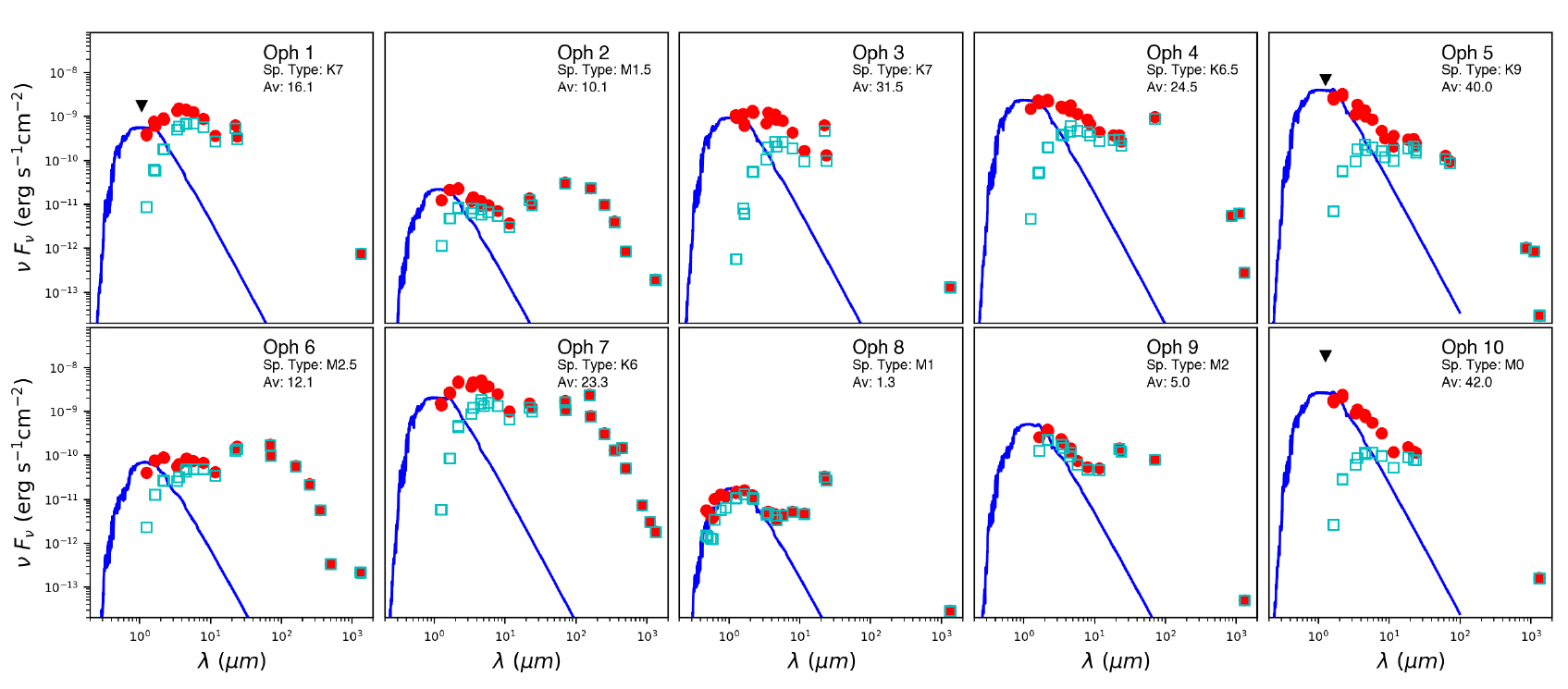}
\figcaption{Spectral energy distributions of the Ophiuchus sources observed in AGE-PRO. The cyan boxes represent the observed optical and IR photometry before correcting for extinction. Red dots show photometric data acquired from the literature and de-reddened using the Av values shown in Table \ref{Table:Sample}. Blue lines are the BT-settl spectra model \citep{Allard2012}, corresponding to the spectral type and normalized to the H-band.  \label{Fig:SED}
}
\end{figure*}

To test whether the AGE-PRO sample is representative of the Class I/FS sources around M3-K6 stars, we perform the two-sample Kolmogorov-Smirnov (KS) test to compare the 1.3 mm flux distributions of the AGE-PRO Ophiuchus sample with those of Class I/FS sources in the same spectral type range in the Ophiuchus region. We find a $P$ = 0.8, indicating the AGE-PRO sample is indistinguishable from the larger sample. Figure \ref{Fig:Hist} displays a histogram of the distribution of 1.3-mm fluxes for the Class I and FS population in the Ophiuchus region with the subset of the spectral type objects between M3 and K6.

Like all the AGE-PRO objects, the Ophiuchus targets are believed to be single (or wide-separation binaries). All known close binaries (a $\lesssim$ 600 au) were excluded. However, as multiplicity surveys are often very incomplete, we provide the available constraints here. 
\textit{Oph 1}, \textit{Oph 4}, \textit{Oph 7}, and \textit{Oph 9} were observed by \citet{Ratzka2005} using the speckle technique. Their survey was sensitive to binaries in the  0.13$''$ to 6.4$''$ separation range and was complete for flux ratios $>$ 0.1. The rest of the Ophiuchus targets were observed with near-IR Adaptive Optics by \citet{Zurlo2020}, and companions with to flux ratios down to $\sim$0.05 can be excluded for separations $>$0.3$''$. We emphasize that these limits correspond to projected separations at the time of observations and that stellar companions can not be ruled out completely from single-epoch data.

\begin{deluxetable*}{l c c c c c c c c c c}
\tabletypesize{\scriptsize}
\label{Table:Sample}
\tablecaption{AGE-PRO Sample: Adopted stellar properties}
\tablehead{
\colhead{2MASS J Name}  & \colhead{Age-Pro ID}  & RA (J2000) & Dec (J2000) & \colhead{SED Class} & \colhead{SpT\tablenotemark{a}} & \colhead{${A}_v$\tablenotemark{b}} & \colhead{$T_{\rm eff}$} & \colhead{$M_\ast$} & \colhead{$\log{L_\ast}$}  &\colhead{Refs.}  
\\
\colhead{}  & \colhead{} &  \colhead{}   & \colhead{} & \colhead{} & \colhead{} & \colhead{(mag.)} &
\colhead{(K)} & \colhead{($M_\odot$)}  & 
\colhead{($L_\odot$)} & \tablenotemark{a} \tablenotemark{b}}
\startdata
 16262357-2424394 & Oph 1\ & 16:26:23.571 & -24:24:40.102 & FS & K7 & 16.1 & 3970 & 0.61 &   -0.06 &  1, 1  \\
16270359-2420054 & Oph 2 & 16:27:03.580 & -24:20:06.018  & FS & M1.5   & 10.1 & 3560 & 0.36 &   -0.38 &  1, 1  \\
16271921-2428438 & Oph 3\ & 16:27:19.196 & -24:28:44.499 & FS & K7   & 31.5  & 3970 & 0.61 &   -0.06 &  2, 2  \\
16272844-2427210 & Oph 4\ & 16:27:28.436 & -24:27:21.790 & FS & K6.5   & 24.5 & 3995 & 0.63&  -0.03 &  3, 3  \\
16273724-2442380 & Oph 5\ & 16:27:37.235 & -24:42:38.543 & I & K9 & 40  & 3880 & 0.55 & -0.12 & 5, 5  \\
16273894-2440206  & Oph 6\ & 16:27:38.931 & -24:40:21.168  & I & M2.5   & 12.1 & 3425 &  0.30 & -0.48 &  1, 1  \\
16313565-2401294 & Oph 7\ & 16:31:35.647 & -24:01:30.048  & I & K6   & 23.3  & 4020 & 0.65 &-0.01 &  4, 4  \\
16230544-2302566 & Oph 8\ & 16:23:05.416 & -23:02:57.553 & I & M1 & 1.3  & 3630 & 0.40 & -0.32 &   5, 5  \\
16262753-2441535 & Oph 9\ & 16:26:27.538 & -24:41:53.527 & FS & M2 &  5.0 & 3490 &  0.33 &   -0.43 &  5, 5  \\
16271838-2439146  & Oph 10\ & 16:27:18.371 & -24:39:15.325  & I & M0   &  44 & 3770 & 0.48 &   -0.21 &   5, 5  \\
\enddata
\tablecomments{Column (1) Lists the 2MASS J name of the target, and Column (2) the AGE-PRO name of the target; Column (3) and (4) RA (right ascension) and Dec (declination); Column (5) SED classification; Column (6) Adopted spectral type; Column (7) Optical extinction used in the spectral classification; Column (8) Effective temperature, Column (10) Stellar masses, and  Column (11) Stellar luminosity values. Stellar masses and luminosities are obtained assuming an age of 1 Myr (see Section \ref{sec:sample}).}
\tablenotetext{a}{Spectral types were taken from (1) \citet{Manara2015}, (2) \citet{Ricci2010}, (3) \citet{Furlan2009}, (4) \citet{McClure2010}, and (5) Ruiz-Rodriguez in Prep. }
\tablenotetext{b}{${A}_v$ values are from (1) \citet{Manara2015}, (2) \citet{Natta2006}, (3) \citet{McClure2009}, (4) \citet{McClure2010}, and and (5) Ruiz-Rodriguez in Prep.}
\end{deluxetable*}

\section{Observations \label{Sec:Obs}}

The AGE-PRO observations of the Ophiuchus star-forming region were conducted in 2022 September-November as part of program 2021.1.00128.L (PI: K. Zhang). This region was observed with the Band 6 receivers using two spectral setups (SS1 and SS2) each covering four basebands. In SS1 and SS2, two SPWs were centered at 234 and 220.5 GHz, each with 128 channels spanning 1.875 GHz, sampling the continuum emission. In SS1, the remaining SPWs were centered at 230.53, 220.40, 219.56, 217.23, 231.32, and 218.22, \,GHz to $^{12}$CO, $^{13}$CO, and C$^{18}$O $J$=2$-$1, and DCN and N$_{2}$D$^{+}$ $J$=3-2, and H$_{2}$CO 3$_{03}-2_{02}$ line, respectively. In SS2, the SPWs were centered at 224.71, 223.88, 239.13, 237.98, 236.45, 235.15, 234.68\,GHz corresponding to the rest frequencies of C$^{17}$O $J$=2$-$1, SO$_{2}$ (6$_{4,2}$-7$_{3,5}$), CH$_{3}$CN (13$_0$--12$_0$), $^{13}$CH$_{3}$OH (5$_{1,4}$--4$_{1,3}$), SO (1$_2$--2$_1$), SO$_{2}$ (4$_{22}$-3$_{13}$), and CH$_{3}$OH (4$_{2,3}$--5$_{1,4}$), respectively. Table \ref{Table:SSs} lists details of these spectral line setups.

\begin{table}
\caption{Beam size parameters using a Briggs parameter of 1.} 
\centering 
\begin{tabular}{l c c c} 
\hline\hline
	   Source  & major axis & minor axis & $\rm PA$  \\
	 \midrule
	   Oph 1 &  $0\farcs{49}$ & $0\farcs{40}$ & $78^\circ$   \\

     Oph 2 & $0\farcs{52}$ & $0\farcs{42}$ & $77^\circ$ \\

     Oph 3\ & $0\farcs{48}$ & $0\farcs{39}$ & $76^\circ$  \\

     Oph 4\  & $0\farcs{46}$ & $0\farcs{37}$ & $76^\circ$ \\

     Oph 5\ & $0\farcs{49}$ & $0\farcs{39}$ & $77^\circ$ \\

     Oph 6\ & $0\farcs{45}$ & $0\farcs{37}$ & $75^\circ$ \\

     Oph 7\ & $0\farcs{43}$ & $0\farcs{36}$ & $76^\circ$ \\

     Oph 8\ & $0\farcs{45}$ & $0\farcs{34}$ & $74^\circ$ \\

     Oph 9\ & $0\farcs{44}$ & $0\farcs{38}$ & $81^\circ$ \\

     Oph 10\ & $0\farcs{46}$ & $0\farcs{36}$ & $73^\circ$  \\
     
     \midrule
	\bottomrule
    \end{tabular}
\label{Table:Beam_size}
\end{table}

To achieve an angular resolution between $\sim$0\farcs3 and $\sim$0\farcs4, each target of our selected sample was observed in a combination of extended (C43-5 and C43-6 hereafter ``LB") and compact (C43-2 or C43-3 hereafter ``SB") configurations. In SS1, it was observed for a total of $\sim$4.3 hrs and $\sim$2.1 hrs in the extended (15m -- 2617m) and compact (14m -- 313 m) configurations, respectively. The SS2 setup required a total of $\sim$9.8 hrs and $\sim$4.9 hrs in the extended (15m -- 1301m) and compact (15m -- 313m) configurations, respectively. \textit{Oph 7} was not part of SS1, instead, we used archival data from the program 2019.A.00034.S. The ALMA large program eDisk supplementary data (2019.A.00034.S) provides $^{12}$CO $J$=2$-$1, $^{13}$CO $J$=2$-$1, C$^{18}$O $J$=2$-$1, and H$_{2}$CO molecular lines, at an angular resolution $\sim$0\farcs4 \citep{Flores2023}. The other observations for SS1 molecular lines, C$^{18}$O $J$=2$-$1, H$_{2}$CO and N$_{2}$D$^{+}$ $J$=3$-$2, were taken from the FAUST large program (2018.1.01205.L), at an angular resolution $\sim$0\farcs3 \citep{Podio2024}. The observational log for AGE-PRO Ophiuchus observations and the Archival data are summarized in Table \ref{Table:Obs_log} in Appendix \ref{App:SSs}.

\section{Data Reduction} \label{sec:Reduction}

To extract continuum and line images, we followed the methodology described in the AGE-PRO overview paper by \citet{Zhang2024}. Using the CASA pipeline version 6.4.3, initial spectrally averaged continuum measurement sets were created by flagging any possible line emission in each spectral window and averaging into a maximum 125 MHz channel width. Flagged channels had velocities $\pm 20$ km s$^{-1}$ from the systemic velocity of each target (on average, the systemic velocity of the Ophiuchus sources is 2.3 km s$^{-1}$). In the case of SS2, we flagged channels with velocities between -140 and 160 km s$^{-1}$ around the $^{13}$CO lines in the continuum SPW. We visually inspected data sets for astrometric alignment to detect any special features in the preliminary continuum images. As we did not find any features, we shifted the averaged continuum measurement sets according to the peak emission based on 2D Gaussian fitting, thus aligning all the EBs to a common phase center. For flux-scale alignment, we first deprojected the continuum visibilities using the inclination and position angle estimated from preliminary images. We then checked that, at short UV-distances which are less affected by atmospheric phase decoherence, the deprojected continuum visibilities did not have a flux scale difference higher than 10$\%$, which was the case for the majority of the observations. In a few cases with differences higher than 10$\%$, we rescaled the flux to the dataset that was closer in time to the flux calibration observations.

Overall, self-calibration of the continuum visibilities was performed following \citet{Zhang2024}, except when noted below, and only for those sources with a continuum of signal-to-noise ratio (SNR) larger than about 20. Initially, we self-calibrated the SB data to concatenate each of the three Execution Blocks (EBs). For SS2, we noticed that the use of the first EB of the SB observations affected the final image by reducing the SNR, therefore, we opted to only concatenate the second and third EBs. We then concatenate the already self-calibrated concatenated SB visibilities with the LB visibilities that only have ALMA pipeline calibration. However, different from the standard procedure outlined in \citet{Zhang2024}, when using the CASA \texttt{gaincal} task in the self-calibration of the SB$+$LB data, we selected the option \texttt{gaintype=`G'}, which computes gain solutions independently for each XX and YY polarizations, instead of the standard \texttt{gaintype=`T'}, which averages both polarizations to increase SNR. We opted for this option to correct phase offsets between polarizations that were not completely removed during the ALMA pipeline calibration. The final combined visibilities were then self-calibrated and imaged with the \texttt{tclean} task using an elliptical mask around the source --and its companions if necessary-- cleaning down to a threshold of $2\times$ the RMS of the image at each iteration, and a Briggs robust parameter between 0.5 and 1, which resulted in average beam sizes of $\sim$ 0\farcs25 and $\sim$ 0\farcs5, respectively. The resulting beam size parameters using a Briggs value of 1 are shown in Table \ref{Table:Beam_size}. The mask aspect ratio and position angle were selected using the inclination and position angle of each disk estimated from a Gaussian fit to the continuum images. Table~\ref{Tab:Continuum} lists the basic parameters and resulting properties of the final 220\,GHz continuum images.  A gallery of the 220 GHz continuum images is shown in Figure \ref{Fig:Mom0}. For simplicity, we opted to present 234 GHz continuum images and their parameters in Appendix \ref{App:Cont_Mom05}.

We also synthesized channel maps of the emission lines outlined above by following the standard procedure adopted in the AGE-PRO program and presented in \citet{Zhang2024}.  The self-calibrated visibilities were prepared by splitting them into individual measurement sets for each targeted line and then continuum-subtracted. The line images were also produced with the tclean task. Unfortunately, due to the nature and characteristics of the very young star-forming region, the Ophiuchus AGE-PRO data generally present high levels of cloud contamination. As a result, the CLEAN masks were built at the source position automatically during an initial cleaning process using the algorithm ``auto-multithresh'' with a noise threshold of 2.5 mJy, sidelobethreshold of 2.0, a noisethreshold of 3.75, a lownoisethreshold of 1.0, a smoothfactor of 1.5, a smallscalebias of 0.45 at scales of 0, 4, and 12. Once the masks were built for each line, we performed another cleaning process using the created masks while applying a threshold of 1 rms. The data sets are generally not sensitive enough to build channel maps of the emission line at the best available resolution, meaning the use of a modest Briggs parameter of 1, while the circularization of the final beam is obtained by controlling the radial weighting of visibilities in the uv-plane. The final data cubes were imaged in LSRK velocity channels at roughly the native channel spacing (0.2 km s$^{-1}$). The resulting clean beams have FWHM $\sim$0.5$^{"}$. Galleries of the CO isotopologues images are shown in Figures \ref{Fig:Mom0}, and their channel maps are shown in Appendix \ref{App:Channel}. Other line observations from SS1 and SS2 are displayed in Appendix \ref{App:Cont_Mom05}. These products can be downloaded from the AGE-PRO official website\footnote{AGE-PRO website: \url{https://agepro.das.uchile.cl}}). There were only a few detections of other molecules beyond CO isotopologues. This paper is focused on the CO gas content.

\section{Results} \label{sec:Results}

In this section, we report the dust continuum results 
(\S  \ref{Sec:Cont}), present the Moment-0 and Moment-1 maps of the CO isotopologue lines and used in the study of the gas (see \S  \ref{Sec:Moments}),  and report disk size measurements from the extracted from the observations of dust and gas (see \S  \ref{Sec:Radial_Profiles} and \S  \ref{Sec:Flux_Radii}).
Moreover,  we derive the corresponding dust masses from the continuum (see \S \ref{Sec:Dust_masses}), and estimate minimum gas masses based on simple slab models of C$^{18}$O and C$^{17}$O line fluxes. For more realistic gas mass estimations based on modeling of disk thermal and CO abundance structures, please see \citet{Trapman2024b}.

\begin{figure*}
\centering
\gridline{\fig{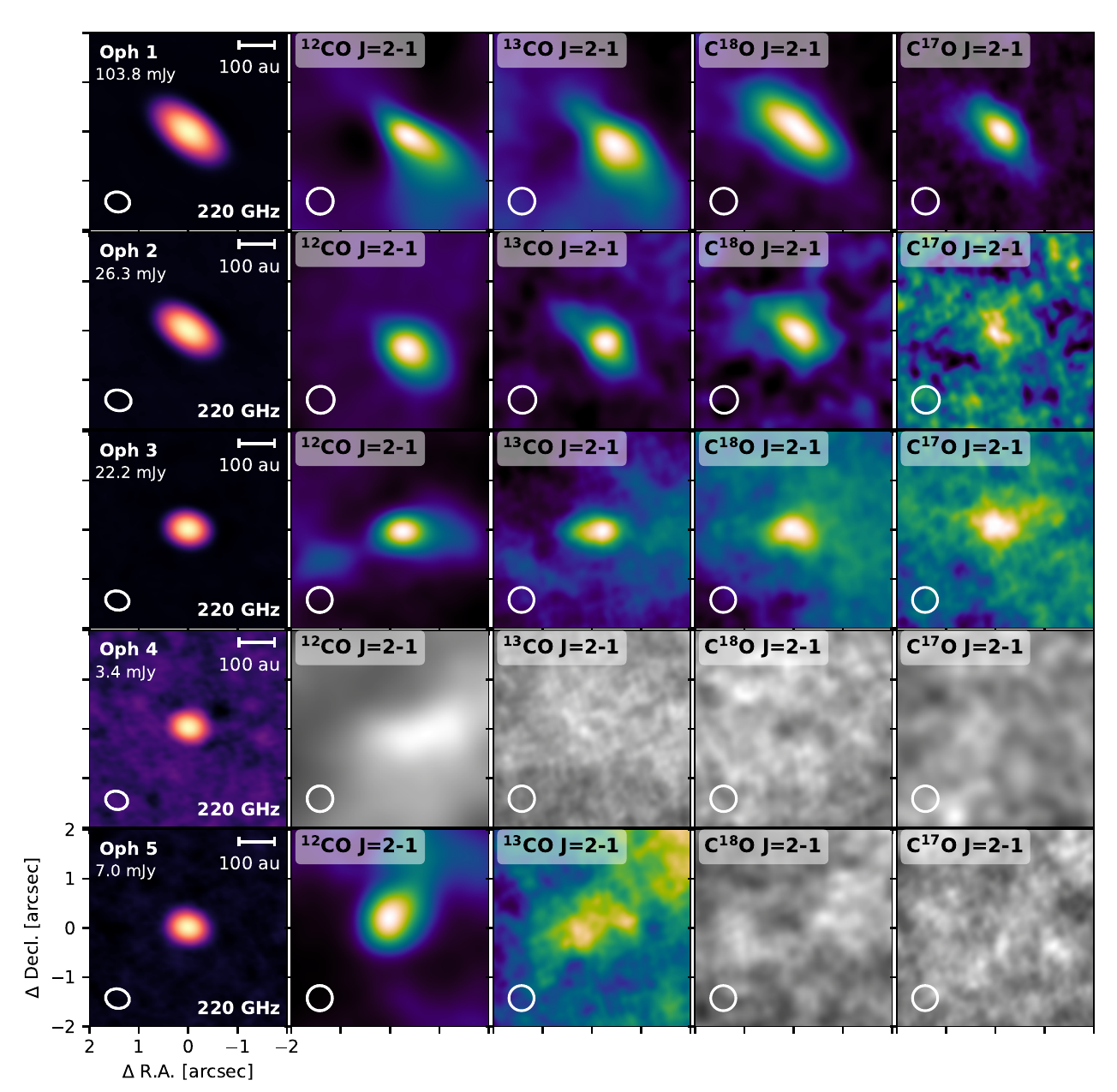}{0.5\textwidth}{}
   \fig{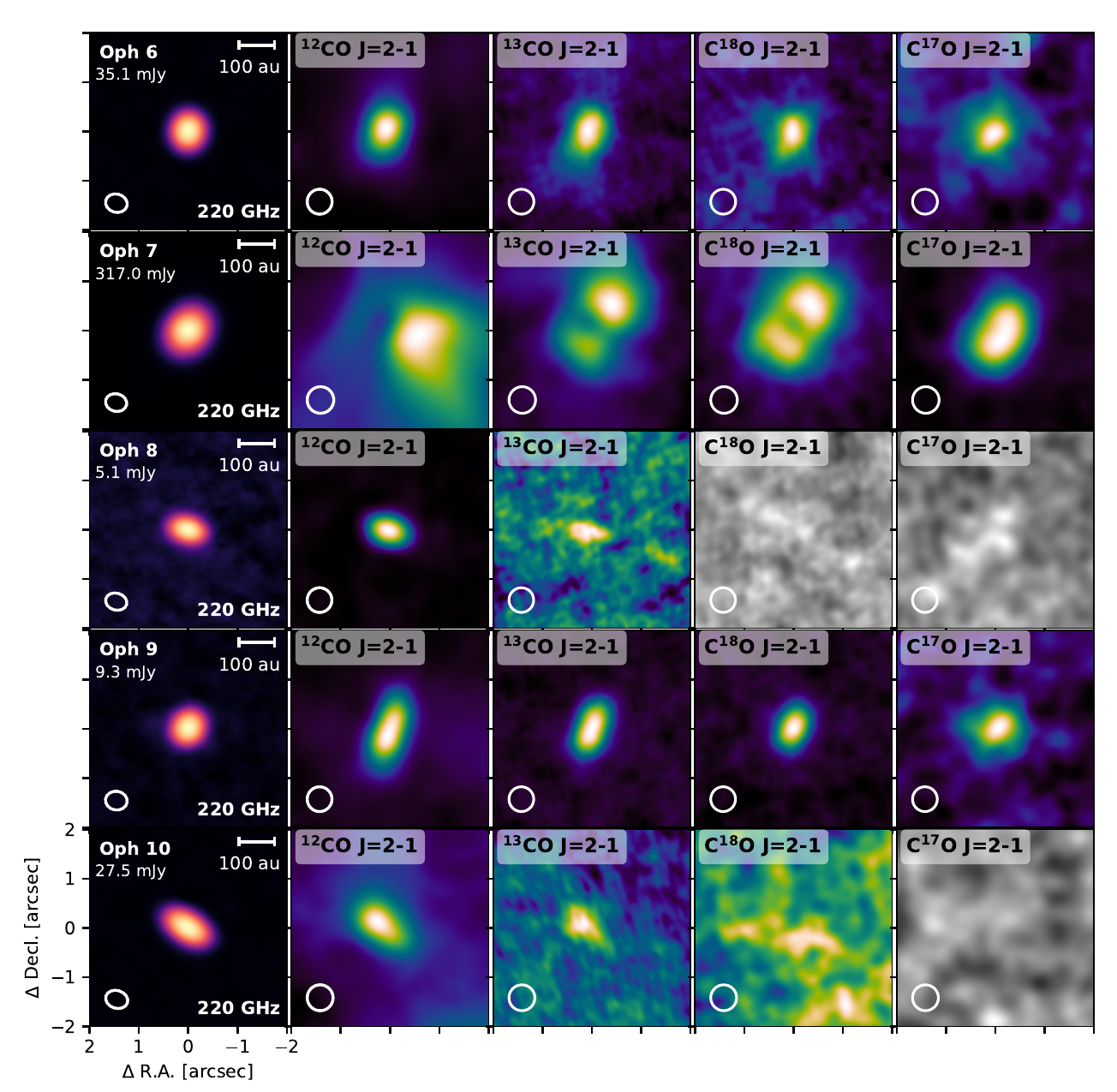}{0.5\textwidth}{}}
\figcaption{220 GHz continuum images and CO isotopologue Moment-0 maps of the AGE-PRO Ophiuchus targets. These ALMA products are obtained with a robust parameter of 1 in the velocity ranges displayed in Table \ref{Table:Mom_parameters} and generate an average beam size of 0.5$^{"}$. Disk-integrated continuum flux is indicated at the left upper side. The resulting beam sizes are represented by the white circles at the lower left corner. Colored and black-and-white images indicate detections and non-detections in our sample, respectively. For \textit{Oph 4}, although $^{12}$CO emission is detected around the source, the moment 1 map shows that it does not show Keplerian rotation. 
\label{Fig:Mom0}
}
\end{figure*}

\subsection{Continuum}\label{Sec:Cont}

\begin{figure}
\centering
\includegraphics[width=0.9\linewidth]{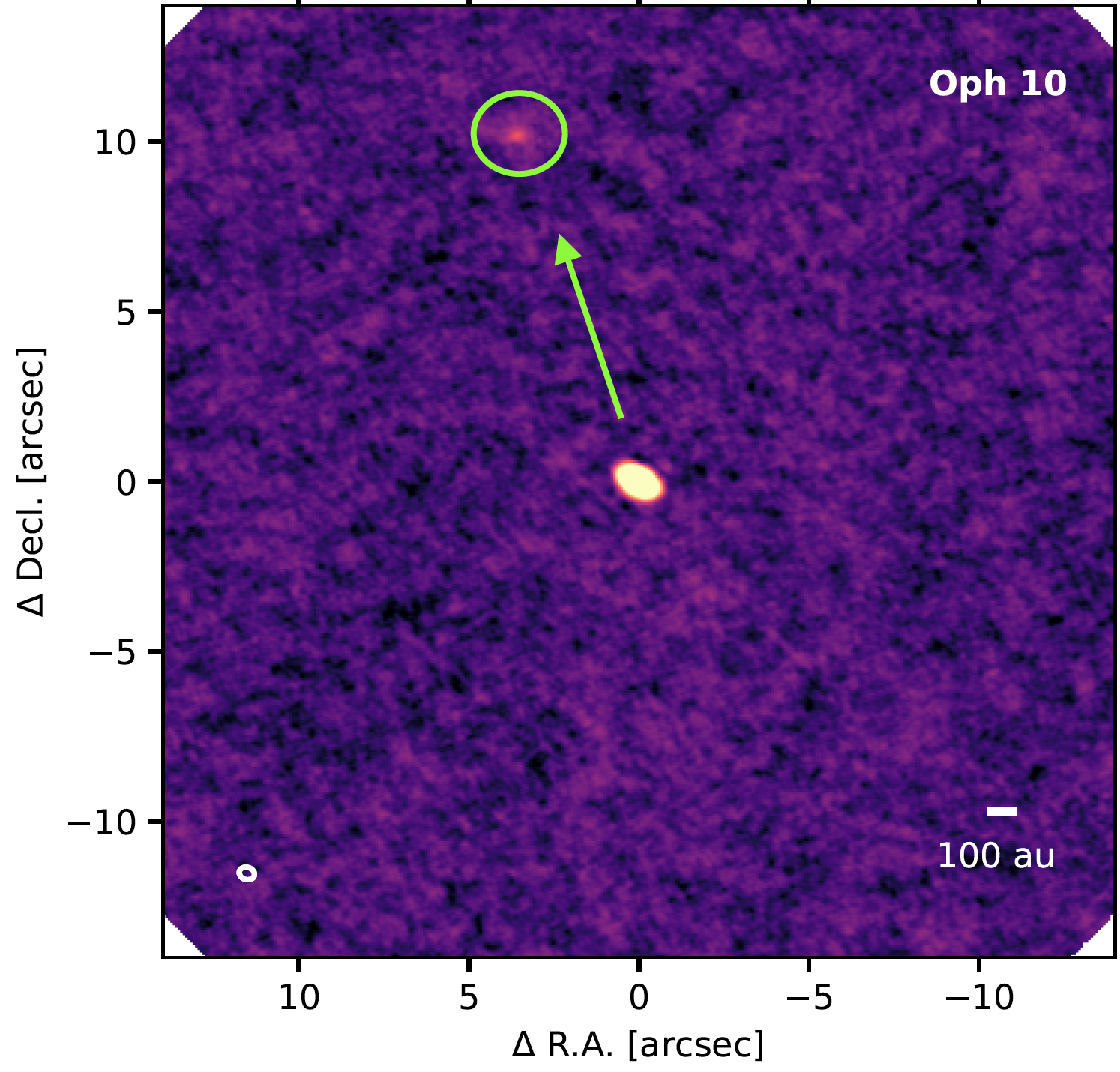}
\figcaption{Zoom out of Fig. \ref{Fig:Mom0} displaying the new object detected in the FOV of \textit{Oph 10} at 220 GHz continuum. The green arrow indicates the position of the new object. The white ellipse in the lower left corner represents the resulting beam size of the continuum image. 
\label{Fig:Binaries}
}
\end{figure}

The 220 GHz continuum images of the AGE-PRO Ophiuchus sources are displayed in the left panel of Figure \ref{Fig:Mom0}. These images are obtained with a robust value of 1 to match the weighting applied in the line data 
cubes, see section \ref{sec:Reduction}. All the 220 GHz continuum emission images have an average beam size of 0.5$^{\prime \prime}$ (70 au). From these observations, one new object was detected close to \textit{Oph 10} at a projected distance of $\sim$11$^{\prime \prime}$ (1600 au) north-east, located at $\sim$ R.A. 16:27:18.63 and Dec. -24:39:05.2, and non-resolved with a peak flux of $\sim$ 0.6 mJy (see Figure \ref{Fig:Binaries}, green circle).
 
Furthermore, observations toward \textit{Oph 1} and \textit{Oph 4} detected two stellar objects already known at distances of $\sim$8$^{\prime \prime}$ (1100 au) south-east \citep[GSS 32;][]{Itoh2015}, and 10$^{\prime \prime}$ (1400 au) north-east \citep[SIO2011;][]{Shirono2011}, respectively; while in the FOV of \textit{Oph 3} two objects are detected at 18.3 $^{"}$ (2500 au) and 16.5$^{\prime \prime}$ (2300 au) south-west \citep[WL 4, WL 5;][]{Wilking2015, Ortiz2018}.

Table \ref{Tab:Continuum} presents inclination and P.A. estimates derived through 2-D Gaussian fittings from the 220 GHz continuum observations in the image plane. We found that 8 disks are well-resolved while \textit{Oph 4} and \textit{Oph 5} are marginally resolved, where both disks are fairly compact. In the Ophiuchus sample, \textit{Oph 1}, \textit{Oph 2}, \textit{Oph 9}, and \textit{Oph 10} display inclinations larger than 70 deg., while \textit{Oph 3}, \textit{Oph 6}, and \textit{Oph 8} are inclined with values larger than 60 deg. 

Since these relatively high inclinations might impact the SED classifications listed in Table \ref{Table:Sample} with a possible misclassification from FS and Class~I objects to later phases, we revisit these objects in Section \ref{Sec:Inclination}.

\subsection{Moment Maps of CO isotopologues}
\label{Sec:Moments}

\begin{figure*}
\centering
\gridline{\fig{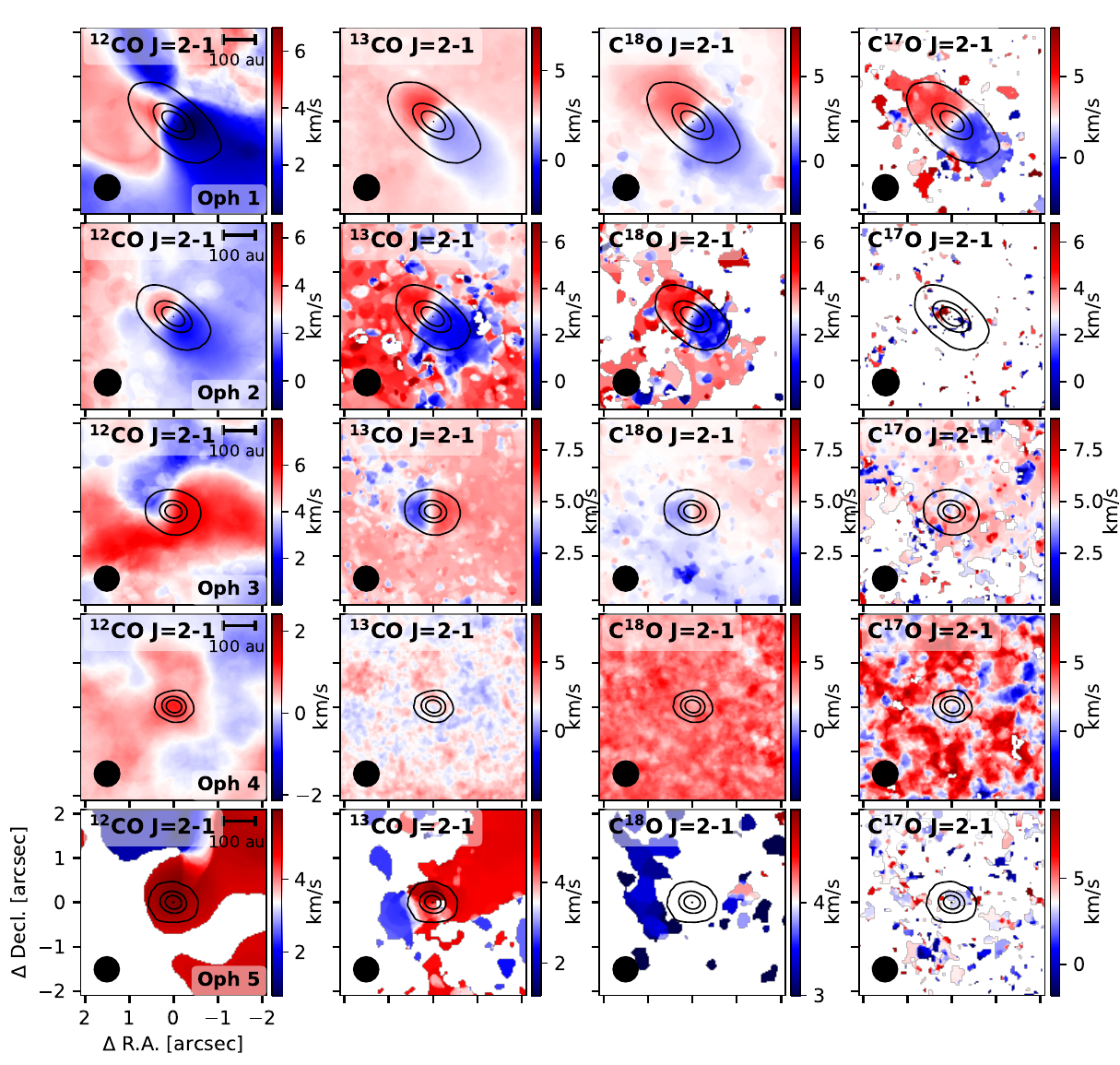}{0.5\textwidth}{}
   \fig{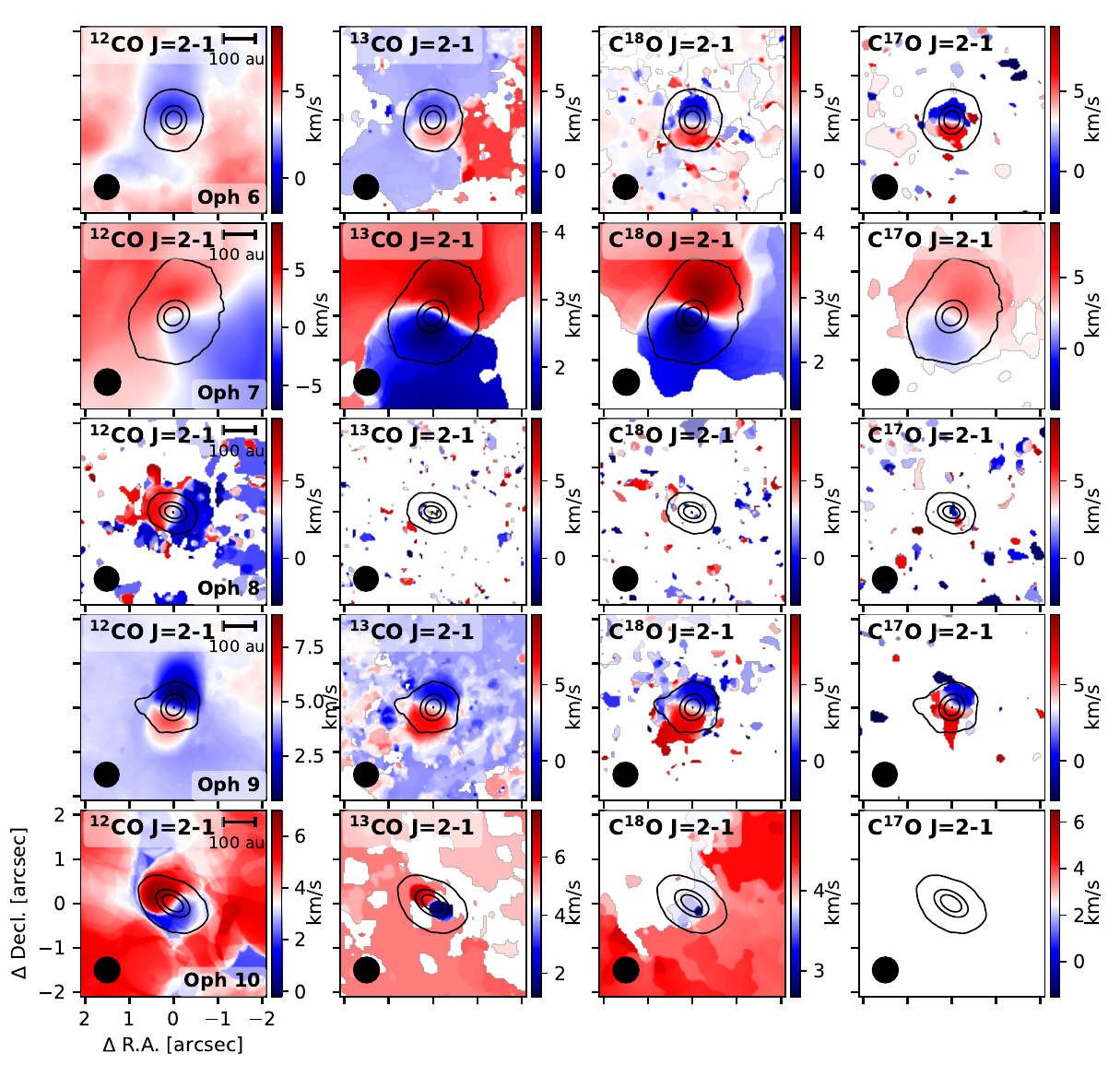}{0.5\textwidth}{}
            }
\figcaption{CO isotopologue Moment-1 maps of the AGE-PRO Ophiuchus targets. These ALMA products are obtained with a robust parameter of 1, generating an average beam size of 0.5$^{"}$. The resulting beam sizes are represented by the black dots at the lower left corner. The moment 1 maps were created by integrating emission having intensity levels as necessary to extract any rotational motions over the same velocity range as the moment 0 maps, see Table \ref{Table:Mom_parameters}. These moment 1 maps were computed with \texttt{bettermoments}. The contours of the 220 GHz continuum are drawn every 3$\sigma$ (1$\sigma$ = 3$\times$10$^{-2}$ mJy beam$^{-1}$).
\label{Fig:Mom1}
}
\end{figure*}

To determine the flux and size of the gaseous disk, we first proceeded to build velocity-integrated intensity (Moment-0) and intensity-weighted velocity (Moment-1) maps of the CO isotopologues. As these sources are highly embedded, we initially extracted CO velocity-stacked spectra using \texttt{GoFish}. We adopted an elliptical region 2$^{\prime \prime}$ in diameter and with  P.A. and inclination derived from the continuum. We then visually inspected the velocity distribution of the cold molecular cloud traced by $^{12}$CO, and any features such as self-absorption, to define velocity ranges for emission integration in each case (see green-shaded areas in Figure \ref{Fig:Spectra} and Table \ref{Table:Mom_parameters} in Appendix \ref{App:VelRange}). We estimate RMS levels from channels free of emission and by assuming a constant noise value (spatially and spectrally) to establish whether or not a given molecule line is detected at levels higher than 5$\sigma$. These values are also used to estimate uncertainties of the Moment-0 and Moment-1 maps and to apply any clipping for each case, if possible. The exact $\rm V_{lsrk}$ values are uncertain, but we integrated over velocity ranges centered at $\sim$2.3 km s$^{-1}$, the average systemic velocity of the Ophiuchus sources. For Moment-0 maps, we integrated without any clipping to make unbiased images by using \texttt{bettermoments} \citep{Bettermoments}. For Moment-1 maps, we used the same velocity ranges as those in Moment-0 maps, but applied clipping values as necessary to extract and/or distinguish any velocity patterns rising from the disk or that emission that belongs to the molecular cloud, see Table \ref{Table:Mom_parameters}. Figure \ref{Fig:Mom0} and Figure \ref{Fig:Mom1} show Moment-0 maps and Moment-1 maps of the CO isotopologues toward a central 2-arcsec region of each Ophiuchus source, respectively.

We emphasize that the line observations are highly affected by the deeply embedded nature of the young stellar objects. A closer look at the channel maps (Appendix \ref{App:Channel}) shows that even though in some cases the molecular line emission coincides with the location of the continuum, it is not straightforward to establish whether this emission is associated with the disk
or with an extended component  (an envelope, an outflow, or the molecular cloud). Furthermore, the surrounding cold material prevents the use of Keplerian masks to generate moment maps from the AGE-PRO Ophiuchus data products. This can be seen in the Moment-1 maps of the CO molecules, Figure \ref{Fig:Mom1}, whose blue- and red-shifted emission features are not as prominent in the kinematics while in a few cases, Keplerian-like rotation is only detected in optically thinner tracers such as C$^{17}$O.

Based on the moment maps, we detect $^{12}$CO $J$=2$-$1 emission (at $\geq$ 7 $\sigma$) and $^{13}$CO $J$=2$-$1 emission (at $\geq$ 5 $\sigma$) in all objects, except for \textit{Oph 4}, which is not detected in any of the molecular lines or it is not clear the emission origin. Similarly, C$^{18}$O $J$=2$-$1 is detected in 6 objects with an additional tentative detection for \textit{Oph 10} of only a $\sim$ 4 $\sigma$. At lower intensity levels, C$^{17}$O $J$=2$-$1 emission is detected in 6 objects with values of $\geq$ 5-7 $\sigma$, except for \textit{Oph 2} with only a $\sim$ 3 $\sigma$ detection. The detections below $\sim$ 3 $\sigma$ were converted to upper limits. The AGE-PRO observations also reveal outflows/streamers associated with \textit{Oph 3}, \textit{Oph 5}, and \textit{Oph 7}. Figure \ref{Fig:Mom0357} in the Appendix \ref{App:Outflows}  shows a zoom-out Moment-0 map of \textit{Oph 3}, \textit{Oph 5}, and \textit{Oph 7} displaying the outflows/streamers associated with these sources.

As individual Ophiuchus targets display different characteristics traced by $^{12}$CO, $^{13}$CO, C$^{18}$O, and C$^{17}$O lines, we briefly describe each source to pave the analysis presented in the next sections:

\paragraph{\textbf{\textit{Oph 1}}}

$^{12}$CO, $^{13}$CO, C$^{18}$O, and C$^{17}$O emission are detected towards \textit{Oph 1}. However, optically thick tracers such as $^{12}$CO and $^{13}$CO lines are contaminated by the surrounding material of this highly embedded YSO, whose emission is mostly detected in the blue-shifted regime. This effect can be seen in the $^{12}$CO and $^{13}$CO channel maps (Appendix \ref{App:Channel}) with absorption by foreground material and envelope emission being more resolved out, see Figures \ref{Fig:Oph112CO} and \ref{Fig:Oph113CO}. In \textit{Oph 1}, optically thinner tracers are less affected by foreground material allowing the detection of the characteristic blue- and red-shifted emission that originates from the disk. That is the case of the C$^{17}$O and C$^{18}$O emission that trace disk material (Figs. \ref{Fig:Mom1}, \ref{Fig:Oph1C18O} and \ref{Fig:Oph1C17O}). In addition, C$^{18}$O is also detected at larger distances from the central source, however, this emission likely belongs to the molecular cloud as is detected at velocities close to the $\sim$$\rm V_{sys}$ (2.3 $\pm$ 0.3 km.s$^{-1}$). 

\paragraph{\textbf{\textit{Oph 2}}}

Similarly to \textit{Oph 1}, \textit{Oph 2} also displays $^{12}$CO, $^{13}$CO, C$^{18}$O, and C$^{17}$O emission but at lower intensity levels and less affected by the colder molecular cloud. The relatively weak $^{12}$CO and $^{13}$CO emission trace disk and envelope materials (Figs. \ref{Fig:Oph212CO} and \ref{Fig:Oph213CO}), while C$^{17}$O and C$^{18}$O originates solely from the disk (Figs. \ref{Fig:Oph2C18O} and \ref{Fig:Oph2C17O}) with C$^{17}$O being detected only at $\sim$4$\sigma$ levels from the integrated flux.

\paragraph{\textbf{\textit{Oph 3}}}

The \textit{Oph 3} disk and surroundings can be traced by $^{12}$CO, $^{13}$CO, C$^{18}$O, and C$^{17}$O emission, however, it is difficult to make a clear distinction from disk' and envelope's emission as can be seen in their channel maps, Figures \ref{Fig:Oph312CO}, \ref{Fig:Oph313CO}, \ref{Fig:Oph3C18O}, and \ref{Fig:Oph3C17O}. As mentioned above, \textit{Oph 3} seems to be associated with an outflow traced mainly by $^{12}$CO with a P.A. of $\sim$ 4 deg at velocities ranging from $\sim$ 5.6 and 8.0 km.s$^{-1}$, see Figure \ref{Fig:Mom0357}a. The red-shifted feature, perpendicular to the disk's PA of $\sim$ 94 deg., appears to be a cavity (with a wide opening angle) that extends up to $\sim$ 1400 au.
In addition, there is extended $^{12}$CO emission, marked by a red arrow in Figure \ref{Fig:Mom0357}a, at $\sim$ 7$^{"}$ (1000 au) southward of the central source, that seems to be linked to the outflow cavity as material being expelled.

\paragraph{\textbf{\textit{Oph 4}}}

From the Ophiuchus sample, \textit{Oph 4} seems to be one of the most embedded YSOs because although $^{12}$CO and $^{13}$CO lines are detected, these lines are tracing only the molecular cloud towards the \textit{Oph's 4} location, see Figs. \ref{Fig:Oph412CO} and \ref{Fig:Oph413CO}). This also might prevent the detection of weaker lines such as C$^{18}$O, and C$^{17}$O in the system, if any (Figs. \ref{Fig:Mom0} and \ref{Fig:Spectra} in Appendix \ref{App:VelRange}). Moreover, \textit{Oph 4} also has the faintest dust continuum of the sample. Given the correlation between continuum emission and C$^{17}$O and C$^{18}$O fluxes, see Section \ref{continuum_vs_gas}, the weak continuum might explain the lack of detection in these
lines.  

\paragraph{\textbf{\textit{Oph 5}}}

Due to the embedded nature of \textit{Oph 5}, the envelope and partially a disk are represented by the structure observed in the $^{12}$CO and $^{13}$CO emission, see Figures \ref{Fig:Oph512CO} and \ref{Fig:Oph513CO}. In particular, the $^{12}$CO and $^{13}$CO emission reveal outflow material associated with this source with a P.A. of $\sim$170 deg, and perpendicular to the disk's PA of $\sim$80 deg, see Figure \ref{Fig:Mom0357}b. These structures are detected in a velocity range between -0.6 and 1.6 km.s$^{-1}$ in the blue-shifted regime, and at higher intensity levels, 4.6 and 6.6 km.s$^{-1}$ in the red-shifted regime. As expected, emission close to the $\sim$$\rm V_{sys}$ (2.6 $\pm$ 0.4 km.s$^{-1}$) are affected by the embedded nature of the source, preventing any detection (Fig. \ref{Fig:Oph512CO}). In addition, the elongated feature located at the southern-east side of the disk and at velocities close to 5.0 km.s$^{-1}$ might be related to accretion streamers feeding the central protostar or disk. Unfortunately, there are no complementary features from other line observations to confirm the nature of the possible streamer.

\paragraph{\textbf{\textit{Oph 6}}}

$^{12}$CO, $^{13}$CO, C$^{18}$O, and C$^{17}$O emission are also detected towards \textit{Oph 6}. In particular, $^{12}$CO and $^{13}$CO emission originate from the central source and envelope with a very deep absorption seen at $\rm V_{LSR}$ between 2.6 and 5.0 km.s$^{-1}$, see Figures \ref{Fig:Oph612CO} and \ref{Fig:Oph613CO}. The absorption is due to cold foreground material which is detected in the structure of the C$^{18}$O, and C$^{17}$O emission at $\rm V_{LSR}$ between $\sim$2.8 and 3.8 km.s$^{-1}$, see Figures \ref{Fig:Oph6C18O} and \ref{Fig:Oph6C17O}. Out of this range, the disk is traced by C$^{18}$O, and C$^{17}$O blue- and red-shifted emission.

\paragraph{\textbf{\textit{Oph 7}}}

From the Ophiuchus sample, \textit{Oph 7} -- a.k.a. IRS 63 -- is the brightest target associated with strong outflows detected in the $^{12}$CO, $^{13}$CO, and C$^{18}$O emission, see Figures \ref{Fig:Oph712CO}, \ref{Fig:Oph713CO}, and \ref{Fig:Oph7C18O}. From $^{12}$CO observations, a shell-like bipolar outflow is revealed with structures alike an expanding shell in the southern cavity traced by the blue-shifted emission. Additionally, the envelope and outflow material (southern cavity) are detected in the $^{13}$CO and C$^{18}$O observations. On the other hand, C$^{17}$O emission is more compact tracing material towards the central object and the disk (Fig. \ref{Fig:Oph7C17O}). Of particular interest is the potential accretion streamers traced by $^{13}$CO, C$^{18}$O, and C$^{17}$O emission. see Fig. \ref{Fig:Mom0_streamer} whose positions coincide with those streamers previously detected by \citet{Flores2023}. A more detailed description and analysis of the revealed structures of \textit{Oph 7} can be found in \citet{Flores2023}.

\paragraph{\textbf{\textit{Oph 8}}}

\textit{Oph 8} is the least embedded Ophiuchus source with compact $^{12}$CO emission tracing the disk. At 3 mJy beam$^{-1}$ $\sigma$ level, weak $^{13}$CO emission is detected at the position of the source, see Figures \ref{Fig:Oph812CO} and \ref{Fig:Oph813CO}, while C$^{18}$O, and C$^{17}$O are not detected in the data cube products. 

\paragraph{\textbf{\textit{Oph 9}}}

In \textit{Oph 9}, CO isotopologues are also affected by the cold foreground material. That is, $^{12}$CO, $^{13}$CO, and C$^{18}$O are absorbed by the extended molecular cloud material in the velocity range between $\sim$1.8 and 4.2 km.s$^{-1}$. This effect is most prominent in the $^{12}$CO observations. Nevertheless, disk material is detected by $^{12}$CO, $^{13}$CO, C$^{18}$O, and C$^{17}$O lines, see Figures \ref{Fig:Oph912CO}, \ref{Fig:Oph913CO}, \ref{Fig:Oph9C18O}, and \ref{Fig:Oph9C17O}.

\paragraph{\textbf{\textit{Oph 10}}}

Observations toward \textit{Oph 10} are highly affected by the Ophiuchus molecular cloud material, where $^{12}$CO, $^{13}$CO, and C$^{18}$O emission trace extended material only allowing a marginal detection of the disk. In contrast, C$^{17}$O only traces foreground material and no disk detection, see Figures \ref{Fig:Oph1012CO}, \ref{Fig:Oph1013CO}, \ref{Fig:Oph10C18O}, and \ref{Fig:Oph10C17O}.

\subsection{Radial Intensity Profiles}
\label{Sec:Radial_Profiles}

\begin{figure*}
\centering
\gridline{\fig{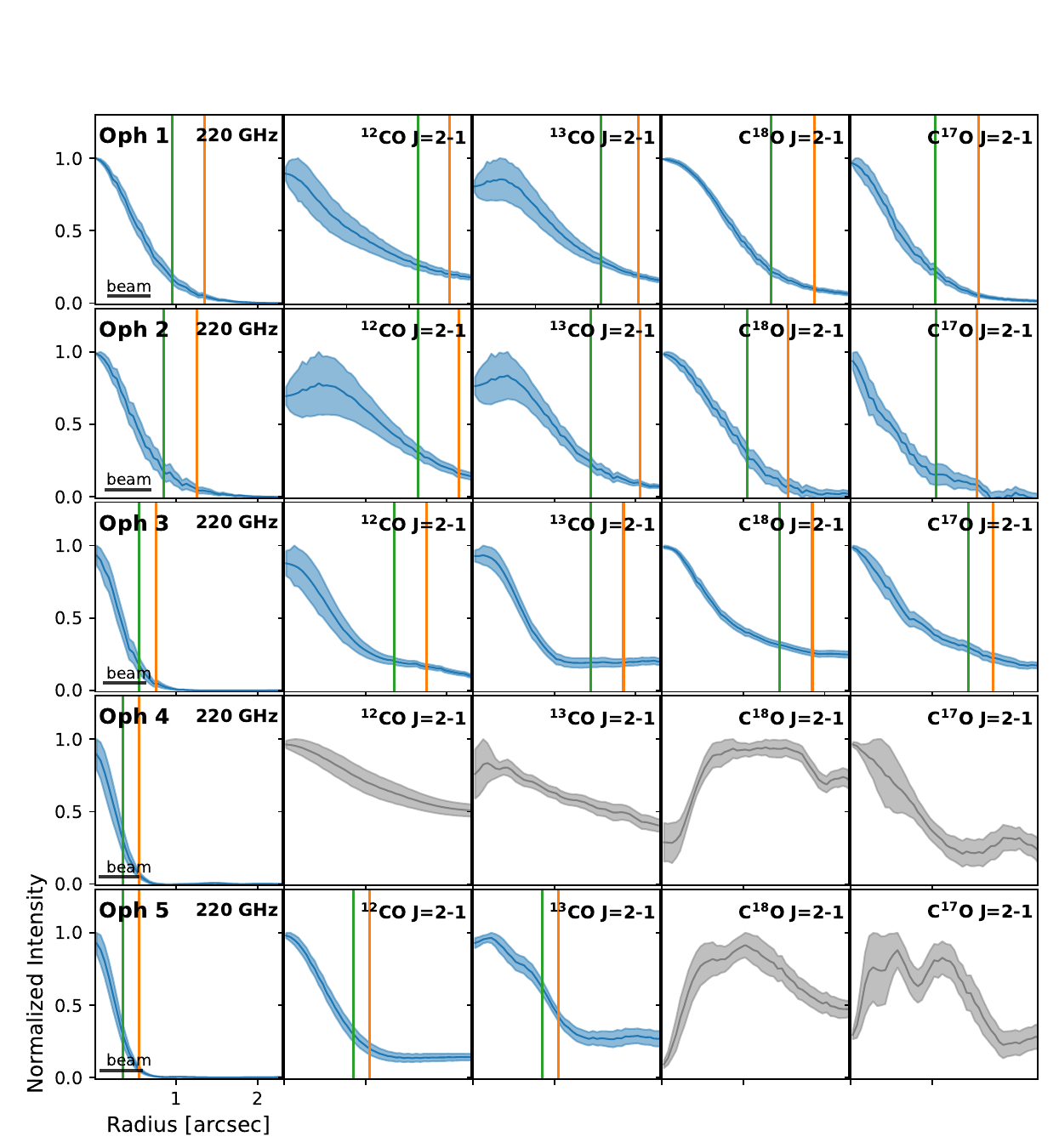}{0.5\textwidth}{}
   \fig{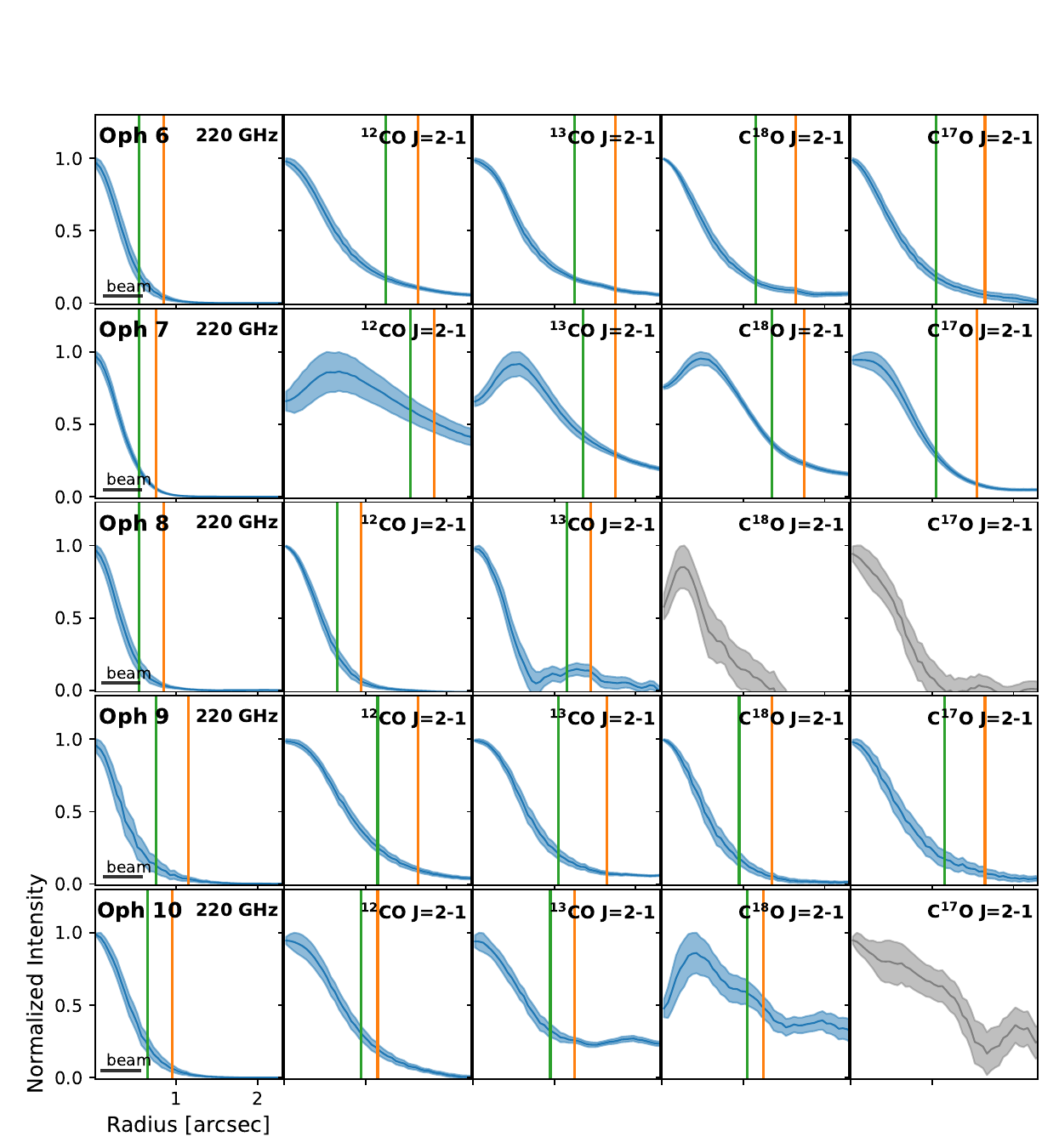}{0.5\textwidth}{}}
\figcaption{Azimuthally averaged deprojected and normalized radial profiles for the 220 GHz continuum emission and CO isotopologues. Blue curves correspond to the disk detections in our sample while grey color curves correspond to the CO lines believed to be mostly associated with extended emission from the envelope and the cloud. Shaded areas are uncertainties estimated as the standard deviation per annulus divided by the square root of the number of beams in the annulus. Green and Orange vertical lines represent radii at 68$\%$ and 90$\%$ of the total flux, respectively.
\label{Fig:RP}
}
\end{figure*}

To uncover possible substructures and compare CO lines and continuum emission in the Ophiuchus sample, we also constructed radial profiles by azimuthally averaging the peak intensities for radial annuli of equal width of one-fourth of the beam major axis using the \texttt{radialprofile} function in the Python package \texttt{GoFish} \citep{GoFish}. In the case of CO lines, Moment-0 maps generated from the resulting cubes are deprojected using the inclination and P.A. estimated from the continuum emission, see Table \ref{Tab:Continuum}. Figure \ref{Fig:RP} shows CO and continuum radial peak intensities normalized to the peak value.

At an average resolution of $\sim$0.5$^{\prime \prime}$, no dust substructures are identified from the AGE-PRO continuum data in the image plane. However, we should stress that dust disk structures have been previously detected at higher resolution data in objects such as \textit{Oph 7} \citep{Flores2023}. Besides the resolution, we should also note that the high inclination values in several Ophiuchus targets could prevent the detection of shallow structures. A more detailed and systematic analysis of continuum emission in the U-V plane is presented by \citet{Vioque2024} to confirm or rule out any substructures that cannot be visually identified in the image plane at the spatial resolution considered in this paper. Conversely, and as mentioned above, some molecular tracers are affected by cloud absorption, which can result in artificial structures in the radial profiles of the disks.

\subsection{Total Flux and Outer Disk Radii Estimates}
\label{Sec:Flux_Radii}

From the detected CO isotopologues and continuum emission, it is possible to obtain estimates of the disk radii and total fluxes. In this work, we define the outer radius that encloses 90$\%$ of the total flux following the curve-of-growth method. This approach measures the outer radius in increasingly larger elliptical apertures, whose shape is built considering the inclination (i) and position angle parameters of the continuum emission (see Table \ref{Tab:Continuum}) until the flux reaches 90$\%$ of the total flux. We also estimate radii at 68$\%$ of the total flux. For each Ophiuchus target, we used the moment 0 maps generated from the continuum-subtracted cubes, see Sec. \ref{Sec:Moments}. However, it should be noted that due to the high levels of contamination of the observations and self-absorption in the region, disk parameters of some of the targets from lines such as $^{12}$CO and $^{13}$CO are challenging to estimate accurately given that there is no a clear distinction between emission from the disk, outflows, or the molecular cloud. Since this process is not trivial, we avoid the use of Keplerian masks and we radially ``truncated" the molecules highly contaminated based on those cases where it was possible to estimate an outer radius from optically thinner tracers such as C$^{18}$O and C$^{17}$O. These rough estimates are only used to provide upper and lower limits for those particular CO species along the disk region. In Appendix \ref{App:Radii_without_mask}, we show that the use of Keplerian masks has only a minimal effect on the gas' radial extension, and therefore, we do not expect an underestimation of the outer radii to affect our results, however, it impacts the integrated flux values. The final CO gas and dust radii and fluxes are summarized in Tables \ref{Tab:Continuum} and \ref{Table:CO}, where $^{12}$CO and $^{13}$CO upper and lower ``truncated" limits are shown in red color. Figure \ref{Fig:Radii} shows the comparison of radii at 90$\%$ measured for 220 GHz continuum, C$^{18}$O and C$^{17}$O line emission. For completeness, Table \ref{Tab:Continuum05} presents disk radial estimates derived from the 220 and 234 GHz continuum images obtained with a Briggs parameter of 0.5.

In the Ophiuchus sample, \textit{Oph 1} and \textit{Oph 2} have the largest continuum emission with an angular extension of $\sim$ $1\farcs{29}$, while \textit{Oph 4} and \textit{Oph 5} are the most compact sources, marginally resolved with an angular extent of $\sim$ $0\farcs{50}$ and $0\farcs{6}$, respectively. Comparing C$^{18}$O and C$^{17}$O with dust extension, \textit{Oph 1} is also the most extended disk with a factor of two gas to dust ratio, see Figure \ref{Fig:Radii}. The largest gas-to-dust size ratio is estimated for \textit{Oph 3} with a factor of $\sim$2.5, however, it is relevant to consider that \textit{Oph 3} is associated with outflow material that might contribute to the gas radius measurements. On the other hand, \textit{Oph 7} is the brightest target in terms of C$^{18}$O, C$^{17}$O, and dust emission even though it is a relatively compact object with a radial dust extension of only $0\farcs{78}\pm0.05$, see Figure \ref{Fig:Flux}.

\subsection{Dust Disk Masses}
\label{Sec:Dust_masses}

To estimate the amount of dust in each object, we compute dust masses assuming optically thin emission and the Rayleigh-Jeans approximation, according to: 

\begin{equation}
     M_{Dust}= \rm \frac{F_{mm}d^{2}}{\kappa _{\nu }B_{\nu }(T_{Dust})},
\end{equation}

where $\rm F_{mm}$ is the observed flux in mJy, $\rm d$ is the distance to the Ophiuchus region of $\sim$138.4$\pm$2.4 pc, $\rm B_{\nu }$ is the Planck function at the dust temperature ($\rm T_{Dust}$) and the observed frequency (220 GHz), while $\kappa _{\nu }$ is the dust opacity calculated by using $\kappa _{\nu }=2.3\left ( \frac{\nu }{230\: \rm GHz} \right )^{\beta }\rm cm^{2}g^{-1}$ = 2.2 $\rm cm^{2}g^{-1}$ with $\beta$ = 1. For Ophiuchus sources, we adopt two different dust temperatures: 1) According to the AGE-PRO approach, which uses a $\rm T_{Dust}$ of 20 K and typically assumed for Class II sources, and 2) A dust temperature according to the luminosity of the system (see Table \ref{Table:Sample}) and scaled using the relation $\rm T_{Dust} = T_{0}(L_{Bol}/L_{\odot })^{0.25}$, where $\rm T_{0} = 43 \ K$ \citep{Tobin2020}. The dependence on the disk dust temperature of highly embedded objects with the luminosity of the system is due to the surrounding envelope also illuminating the disk; hence, higher disk dust temperatures are expected at young ages \citep{Osorio2003}. Table \ref{Tab:Continuum} shows disk dust masses estimated from both approaches\footnote{A comparison with previous B6 observations can be found in Appendix \ref{App:Masses}.}.

\begin{deluxetable*}{l c  c c c c c c c c c}
\tabletypesize{\scriptsize}
\label{Tab:Continuum}

\tablecaption{Disk Parameters from  220 GHz Continuum.}
\tablehead{\colhead{Source} & \colhead{$i$} & \colhead{$PA$} & \colhead{Flux} & \colhead{Peak} &  \colhead{R 90$\%$} & \colhead{R 68$\%$}& \colhead{$M_{Dust}(20 \ \rm K)$} & \colhead{$M_{Dust}$\tablenotemark{a}}\\
 & & & [mJy] & [mJy beam$^{-1}$] & & & [$M_{\oplus}]$ & [$M_{\oplus}$]
}
\startdata
	   Oph 1 &  $73.8^\circ\pm0.2^\circ$ & $48.2^\circ\pm0.30^\circ$  & 103.5 $\pm$ 0.5 & 50 $\pm$ 0.2 & $1\farcs{29}\pm0.02$ & $0\farcs{88}\pm0.01$ & 61 & 25 (42) \\
     \midrule
     Oph 2   & $74.5^\circ\pm3.0^\circ$ & $48.7^\circ\pm0.4^\circ$ & 27.5 $\pm$ 0.14 & 18.3 $\pm$ 0.1 & $1\farcs{20}\pm0.04$ & $0\farcs{80}\pm0.01$ & 17 & 9 (35)\\
     \midrule
     Oph 3\   & $69.0^\circ\pm3.8^\circ$ & $94.1^\circ\pm3.5^\circ$ & 20.4 $\pm$ 0.08 & 19.8 $\pm$ 0.04 & $0\farcs{87}\pm0.10$ & $0\farcs{60}\pm0.07$ & 12 & 5 (42)& \\
     \midrule
     Oph 4\ & $53.2^\circ\pm7.5^\circ$ & $90.3^\circ\pm15^\circ$ & 1.8 $\pm$ 0.05 & 1.8 $\pm$ 0.03 & $0\farcs{51}\pm0.18$ & $0\farcs{38}\pm0.10$ & 1 & 0.5 (42)\\
     \midrule
     Oph 5\  & $43.0^\circ\pm9.3^\circ$ & $80.4^\circ\pm0.9^\circ$ & 4.3 $\pm$ 0.04 & 3.9 $\pm$ 0.2 & $0\farcs{57}\pm0.03$ & $0\farcs{40}\pm0.20$ & 3 & 1 (40)\\
     \midrule
     Oph 6\ & $68.1^\circ\pm0.4^\circ$ & $168.4^\circ\pm0.64^\circ$ & 34.5 $\pm$ 0.1 & 23.7 $\pm$ 0.04 & $0\farcs{97}\pm0.04$ & $0\farcs{64}\pm0.10$ & 20 & 11 (33)\\
     \midrule
     Oph 7\  & $46.2^\circ\pm0.4^\circ$ & $149.3^\circ\pm1.7^\circ$ & 306 $\pm$ 2.5 & 128.7 $\pm$ 0.8 & $0\farcs{78}\pm0.05$ & $0\farcs{55}\pm0.04$ & 180 & 72 (43)\\
     \midrule
     Oph 8\ & $66.1^\circ\pm2.3^\circ$ & $73.2^\circ\pm2.1^\circ$ & 4.0 $\pm$ 0.05 & 3.3 $\pm$ 0.03 & $0\farcs{80}\pm0.30$ & $0\farcs{55}\pm0.30$ & 2.5 & 1.2 (36)\\
     \midrule
     Oph 9\ & $72.0^\circ\pm3.3^\circ$ & $155.2^\circ\pm3^\circ$ & 7.7 $\pm$ 0.08 & 5.8 $\pm$ 0.04 & $1\farcs{14}\pm0.10$ & $0\farcs{73}\pm0.06$ & 5 & 2.5 (34)\\
     \midrule
     Oph 10\ & $70.5^\circ\pm0.22^\circ$ & $53.2^\circ\pm0.3^\circ$ & 26.5 $\pm$ 0.11 & 16.2 $\pm$ 0.04 & $0\farcs{94}\pm0.03$ & $0\farcs{65}\pm0.14$  & 16 & 7 (38)\\
     \enddata

\tablenotetext{a}{$M_{Dust}$ adopting $\rm T_{Dust}$ -- in parenthesis -- scaled using the luminosity of the system \citep{Tobin2020}, see Section \ref{Sec:Dust_masses}.  }
\end{deluxetable*}

\subsection{Gas Masses from C$^{18}$O and C$^{17}$O}
\label{Sec:Gas_masses}

Here we estimate the minimum gas disk masses from the integrated line fluxes of the C$^{18}$O and C$^{17}$O $J$=2-1 lines. These lines are optically thinner than the more abundant $^{12}$CO and $^{13}$CO lines, and therefore are less affected by contamination from the inner envelope and cloud. We adopt the simple slab model assumptions, approximating the emitting region as a flat, uniform slab characterized by a single temperature. The model inputs are temperature, column density, and emitting area. For temperature, we use the luminosity-scaled average dust temperature listed in Table~\ref{Tab:Continuum}, typically between 33-43\,K. The emitting area is defined as $\pi$R$_{90\%}^2d^2$cos($i$), where R$_{90\%}$ is the angular radius enclosing 90\% of the total line flux, $d$ is the source distance, and $i$ is the disk inclination angle. We employ the public slab model code \texttt{spectools\_ir}\footnote{\url{https://github.com/csalyk/spectools_ir}} to compute the line fluxes. Starting with an initial guess of the column density, the line flux is calculated. The ratio of the observed to model line flux is then used to scale the column density iteratively until the difference between the model and observed fluxes is within 1$\%$. Finally, the total gas disk masses are derived using the abundance ratio of the species relative to H$_2$, as described by the following equation
\begin{equation}
  \rm  M_{gas}=N(X) \pi R_{90\%}^2d^2 \left [ \frac{^{12}CO}{X} \right ]\left [\frac{H_{2}}{^{12}CO}\right ]\mu m_{H}
\end{equation}
where $\mu$=2.8, which is the mean molecular weight. 
We adopt the canonical abundance ratios of CO/H$_{2}$ = 10$^{-4}$, CO/C$^{18}$O = 557 and C$^{18}$O/C$^{17}$O = 3.6 \citep{Wilson1999}. The gas disk mass estimates are presented in Table \ref{Table:CO}. These estimates should be considered as lower limits of the gas disk masses, because a large fraction of CO may be in the ice phase in the cold disk region, and C$^{18}$O and C$^{17}$O may be highly optically thick in the inner disk region \citep[e.g.,][]{vantHoff2018,Zhang2020}. Given the relatively low stellar luminosity of our sample (Table 1), the CO freeze-out fraction is expected to be large. More accurate gas mass estimations of the Ophiuchus sources are presented in \citet{Trapman2024b}, which uses 2-dimensional thermo-chemical models to model disk temperature, density, and abundance structures. Indeed, the gas masses from our simple slab models are 2-50 times lower than the estimates of \citet{Trapman2024b}.

\begin{table*}
\centering

    \caption{Disk Parameters CO lines.}
    \begin{tabular}{c c c c c c c}\toprule
	   Source & Line &Radius (68$\%$) & Radius (90$\%$) & Integrated Flux & Min M$_{\rm gas}$ & N(X) \\
            &        & arcsec & arcsec & [mJy km s$^{-1}$] & [$\rm M_{\odot}$] & [$\rm cm^{-2}$] \\
	 \midrule
  Oph 1 &$^{12}$CO & \textcolor{red}{2.13 $_{0.007}^{0.007}$} & \textcolor{red}{2.70 $_{0.011}^{0.011}$} & \textcolor{red}{6063 $\pm$ 17\tablenotemark{L}} \\
     &$^{13}$CO & \textcolor{red}{2.05 $_{0.006}^{0.006}$} & \textcolor{red}{2.65 $_{0.010}^{0.010}$} & \textcolor{red}{2165 $\pm$ 12\tablenotemark{L}}\\
      & C$^{18}$O  & 1.70 $_{0.007}^{0.007}$  & 2.45 $_{0.017}^{0.017}$  &   1120 $\pm$ 9  & 6.3$\times$10$^{-3}$ & 5.8$\times$10$^{15}$ \\
     &C$^{17}$O & 1.42 $_{0.018}^{0.018}$ & 2.10 $_{0.055}^{0.048}$ &  444 $\pm$ 10 & 7.2$\times$10$^{-3}$ & 2.5$\times$10$^{15}$ \\
     \midrule

     Oph 2 &$^{12}$CO & \textcolor{red}{1.72 $_{0.018}^{0.018}$} & \textcolor{red}{2.30 $_{0.040}^{0.037}$} &  \textcolor{red}{1410 $\pm$ 8\tablenotemark{L}}\\
     &$^{13}$CO & \textcolor{red}{1.50 $_{0.020}^{0.020}$} & \textcolor{red}{2.22 $_{0.058}^{0.053}$} & \textcolor{red}{382 $\pm$ 7\tablenotemark{L}}\\
      & C$^{18}$O  & 1.10 $_{0.041}^{0.039}$  & 1.55 $_{0.137}^{0.103}$  &   91.0 $\pm$ 5 & 4.0$\times$10$^{-4}$ & 9.2$\times$10$^{14}$ \\
     &C$^{17}$O & 1.15 $_{0.128}^{0.120}$ &  1.45 $_{0.270}^{0.192}$  &  23.0 $\pm$ 5 & 3.2$\times$10$^{-4}$ & 2.3$\times$10$^{14}$ \\
     \midrule

     Oph 3 &$^{12}$CO\tablenotemark{L} & \textcolor{red}{1.40 $_{0.009}^{0.009}$} & \textcolor{red}{1.78 $_{0.017}^{0.017}$} &  \textcolor{red}{1662.8 $\pm$ 8\tablenotemark{L}} \\
     &$^{13}$CO\tablenotemark{L} & \textcolor{red}{1.43 $_{0.018}^{0.018}$} & \textcolor{red}{1.80 $_{0.066}^{0.054}$} & \textcolor{red}{577.5 $\pm$ 9\tablenotemark{L}}\\
      & C$^{18}$O & 1.48  $_{0.029}^{0.029}$ & 1.85 $_{0.119}^{0.105}$  & 410 $\pm$ 6 & 1.6$\times$10$^{-3}$ & 2.5$\times$10$^{15}$  \\
     &C$^{17}$O & 1.42 $_{0.094}^{0.083}$ & 1.75 $_{0.269}^{0.191}$ &  196 $\pm$ 8  & 2.3$\times$10$^{-3}$ & 1.2$\times$10$^{15}$\\
     \midrule

      Oph 4 &$^{12}$CO & -- & -- &  -- \\
     &$^{13}$CO & -- & -- & -- \\
      & C$^{18}$O & -- & --  & --  \\
     &C$^{17}$O & -- & -- &  -- \\
     \midrule

     Oph 5 &$^{12}$CO & \textcolor{red}{0.85 $_{0.015}^{0.012}$} & \textcolor{red}{1.05 $_{0.042}^{0.030}$} &  \textcolor{red}{3854 $\pm$ 46\tablenotemark{U}} \\
     &$^{13}$CO & \textcolor{red}{0.85 $_{0.100}^{0.112}$} & \textcolor{red}{1.05 $_{0.042}^{0.030}$} & \textcolor{red}{294.6 $\pm$ 10\tablenotemark{U}}\\
     & C$^{18}$O & --  &  -- & --   \\
     &C$^{17}$O & -- &  -- &  --      \\
     \midrule

       Oph 6 &$^{12}$CO & \textcolor{red}{1.26 $_{0.005}^{0.005}$}  &  \textcolor{red}{1.71 $_{0.013}^{0.013}$} &  \textcolor{red}{4019 $\pm$ 10} \\
     &$^{13}$CO & \textcolor{red}{1.27 $_{0.023}^{0.019}$}  &  \textcolor{red}{1.72 $_{0.036}^{0.036}$} & \textcolor{red}{710 $\pm$ 11}\\
      & C$^{18}$O  & 1.20 $_{0.047}^{0.044}$  &  1.70 $_{0.064}^{0.063}$  & 229 $\pm$ 7 & 7.3$\times$10$^{-4}$ & 1.4$\times$10$^{15}$ \\
     &C$^{17}$O & 1.10 $_{0.064}^{0.061}$ &  1.60 $_{0.191}^{0.138}$ &  199 $\pm$ 9  & 2.1$\times$10$^{-3}$ & 1.3$\times$10$^{15}$\\
     \midrule
     
       Oph 7 &$^{12}$CO & \textcolor{red}{1.55$_{0.035}^{0.036}$} & \textcolor{red}{1.85$_{0.065}^{0.064}$} &  \textcolor{red}{16780 $\pm$ 100\tablenotemark{U}} \\
     &$^{13}$CO & 1.40$_{0.007}^{0.007}$ & 1.77$_{0.014}^{0.014}$ & 5114 $\pm$ 20\\
      & C$^{18}$O  & 1.32$_{0.010}^{0.010}$ & 1.73$_{0.016}^{0.016}$ &  2542 $\pm$ 13 & 7.7$\times$10$^{-3}$ & 1.4$\times$10$^{16}$\\
     &C$^{17}$O & 1.08$_{0.016}^{0.016}$ & 1.54$_{0.060}^{0.060}$ &  1595 $\pm$ 21   & 1.3$\times$10$^{-2}$ & 8.7$\times$10$^{15}$\\
     \midrule

       Oph 8 &$^{12}$CO & 0.67$_{0.010}^{0.010}$ & 0.96$_{0.022}^{0.022}$ & 452 $\pm$ 8 \\
     &$^{13}$CO & 0.58$_{0.230}^{0.300}$ & 0.91$_{0.322}^{0.355}$ & 36 $\pm$ 10\\
     & C$^{18}$O  & 0.79$_{0.200}^{0.200}$  &  1.10$_{0.300}^{0.200}$ &  17 $\pm$ 5.0\tablenotemark{U} & 4.7$\times$10$^{-5}$ & 2.2$\times$10$^{14}$\\
     &C$^{17}$O & 0.71$_{0.100}^{0.200}$  &  0.92$_{0.300}^{0.100}$ &  26 $\pm$ 7.0\tablenotemark{U} & 2.4$\times$10$^{-4}$ & 4.4$\times$10$^{14}$ \\
     \midrule

       Oph 9 &$^{12}$CO & 1.15 $_{0.008}^{0.008}$  &  1.62 $_{0.018}^{0.021}$ &  1946 $\pm$ 9 \\
     &$^{13}$CO & 1.10 $_{0.017}^{0.016}$  &  1.62 $_{0.045}^{0.045}$ & 800 $\pm$ 7\\
      & C$^{18}$O  & 0.91 $_{0.020}^{0.023}$  &  1.36 $_{0.100}^{0.131}$ &  267 $\pm$ 7 & 1.2$\times$10$^{-3}$ & 3.6$\times$10$^{15}$ \\
     &C$^{17}$O & 1.14 $_{0.042}^{0.042}$  &  1.62 $_{0.142}^{0.117}$ &  200 $\pm$ 5  & 2.6$\times$10$^{-3}$ & 1.5$\times$10$^{15}$ \\
       \midrule

       Oph 10 &$^{12}$CO & 0.97 $_{0.015}^{0.016}$  &  1.27 $_{0.034}^{0.034}$ & 513 $\pm$ 6 \\
     &$^{13}$CO & 1.00 $_{0.045}^{0.045}$  &  1.25 $_{0.085}^{0.096}$ & 130 $\pm$ 7\\
      & C$^{18}$O  & 1.10 $_{0.183}^{0.215}$  &  1.26 $_{0.200}^{0.297}$ & 42 $\pm$ 4 & 1.5$\times$10$^{-4}$ & 5.2$\times$10$^{14}$  \\
     &C$^{17}$O & 1.08$_{0.100}^{0.100}$  &  1.25$_{0.100}^{0.100}$ &  30 $\pm$ 6.1\tablenotemark{U} & 3.5$\times$10$^{-4}$ & 4.6$\times$10$^{14}$ \\

     \midrule
	\bottomrule
    \end{tabular}
\label{Table:CO}
\tablenotetext{L}{Lower Limits}
\tablenotetext{U}{Upper Limits}
\end{table*}

\section{Discussion}
\label{Sec:Discussion}

\subsection{Inclination and Extinction impacting SED classification}

It is known that in the spectral classification of YSOs based on their SEDs and spectral index, edge-on objects could resemble highly embedded objects, while face-on embedded sources may appear to be more evolved \citep{Furlan2016}. Since half of the AGE-PRO Ophiuchus sources are highly inclined ($i\gtrsim$ 70 deg.), the potential misclassification of Class I and FS YSOs viewed at high inclination needs to be evaluated due to its impact in our analysis and understanding the evolution of the Class I and FS disks to Class II disks, an important purpose of the AGE-PRO program. Highly inclined  objects in our sample include \textit{Oph 1}, \textit{Oph 2}, \textit{Oph 9}  and \textit{Oph 10} (i $>$  70 deg) and to a lesser extent \textit{Oph 3}, \textit{Oph 6}, and \textit{Oph 8} (i $\sim$ 65 -- 70 deg). A related concern is the potential misclassification of some of the AGE-PRO Ophiuchus targets, especially the flat-spectrum sources, since the slope at shorter wavelengths than 5 $\mu$m can be significantly influenced by foreground reddening, and self-extinction due to inclination \citep{McClure2009, McClure2010}.

\subsubsection{Extinction}

To evaluate whether our Ophiuchus targets are bona fide embedded sources, we first consider the column densities of their surrounding regions using the column density maps of the Ophiuchus cloud from \textit{Herschel} \citep{Ladjelate2020}. We find that most of the AGE-PRO targets are located in dense regions with $\gtrsim$ 5 $\times$ 10$^{22}$ cm$^{-2}$(Fig. \ref{Fig:Density}), corresponding to extinction levels ranging from $\sim$10 to $\sim$60 mag in the V band, see Table \ref{Table:Extinction}, using the standard conversion $N_{\rm H_{2}}$ (cm$^{-2}$)=0.94 $\times 10^{21}A_{V}$ \citep{Bohlin1978}. The exceptions to the above are \textit{Oph 8} and \textit{Oph 9} with column densities of 2 $\times$ 10$^{21}$ cm$^{-2}$ and 5.5 $\times$ 10$^{21}$ cm$^{-2}$, respectively,  which translate to $A_{v}$ $\sim$ 2.3 and 6.3 mag. These \textit{Herschel} results are consistent with the line-of-sight extinctions calculated as part of the spectral type classification, according to which the Ophiuchus sample presents foreground extinction from 10 to 40 mag in the V band, except for \textit{Oph 8} and \textit{Oph 9}, which have relatively low levels of $A_{v}$, $\sim$1 and 5 mag., respectively (see Table \ref{Table:Sample}). 
 
As an additional check,  we also consider the spectral index between 5 and 12 $\mu$m ($\alpha_{5-12}$), which is sensitive to the evolutionary state of YSOs, and adopting the definitions proposed by \citet{McClure2010}, we have:

\begin{itemize}
    \item Envelope dominated: $\alpha_{5-12}$ $>$ -0.2
    \item Disk dominated: $\alpha_{5-12}$ $<$ -0.2
    \item Photospheric dominated: $\alpha_{5-12}$ $<$ -2.25
\end{itemize}

Upon deriving $\alpha_{5-12}$ values using dereddened photometry data and listed in Table \ref{Table:Extinction}, we find once again that all AGE-PRO Ophiuchus targets are envelope-dominated objects, except for \textit{Oph 8} and \textit{Oph 9} whose SEDs are more likely disk dominated.

From everything considered, we conclude that all of our Ophiuchus targets are still embedded objects associated with dense envelope material, with the possible exception of \textit{Oph 8} and \textit{Oph 9}, which might be in a more evolved evolutionary phase.

\subsubsection{Inclination}
\label{Sec:Inclination}

\begin{figure*}[h!]
\centering
\fig{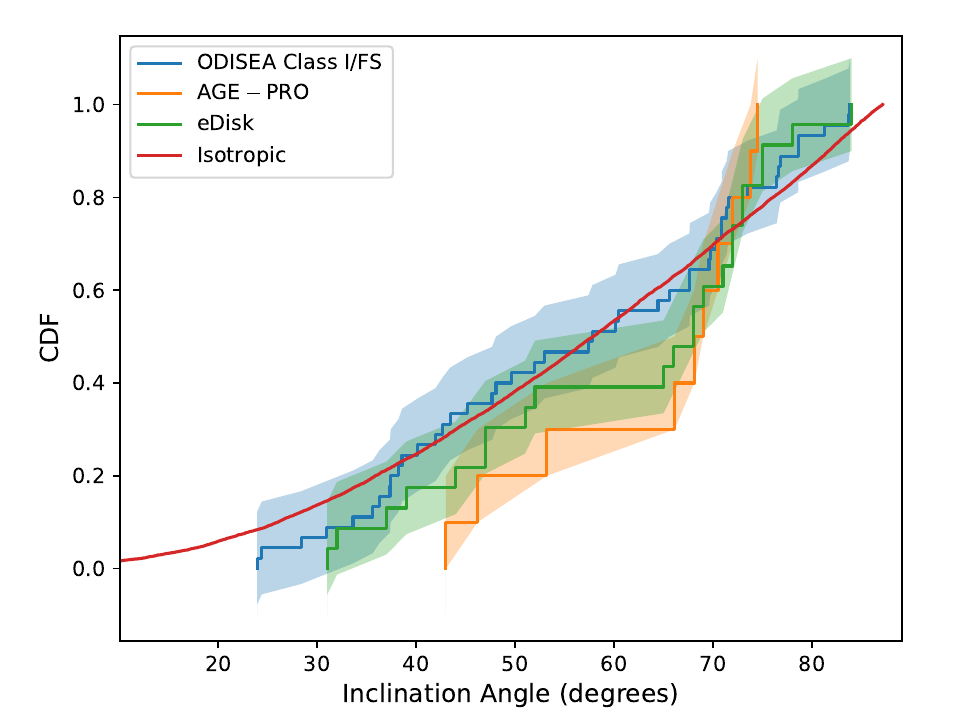}{0.49\textwidth}{}
\fig{Inc_ClassII.pdf}{0.49\textwidth}{}
\figcaption{Left panel: cumulative distributions of the disk inclinations for embedded sources in the Ophiuchus star forming region and those studied in the eDisk Large Program \citep{Ohashi2023}. Class I and FS disks studied in the ODISEA project \citep{Dasgupta2025} are shown with a blue shaded region, disk inclinations from AGE-PRO and eDisk are shown with orange and green shaded regions, respectively. 
The red line represents the isotropic distribution,  against which we compared each sample, see Section \ref{Sec:Inclination} and Table \ref{Table:KS}. The right panel compares the distribution of inclinations of the  Class II sources in Ophiuchus to the isotropic distribution. 
\label{Fig:CDF_inc}
}
\end{figure*}

\begin{table*}[ht]
\caption{K-S Tests} 
\centering 
\begin{tabular}{l c c} 
\hline\hline
	   Sample  & D-Value & p-Value \\

   \midrule
  	 AGE-PRO Oph. vs. Isotropic Function &  0.33 & 0.19   \\
      
	\midrule
      ODISEA Embedded vs. Isotropic Function &  0.09 & 0.83   \\

     \midrule
      ODISEA Embedded vs. AGE-PRO Oph.& 0.31 & 0.35  \\

    \midrule
      eDisk vs. Isotropic Function &  0.22 & 0.20   \\

\midrule
      ODISEA Class II vs. Isotropic Function &  0.28 & 0.003   \\
    


     \midrule
	\bottomrule
    \end{tabular}
\label{Table:KS}
\end{table*}

The fact that half of the AGE-PRO Ophiuchus sources are highly inclined ($i\gtrsim$ 70 deg.) begs the question of whether samples of Class I and Flat Spectrum sources are significantly contaminated by more evolved sources that have shallower SEDs due to inclination effects. The answer to this question has direct implications for the lifetimes of different SED stages since they are usually estimated from the relative incidence of different SED classes in molecular clouds \citep{Evans2009}. In particular, if the occurrence of embedded sources has been overestimated due to inclination effects, this would imply that their lifetimes have also been overestimated. To test this hypothesis, we compare the empirical cumulative distribution function (ECDF) of the inclinations of the AGE-PRO Ophiuchus sample to that of an isotropic distribution of disk orientation (See Fig. \ref{Fig:CDF_inc}). We also compare the isotropic distribution to the observed distribution of the 23 deeply embedded (Class 0 and I) disks studied in the eDisk Large Program \citep{Ohashi2023}.

Since we are only interested in the inclination projected along the line of sight, we generate a uniform distribution with 10,000 random values of the inverse cosine from 0 to 1, i.e. 0 deg. $<$ $i$ $<$ 90 deg. The red line in Fig. \ref{Fig:CDF_inc} is the resulting isotropic distribution. To statistically evaluate the distributions, we used the Kolmogorov-Smirnov (K-S) test to compare the observed distributions with the isotropic distribution.  The D and p values are listed in Table \ref{Table:KS}. The D value corresponds to the maximum distance between the cumulative distributions that are being compared, while the p value indicates the probability that both distributions have been drawn from the same parent population, given the D value and the sample sizes. We find that both the AGE-PRO Ophiuchus and eDisk distributions are shifted 
towards high inclinations with respect to the isotropic ddistribution (D values of 0.33 and 0.31, respectively). However, given the small sample sizes, the p values are moderate and far from statistically significant (which would require p values $<$ 0.05). This implies that the highly inclined disk in AGE-PRO Ophiuchus and eDisks could be the result of pure chance. For instance, the distribution of inclinations for the  AGE-PRO targets in Ophiuchus is statistically indistinguishable from that of the full sample of embedded disks in Ophiuchus (see Table 5).

To test whether the full population of embedded sources in Ophiuchus is biased toward highly inclined disks, we compare the inclinations of the $\sim$50 embedded (Class I and FS) disks studied in the ODISEA project \citep{Dasgupta2025}, against the isotropic distribution (also included in Fig. \ref{Fig:CDF_inc} and Table \ref{Table:KS}). We find that a D value of just 0.09 and a p value of 0.83,  indicating that the inclinations of embedded sources in ODISEA (from which the AGE-PRO sample was selected) are fully consistent with random orientations. 

We thus conclude that there is no statistical evidence that the embedded objects are significantly contaminated by Class II sources that mimic the SED of Class I and FS objects due to inclination effects.

For completeness, we also investigated the distribution of inclination of Class II disks in Ophiuchus (also from \citet{Dasgupta2025}. We find that the inclinations of the Class II disks in Ophiuchus are biased toward low inclinations (see Figure 8, right panel) at a statistically significant level (D value = 0.28; P value = 0.003).   This bias is interesting,  and we speculate that it might be related to the fact that embedded sources tend to drive outflows, which can clear the envelope along the line of sight of a low-inclination disk. Therefore, the SED of an embedded YSO might look like that of a Class II source if the disk is close to face-on.  If this interpretation is correct, it would mean that the number of embedded sources relative to Class II sources has been underestimated and that the embedded phase might last somewhat longer than currently believed. Measuring disk inclinations for larger samples of YSOs across SED classes is needed to quantify this effect. The existence of this bias also means that the SED classification of face-on objects should be taken with caution, as in the case of highly inclined sources.

\subsection{Dust and Gas extension at the earliest phases}

\begin{figure*}[h!]
\centering
\fig{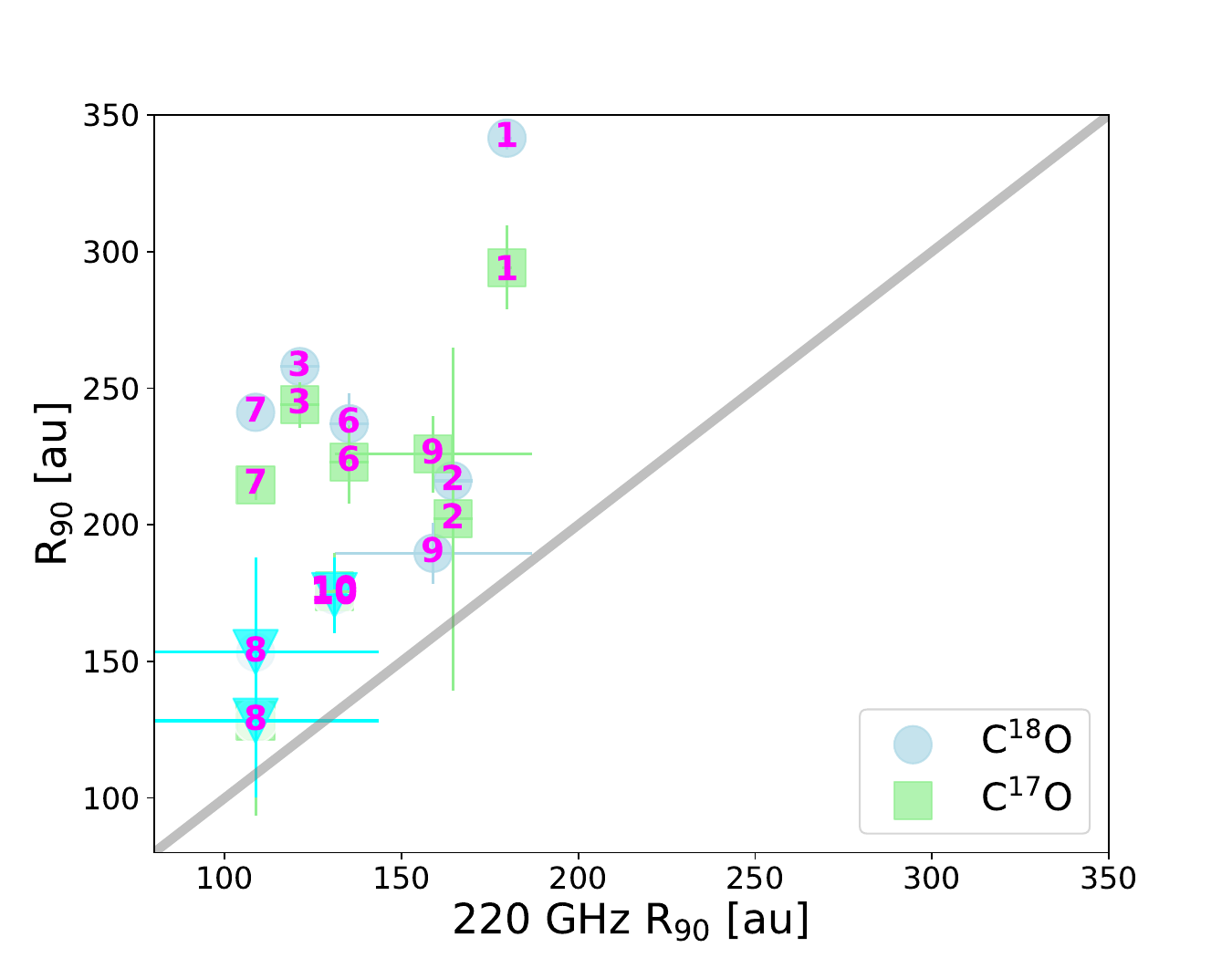}{0.45\textwidth}{}
\fig{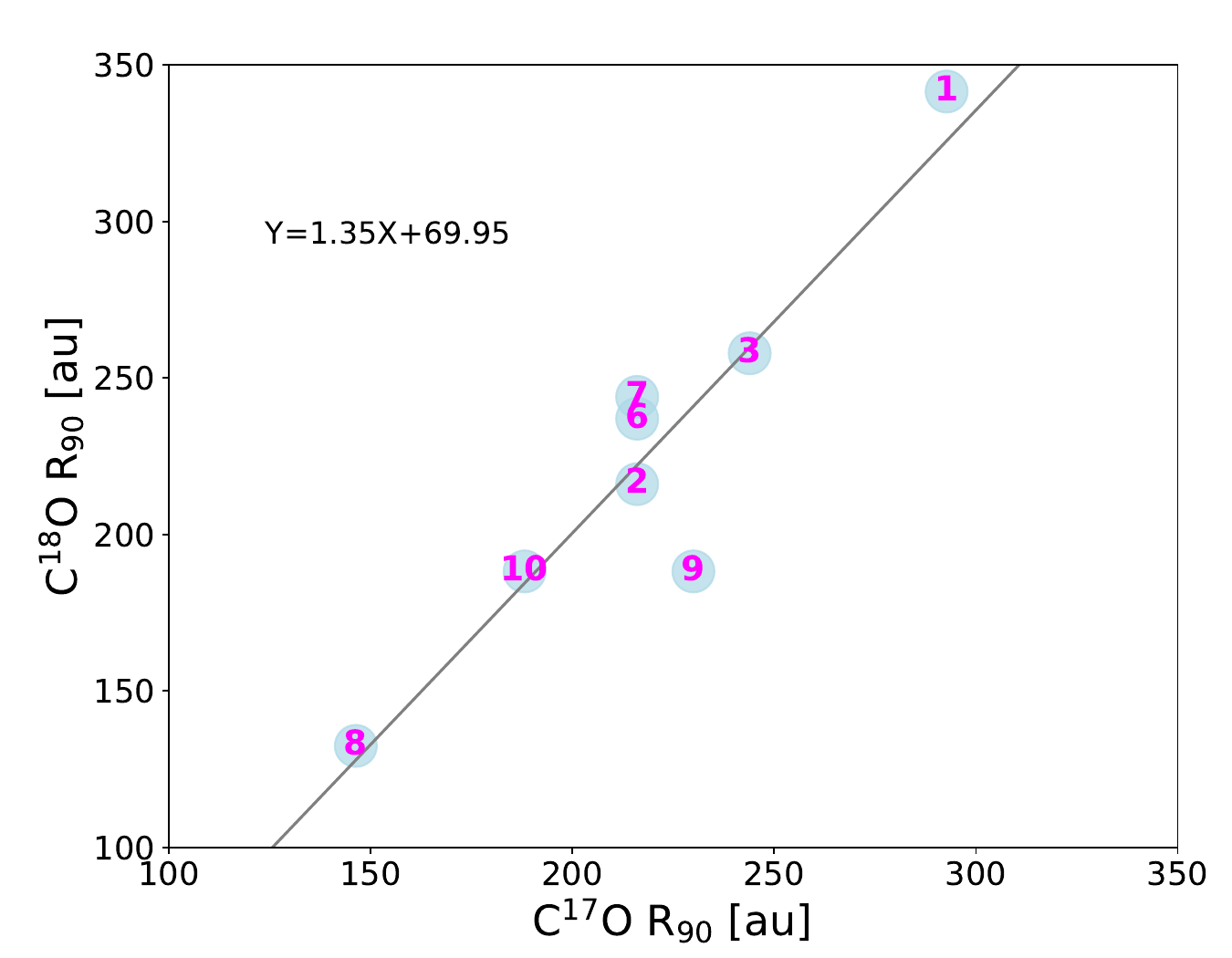}{0.45\textwidth}{} 
\figcaption{Comparisons between radii estimated at R$_{90\%}$ of the total flux from 220 GHz continuum and C$^{18}$O and C$^{17}$O products. The gray solid line in the left panel represents a visual guide for the flux radial extension of dust and C$^{18}$O and C$^{17}$O lines. The best fit shown in the right panel results from linear regression in a log-log scale and is plotted in a gray solid line as a positive relation. 
\label{Fig:Radii}
}
\end{figure*}

One observational signature of dust evolution in protoplanetary disks is the size dichotomy between the gas and dust, where the radii traced by the $^{12}$CO emission are generally more extended than the continuum disk \citep[e.g.][]{Long2022}. This dichotomy is well
established in Class II disks, which are older than our AGE-PRO Ophiuchus sources. Unfortunately, since the $^{12}$CO emission of our Ophiuchus sources is highly contaminated by extended emission,  we cannot compare the sizes between $^{12}$CO and continuum disk radii for all sources as it is usually done for Class II sources. Tables \ref{Tab:Continuum} and \ref{Table:CO} show dust and CO radius estimates, where some of these values are affected by the presence of outflow and envelope structures with only 3 objects (\textit{Oph 8}, \textit{Oph 9}, and \textit{Oph 10}) showing $^{12}$CO emission that can be distinguished reasonably well from surrounding material and not affected by self-absorption effects. Note, however, that in general, we measure $^{12}$CO and $^{13}$CO emission extend further out than continuum emission. On the one hand, if we consider grain growth to millimeter sizes and subsequent inward radial drift of the grains, it would mean that the large grains have already decoupled from the gas at this early stage. On the other hand, this radial difference may also be due to the line optical depth being much higher than the continuum optical depth, generating the effect of a more extended gas disk.

In contrast to the optical thicker tracers, C$^{18}$O and C$^{17}$O are less contaminated and can provide gas radii for most of our sources. Figure \ref{Fig:Radii} shows R$_{90\%}$ (the radius enclosing 90$\%$ of the flux) as traced by C$^{18}$O and C$^{17}$O as a function R$_{90\%}$ traced by the dust. Overall, C$^{18}$O and C$^{17}$O present similar radial extensions, ranging from $\sim$150 to 350 au with only subtle differences (see also Figure \ref{Fig:Radii}, right panel), while the continua trace smaller R$_{90\%}$ radii,  in a range between 100 and 200 au. Considering that these molecules are optically thinner tracers, the observed gas-dust size ratios (ranging from $\sim$1.5 to $\sim$2.5) should not be affected as much by the difference in optical depth with respect to the continuum. However, the low S/N and the extended emission might still impact the gas measurements somewhat.

In more evolved Class II disks, it has been suggested that a gas radius traced by $^{12}$CO that is a factor of 4 larger than the continuum disk represents strong evidence for radial drift of dust in the disk \citep{Trapman2019}. However, considering all observational caveats mentioned above, our results cannot be directly compared to $^{12}$CO studies in Class II sources. At face value, the smaller continuum disk radii with respect to the gas suggest that significant dust growth and radial drift have already begun in embedded objects with ages 
$\lesssim$1 Myr, but detailed modeling of the dust and gas around embedded disks is needed to test this scenario. 

\subsection{Dust emission and gas line fluxes: A Challenging study in embedded disks}\label{continuum_vs_gas}

\begin{figure*}
\centering
\gridline{\fig{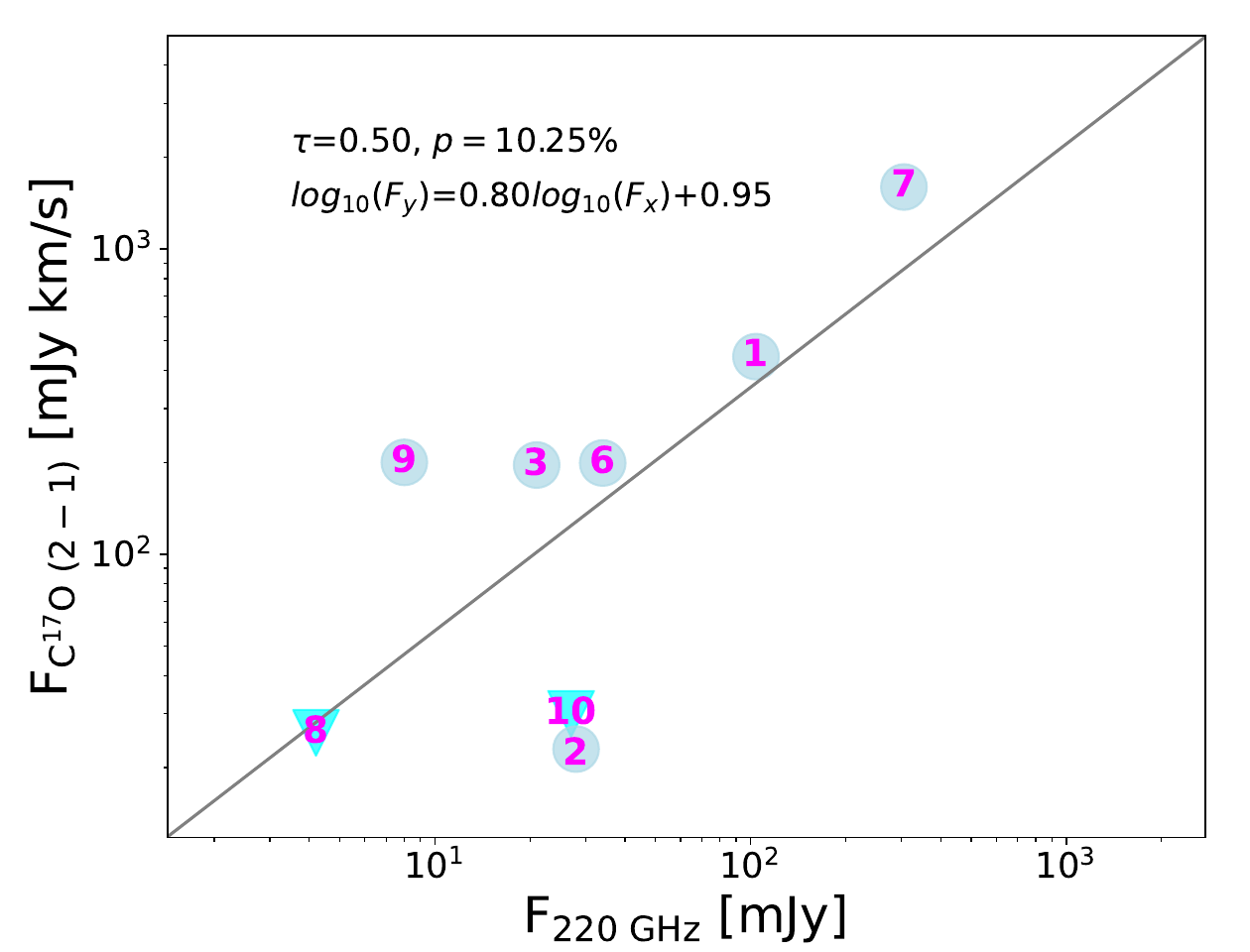}{0.33\textwidth}{}
   \fig{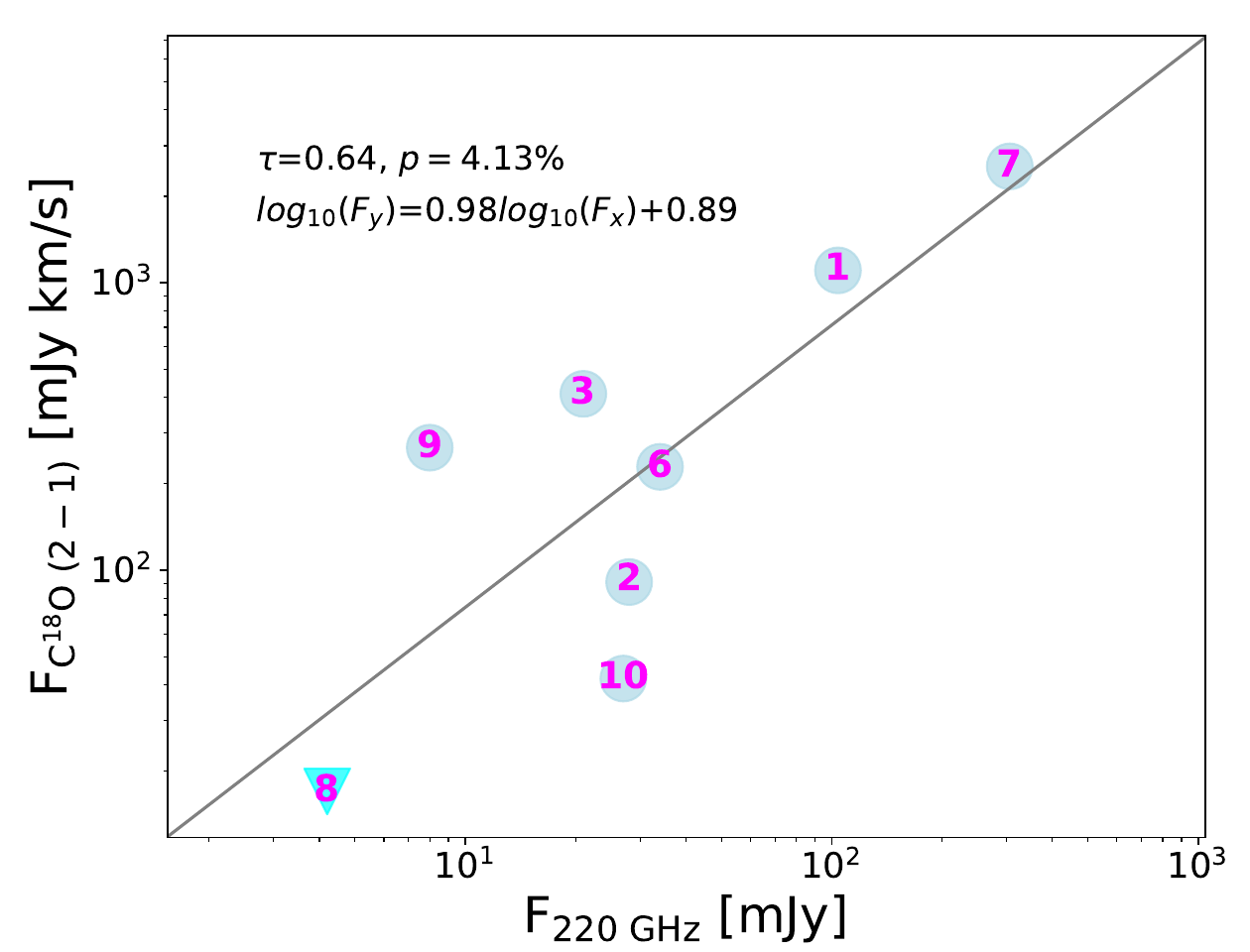}{0.33\textwidth}{}
   \fig{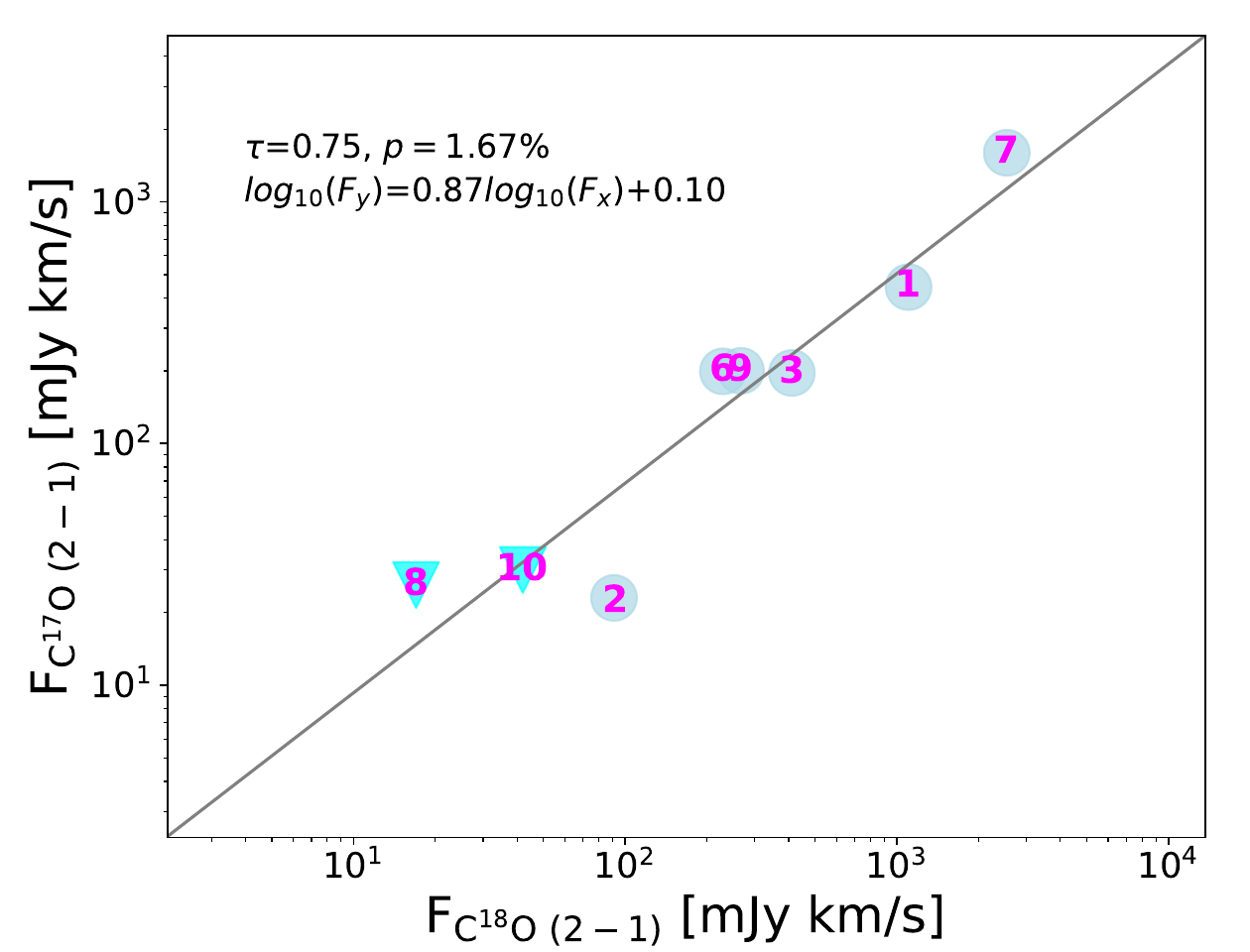}{0.33\textwidth}{}
            }
\figcaption{Comparison of the 220 GHz emission and thin CO line fluxes. Cyan triangles represent upper limits. The results of \textsc{pymccorrelation} Kendall's $\tau$ tests for the Ophiuchus sample are reported in each panel. We find positive correlations between the C$^{18}$O and C$^{17}$O fluxes with the Band 6 220 GHz continuum flux. 
\label{Fig:Flux}
}
\end{figure*}

Measuring the gas mass of protoplanetary disks has been proven to be difficult, especially in highly embedded disks due to the chaotic environment characteristic of these early evolutionary phases. Comparing AGE-PRO continuum maps with CO lines, ideally, could have been performed with $^{12}$CO lines tracing low-density material in the disk. However, as mentioned in previous sections, $^{12}$CO measurements are not reliable for at least seven out of ten sources in the AGE-PRO Ophiuchus target list, see Table \ref{Table:CO}. Hence, we relied on any correlations that can be found between the dust continuum and C$^{18}$O and C$^{17}$O detections. To that end, we follow Kendall's rank correlation approach, a non-parametric test procedure to measure the strength of the relationship between two variables. In this case, we use the \textsc{pymccorrelation} script\footnote{https://github.com/privong/pymccorrelation.} that provides the Kendall's $\tau$ index and the percent probability $p$ of non-correlation. Overall, Kendall's $\tau$ index ranges from -1 to 1, where -1 represents a perfect negative association, 0 represents no association, and 1 represents a perfect positive association. To facilitate a direct comparison between 220 GHz emission and C$^{18}$O and C$^{17}$O line fluxes, we have also performed linear regressions that are displayed in Figure \ref{Fig:Flux} together with $\tau$ and $p$ values. From these values, we find that C$^{18}$O and C$^{17}$O weakly correlate with the 220 GHz emission, while C$^{18}$O and C$^{17}$O line fluxes present a stronger correlation. This latter correlation between C$^{18}$O and C$^{17}$O is expected because of their similar mass and molecular structure probing gas under similar physical conditions, whose slight difference in the emission from these two isotopes could be related to opacity effects and the embedded nature of the objects, see below.

From the direct comparison between C$^{18}$O and C$^{17}$O line and continuum emission shown in Figure \ref{Fig:Flux}, \textit{Oph 1} and \textit{Oph 7} are the brightest objects in the sample. Interestingly, we note that in the continuum \textit{Oph 1} is the most extended source, while \textit{Oph 7} is one of the most compact objects, see Table \ref{Tab:Continuum} and Figure \ref{Fig:Radii}. These observational features could be explained if \textit{Oph 1} and \textit{Oph 7} -- of similar stellar mass with moderate mass accretion rates of $\dot{M}_{acc} = 3 \times 10^{-9}M_{\odot } \ yr^{-1}$ and $\dot{M}_{acc} = 5 \times 10^{-8}M_{\odot } \ yr^{-1}$, respectively -- are in slightly different quiescent phases, where \textit{Oph 7} likely approaches an outburst episode \citep{Manara2015, Flores2023}. This scenario is supported by an imbalance between mass infall and mass accretion rates in the \textit{Oph 7} system with an infall rate of $\dot{M}_{acc} = 4 \times 10^{-6}M_{\odot} \ yr^{-1}$ \citep{Flores2023}. This suggests that the \textit{Oph 7} disk could be in a mass buildup phase that eventually becomes gravitationally unstable and can trigger an outburst accretion episode \citep{Armitage2001}. If this is the case, \textit{Oph 7} with a compact and massive disk is expected to suffer a sudden increase in mass accretion levels relatively soon \citep[$\sim$10$^{4}$ yr;][]{Flores2023}. This phenomenon is known as FU Ori-type outbursts, and whose disk radii are found to be smaller, for a given mass, than their quiescent Classical T Tauri counterparts \citep[e.g.][]{Cieza2018}.

In addition, AGE-PRO also provides N$_{2}$D$^{+}$ $J$=3$-$2 observations that, together with C$^{18}$O line fluxes, potentially can be used to estimate CO-to-H$_{2}$ ratios and gas masses  \citep{Trapman2022}. However, N$_{2}$D$^{+}$ $J$=3$-$2 is not detected in any of the Ophiuchus sources, see Figure \ref{Fig:Mom0SS1}, preventing us from performing similar analysis as in other star-forming regions part of the AGE-PRO LP, see \citet{Deng2024} and \citet{Agurto2024}. It is worth noting that the lack of N$_{2}$D$^{+}$ detections might be one of the consequences of the abundant CO emission in embedded objects expected to have higher temperatures than older disk counterparts \citep[e.g.][]{Jorgensen2004}.

\subsubsection{Disk Gas Mass Estimates}

\begin{figure*}
\centering
\gridline{\fig{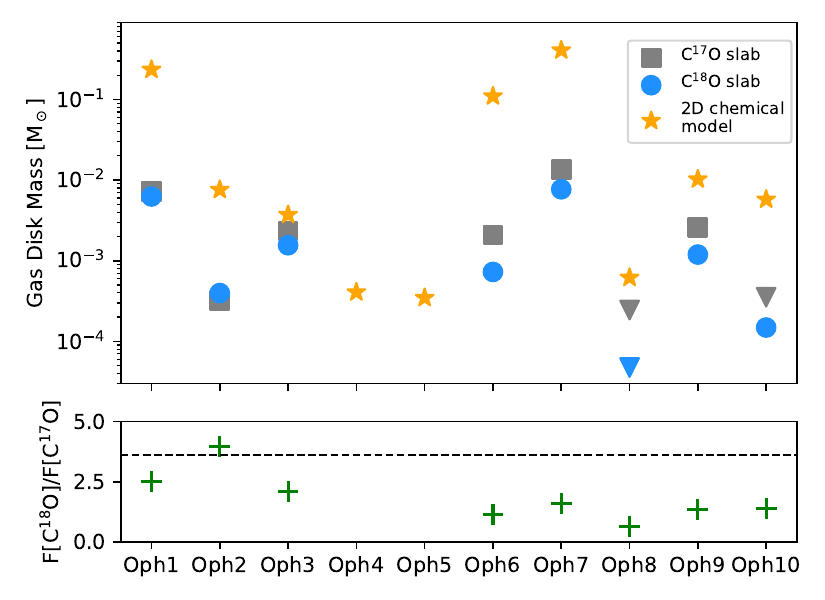}{0.6\textwidth}{}
            }
\vspace{-0.6cm}
\figcaption{Top panel: Gas masses of the Ophiuchus disks estimated from the slab model fitting of C$^{18}$O and C$^{17}$O $J$=2-1 line fluxes and gas masses estimated using 2-dimensional chemical models in \citet{Trapman2024b}.  Upper limits of Oph 8 and 10 are presented as downward-pointing triangles. Bottom panel: Ratios of C$^{18}$O to C$^{17}$O integrated line fluxes. The solid green line represents the ISM abundance ratio of 3.6 \citep{Wilson1999}, the expected value if both CO isotopologues are optically thin. 
\label{Fig:GM}
}
\end{figure*}

In Section \ref{Sec:Gas_masses}, we derived disk mass estimates from C$^{18}$O and C$^{17}$O fluxes, which we plot in  Figure \ref{Fig:GM}. For most disks, we find that the total disk mass estimated from the C$^{17}$O emission is slightly higher than that estimated from C$^{18}$O emission with an average factor of 1.5. This difference can be attributed to C$^{18}$O being optically thicker than C$^{17}$O since the ratio of the integrated line intensities of the two species is lower than the expected value of 3.6 \citep{Penzias1981, Wilson1999}, except for \textit{Oph 2} with a C$^{18}$O to C$^{17}$O ratio of $\sim$ 4.4, see bottom panel in Figure \ref{Fig:GM}.

In general, we find that the gas mass decreases with the emission size of the C$^{17}$O and C$^{18}$O. Thus, the lowest gas mass estimates correspond to those objects with the smallest radii, while the most massive disk corresponds to the largest disk, i.e. \textit{Oph 1}. The fact that the gas mass estimates from C$^{17}$O are higher than those estimates from C$^{18}$O is an indication of the optically thinner nature of C$^{17}$O compared with C$^{18}$O in highly embedded objects, even though C$^{18}$O radii are slightly larger than C$^{17}$O radii (see Table \ref{Table:CO}). Given these points, we conclude that C$^{17}$O emission is a more suitable gas tracer to estimate disk gas masses for highly embedded objects, but such estimates require very deep observations to be sensitive to the outer regions of the disk. A more detailed study of the AGE-PRO disk gas masses is presented in \citet{Trapman2024b}.

We find that \textit{Oph 7} is a particular case in terms of disk size and mass, since it has a compact but massive gas disk. As mentioned above, this could indicate that \textit{Oph 7} is in a mass accumulation phase that will eventually trigger a FU Ori-type outburst, which is believed to be a short-lived but important event on the evolution of a Class~I system.

\subsubsection{Non-Detections in Embedded Objects}

In addition to the CO isotopologues, the AGE-PRO observations of embedded sources also targeted molecular lines such as H$_{2}$CO, DCN, N$_{2}$D$^{+}$, SO$_{2}$, CH$_{3}$CN, $^{13}$CH$_{3}$OH, SO, and CH$_{3}$OH to investigate the physical and chemical structure of disks during the embedded stage. Surprisingly, only a few lines were detected in some Ophiuchus sources, see Appendix \ref{App:Cont_Mom05}. That is the case of H$_{2}$CO detected towards the disks of \textit{Oph 2}, \textit{Oph 6}, \textit{Oph 7} \citep[reported in ][]{Flores2023}, \textit{Oph 9}, and \textit{Oph 10}. Also, DCN is detected in \textit{Oph 8} and \textit{Oph 9}, while SO$_{2}$ is only detected in \textit{Oph 7}.

The molecules covered by AGE-PRO are known to trace emission from the different components in embedded systems: disks, outflows, and envelopes \citep[e.g.][]{Harsono2021}. It is then puzzling that despite the strong CO emission detected in the Ophiuchus target list, our ALMA observations result in a low rate of detections from higher density, kinetic temperature, and high energy tracers associated with different physical phenomena characteristic of the Class I and FS objects. Given the general lack of detections, we report the upper limits on the different emission lines by taking the integrated noise within a window of width 10 km s$^{-1}$. Table \ref{Table:Upper_Limits} shows the average 3-$\sigma$ upper limit of the lines. Unfortunately, the lack of detections prevents a detailed analysis of the physical and chemical parameters of the disks and their surroundings as previously done in a few embedded Class I objects \citep[e.g.][]{Artur2019, Harsono2021, Mercimek2022, Artur2023}. One of the possible explanations for the non-detection of these molecules is the relatively low luminosity of the AGE-PRO targets with respect to objects previously studied.

\begin{deluxetable*}{l c c c c c c c c}
\tabletypesize{\scriptsize}
\label{Table:Upper_Limits}
\tablecaption{AGE-PRO Sample: Line Upper limits}
\tablehead{
\colhead{3-$\sigma$ level}  &   \colhead{H$_{2}$CO} & \colhead{DCN}   &  \colhead{N$_{2}$D$^{+}$} & \colhead{SO$_{2}$} &\colhead{CH$_{3}$CN}  &\colhead{$^{13}$CH$_{3}$OH} &\colhead{SO} &\colhead{CH$_{3}$OH} }
\startdata
mJy.km s$^{-1}$ & 7 & 8 & 9 & 8 & 9 & 7 & 6 & 7  \\
\enddata

\end{deluxetable*}

\subsection{Observational limitations in embedded sources}

Within the AGE-PRO sample, the analysis of the embedded sources has proven to be particularly challenging given some important limitations. As previously mentioned,  we were unable to measure disk radii with high accuracy from optical thick molecules such as $^{12}$CO and $^{13}$CO. Although not desirable, this is expected as objects such as \textit{Oph 1}, \textit{Oph 4}, \textit{Oph 5}, and \textit{Oph 6}  are still deeply embedded in the envelope, which is a natural consequence of a forming stellar object. This makes it challenging to unambiguously disentangle signals of early disk evolution directly. Similarly, the overlapping emission between the envelope, outflows, and disk emission also makes it challenging to obtain disk gas sizes in objects such as \textit{Oph 5} and \textit{Oph 7}, because some of the species show velocity structures that appear to trace infalling and rotating material rather than a Keplerian rotation. Furthermore, high extinction levels characteristic of this region also prevent the use of Keplerian rotation masks to distinguish embedded disks from envelopes and outflows, see Fig. \ref{Fig:Mom1}. Notably, direct measurements of size and flux values from embedded disks are not as straightforward as more evolved objects like Class ~II and ~III objects, due to their structures being affected, especially at optically thick tracers.

Although ring and spiral structures in still embedded disks have been successfully observed in Class~0 and ~I objects \citep[e.g.][]{Lee2020, Segura2020, Sheehan2020}, suggesting ongoing planet formation, angular resolutions of $\lesssim$0.1 arcsec are required to detect any of these features in embedded disks. The average angular resolution of AGE-PRO towards Ophiuchus reaches values of $\sim$0.3 arcsec; hence, limiting our capability of searching for scientifically interesting substructures. We should also note that a possible explanation for the lack of visible structures at the 220 GHz continuum emission images could be that the continuum emission is optically thick, as is the case of \textit{Oph 7}, and some other sources present highly inclined disk-like structures as \textit{Oph 1}, \textit{Oph 2}, \textit{Oph 8}, and \textit{Oph 9} or more compacted structures as \textit{Oph 4} and \textit{Oph 5}. Certainly, follow-up programs are crucial to observe these and more embedded sources at $\lesssim$0.1 arcsec angular resolutions to search for any substructures that link embedded disks with the first steps of planet formation and thus, be able to constrain the timescales to build up a planet after the birth of the disk.

\section{Summary and Conclusions}
\label{Sec:Summary}

The AGE-PRO Large Program aims to systematically trace the evolution of gas disk mass and size throughout the lifetime of protoplanetary disks. To this end, a total of 30 disks from the Ophiuchus (10), Lupus (10), and Upper Sco (10) star-forming regions have been included in the target list. This paper describes the Band-6 ALMA observations of the Ophiuchus star-forming region, which harbors one of the youngest stellar objects in the Solar neighborhood at a distance of $\sim$140 PC. Due to the high extinction levels towards our targets, stellar luminosities are highly uncertain. This prevented us from deriving age estimates from comparisons to evolutionary models. Therefore,  for the Ophiuchus sample,   AGE-PRO adopts the statistical ages previously derived for embedded sources (Class I and FS) in nearby star-forming regions ($\lesssim$1 Myr). For our purposes, the selection criteria require YSOs that are still highly embedded in their natal molecular envelope.  Our sample comprises initially classified Class I and FS YSOs with SpTs between M3 and K6. We have extracted (sub-)millimeter data products, including $\sim$1.3 mm continuum and line images for $^{12}$CO, $^{13}$CO, C$^{18}$O, and C$^{17}$O  $J$=2$-$1, and H$_{2}$CO, DCN, N$_{2}$D$^{+}$ $J$=3$-$2 lines, and additional line images of SO$_{2}$, CH$_{3}$CN, $^{13}$CH$_{3}$OH, SO, and CH$_{3}$OH). 
From these ALMA products, we have estimated disk parameters from continuum and CO isotopologues data, and here is the summary of our main findings:

\begin{enumerate}
   \item With an angular resolution of $\sim$0.3 arcsec, the 1.3 mm dust continuum emission reveals 4 nearly edge-on disks ($\geq$70 deg.), and 3 highly inclined disks ($\geq$60 deg.) in our sample. This suggests a potential initial SED misclassification. While we can not rule out that some individual objects 
   (e.g, \textit{Oph 8} and \textit{Oph 9}) might be misclassified due to high inclination, we find that the population of embedded sources (Class I and Flat Spectrum objects) in Ophiuchus is not significantly contaminated by highly inclined Class II sources.

   \item From continuum data, dust disk sizes at 90$\%$  of the total flux  were measured with radii ranging from 
   $\sim$0.5$^{"}$ and 1.2$^{"}$.

   \item Due to the highly embedded nature of the targets, optically thicker lines ($^{12}$CO and $^{13}$CO) are severely contaminated by extended emission (cloud, envelope, and/or outflows). This prevented us from estimating with high accuracy disk radii and flux values from these molecules.

   \item Optically thinner lines, C$^{17}$O and C$^{18}$O, are detected in 7 Ophiuchus objects. In contrast to the optical thicker tracers ($^{12}$CO and $^{13}$CO), C$^{18}$O and C$^{17}$O are less contaminated allowing us to estimate gas radii for most of our sources. We noticed a strong correlation between these isotopologues whose extension traces similar disk radii with a slight difference in their emission levels.

   \item C$^{18}$O and C$^{17}$O radii are generally larger than dust disk sizes. This outcome suggests that dust growth and radial drift have already begun in embedded objects with ages $\lesssim$1 Myr. However, detailed modeling of the dust and gas around embedded disks is needed to test this scenario.

    \item C$^{18}$O and C$^{17}$O fluxes were used to estimate gas disk masses. The estimates from the former are slightly lower than those from the latter,  suggesting  C$^{18}$O emission might be partially optically thick. We conclude that the C$^{17}$O line a more suitable to estimate gas disk masses for highly embedded objects, but that such estimates require very deep observations to be sensitive to the outer regions of the disk.

   \item From $^{12}$CO observations, it has been detected outflows associated with \textit{Oph 3}, \textit{Oph 5}, and \textit{Oph 7}. In addition, \textit{Oph 5} might be associated with accretion streamers, however, this has to be confirmed with future observations and studies.

   \item Among the observed disks, \textit{Oph 7} and \textit{Oph 1} exhibit contrasting characteristics. \textit{Oph 7} is the brightest and one of the most compacted targets in the sample. In contrast, \textit{Oph 1} is the most extended source with a relatively fainter disk.

 \end{enumerate}







\paragraph{Acknowledgement.}

This paper makes use of the following ALMA data: ADS/JAO.ALMA$\#$2021.1.00128.L. ALMA is a partnership of ESO (representing its member states), NSF (USA) and NINS (Japan), together with NRC (Canada), MOST and ASIAA (Taiwan), and KASI (Republic of Korea), in cooperation with the Republic of Chile. The Joint ALMA Observatory is operated by ESO, AUI/NRAO, and NAOJ. The National Radio Astronomy Observatory is a facility of the National Science Foundation operated under cooperative agreement by Associated Universities, Inc.

L.A.C and C.G.R. acknowledge support from the Millennium Nucleus on Young Exoplanets and their Moons (YEMS), ANID - Center Code NCN2021\_080 and L.A.C. also acknowledges support from the FONDECYT grant \#1241056. J.M. acknowledges support from FONDECYT de Postdoctorado 20241134 No 3240612.
D.D. and I.P. acknowledge support from Collaborative NSF Astronomy \& Astrophysics Research grant (ID: 2205870).
L.P. acknowledges support from ANID BASAL project FB210003 and ANID FONDECYT Regular \#1221442.
K.Z. and L.T.acknowledges the support of the NSF AAG grant \#2205617.
N.T.K. acknowledges support provided by the Alexander von Humboldt Foundation in the framework of the Sofja Kovalevskaja Award endowed by the Federal Ministry of Education and Research.
C.A.G. acknowledges support from FONDECYT de Postdoctorado 2021 \#3210520. 
P.P. and A.S. acknowledge the support from the UK Research and Innovation (UKRI) under the UK government’s Horizon Europe funding guarantee from ERC (under grant agreement No 101076489).
A.S. also acknowledges support from FONDECYT de Postdoctorado 2022 $\#$3220495.
B.T. acknowledges support from the Programme National “Physique et Chimie du Milieu Interstellaire” (PCMI) of CNRS/INSU with INC/INP and co-funded by CNES.
G.R. acknowledges funding from the Fondazione Cariplo, grant no. 2022-1217, and the European Research Council (ERC) under the European Union’s Horizon Europe Research \& Innovation Programme under grant agreement no. 101039651 (DiscEvol). Views and opinions expressed are however those of the author(s) only, and do not necessarily reflect those of the European Union or the European Research Council Executive Agency. Neither the European Union nor the granting authority can be held responsible for them. K.S. acknowledges support from the European Research Council under the Horizon 2020 Framework Program via the ERC Advanced Grant Origins 83 24 28.

All figures were generated with the \texttt{PYTHON}-based package \texttt{MATPLOTLIB} \citep{Hunter2007}. This research made use of Astropy,\footnote{http://www.astropy.org} a community-developed core Python package for Astronomy \citep{astropy:2013, astropy:2018}, and Scipy \citep{2020SciPy-NMeth}.

\clearpage


\appendix

\restartappendixnumbering

\section{Extinction and Luminosity}
\label{App:Extinction}

\begin{deluxetable*}{l c c c c c c c c}
\tabletypesize{\scriptsize}
\label{Table:Extinction}
\tablecaption{AGE-PRO Sample: Extinction and luminosities}
\tablehead{
\colhead{Age-Pro ID}  &   \colhead{${A}_v$} & \colhead{$\log{L_\ast}$}   &  \colhead{${A}_v$}\tablenotemark{a} & \colhead{$\log{L_\ast}$} &\colhead{${A}_v$}\tablenotemark{b}  &\colhead{${A}_v$}\tablenotemark{b} &\colhead{BCK} &\colhead{$\alpha_{5-12}$}  \\
& [mag] & [$L_\odot$] & [mag.] & [$L_\odot$] & [mag.] & [mag.] }
\startdata
Oph 1\ &  16.1 &  $-0.34\pm0.1$ & 27.3 & $0.07\pm0.3$ &   24.5  & 33.3 & 2.53 & 0.76  \\
Oph 2 &   10.1 & $-1.94\pm0.1$  & 21.2 & $-1.54\pm0.3$ &   9.6 & 13.0 & 2.67 & 0.39  \\
Oph 3\ &  31.5  & $-0.29\pm0.1$ &28.8 & $-0.42\pm0.3$ &   28.7 & 39.1 & 2.53 & 1.23  \\
Oph 4\ &  24.5 &  $0.02\pm0.1$ & 32.6 & $0.32\pm0.3$ &   63.8  & 87.0 & 2.52 & 0.37  \\
Oph 5\ &  40  &   $0.01\pm0.1$  & 39.7 & $0.01\pm0.3$ &   26.6 & 36.2 & 2.57 & 1.03  \\
Oph 6\ &  12.1 &  $-1.37\pm0.1$ & 24.1 & $-0.94\pm0.3$ &   11.7 & 15.9 & 2.70 & -0.19  \\
Oph 7\ &  23.3  & $0.33\pm0.1$ & 35.1 & $0.76\pm0.3$ &   29.8 & 40.6b& 2.52 & 0.83  \\
Oph 8\ &  1.3  &  $-2.15\pm0.3$ & 7.6 &$-1.92\pm0.3$ &   2.34 & 3.20 & 2.66 &  -1.10  \\
Oph 9\ &  5.0 &  $-0.7\pm0.1$  & 24.1 &$-0.01\pm0.3$ &   6.4 & 8.6 & 2.68 &  -0.46  \\
Oph 10\ & 44 &  $-0.17\pm0.1$  & 29.5 & $-0.69\pm0.3$ &   23.4  & 31.9& 2.62 & 1.22  \\
\enddata
\tablecomments{Column (1) lists the AGE-PRO name of the target; Column (2) Visual extinction values used in the spectral classification, see Section \ref{sec:sample}; Column (3) Stellar luminosities from 2MASS K photometry and using ${A}_v$ from Table \ref{Table:Sample}, Column (4) Optical extinctions estimated by JHK colors; Column (5) Stellar luminosities from 2MASS K photometry and using ${A}_v$ from JHK colors; Columns (6), and (7) Upper limits from \textit{Herschel} observations for thin and thick media, respectively; Column (8) Bolometric corrections computed from BT-Settl theoretical evolutionary models of \citep{Baraffe2015}; Column (9) Spectral index between 5 and 12 $\micron$ ($\alpha_{5-12}$) to evaluate evolutionary state of YSOs according to \citep{McClure2010}.}
\tablenotetext{a}{${A}_v$ values estimated using $ A_{J} = 2.62[(J-H) - (J-H)_{o}]$}
\tablenotetext{b}{${A}_v$ values using the standard conversions $N_{H_{2}} (cm^{-2})=0.94 \times 10^{21}A_{V}$ and $N_{H_{2}} (cm^{-2})=0.69 \times 10^{21}A_{V}$ for diffuse and dense molecular medium, respectively \citep{Bohlin1978, Draine2003}. 
}
\end{deluxetable*}

Accuracy in the estimates of stellar parameters for embedded YSOs is significantly affected by the line-of-sight extinction levels toward each object. Extinction values measured from observed color excesses relative to an intrinsic color depend on the wavelength regime, extinction curve \citep[e.g.][]{Mathis1990, Fitzpatrick1999}, and a specific spectral type, but work relatively well for bare stellar photospheres and Class II sources. However, for embedded YSOs, methods based on color excesses are particularly problematic because all optical and near-IR wavelengths are highly contaminated by non-photospheric emission. While the accretion shock mostly affects the optical wavelengths, the thermal emission from the inner disk overwhelms the stellar photosphere in the near-IR. These non-photospheric components impose strong limitations on these methods and lead to a degeneracy between the extinction and spectral type \citep[e.g][]{Ricci2010, Furlan2009}. Still, we have estimated extinction values from observed 2MASS J - K colors ($J-H$) by using $ A_{J} = 2.62[(J-H) - (J-H)_{o}]$, where $(J-H)_{o}$ is the expected PMS color from \citet{Pecaut2013}, and an extinction relation $A_{J}/A_{V} = 0.26$ from \citet{Cieza2005}. Table \ref{Table:Extinction} lists $A_{V}$ values for each AGE-PRO Ophiuchus target obtained from the J-H color excess, and the corresponding ${L_\ast}$ values estimated following a similar approach as mentioned in Section \ref{sec:sample} (applying a bolometric correction to the extinction-corrected K-band). As seen in Table \ref{Table:Extinction}, the extinctions calculated from the J-H color excesses (fourth column) can be significantly different from the values adopted by AGE-PRO (second column).

Although there have been different efforts to reduce the extinction uncertainties by fitting observed spectra to artificially reddened spectral templates, thus providing simultaneously SpT and $A_{V}$ \citep[e.g.][]{Manara2015}, this approach unfortunately still suffers from large uncertainties that depend on the signal to noise, the adopted reddening laws and the wavelength coverage.  As a result, the extinction measurements for the same YSO can vary dramatically in the literature \citep{Carvalho2022},  which in turn affects the estimates of its bolometric luminosity.

\section{AGE-PRO Ophiuchus Correlator Set-ups and ALMA Observational log}
\label{App:SSs}

The correlator setups and observational log for AGE‐PRO observations and the Archival data are summarized in Tables \ref{Table:SSs} and \ref{Table:Obs_log}.

\begin{deluxetable*}{llccccccc}
\tablecaption{AGE-PRO Correlator Set-ups}
\label{Table:SSs}
\tablehead{
\colhead{Set-up}   &\colhead{Center Freq.}  &\colhead{Line Targets }   & \colhead{Vel. Res.}    & \colhead{Bandwidth}\\
 \colhead{}   & \colhead{[GHz]} &  \colhead{} &\colhead{[km~s$^{-1}$]}    & \colhead{[MHz]} 
}
\startdata
 &217.238530  &DCN J=3--2 &0.097  &58.59\\
 &218.222192  &H$_2$CO $3_{03}-2_{02}$  &0.097  &58.59\\
&219.560358  &C$^{18}$O J=$2-1$  &0.193  &58.59\\
&220.398684  &$^{13}$CO J=$2-1$  &0.192  &58.59\\
SS1 &220.679320  &CH$_3$CN  12$_1$--11$_1$   &0.192  &58.59\\
&220.742990  &CH$_3$CN   12$_4$--11$_4$    &0.192  &58.59\\
&230.538000  &CO J=2--1  &0.092  &58.59\\
&231.321828  &N$_2$D$^+$ J=3--2  &0.091  &58.59\\
&234.000000  &Continuum band  &1.446  &1875\\
\hline
&220.500000  &Continuum band  &1.535  &1875\\
&223.883569  &SO$_2$ 6$_{4,2}$-7$_{3,5}$  &0.096  &58.59\\
&224.714373  &C$^{17}$O J=$2-1$  &0.094  &58.59\\
SS2 &234.683390  &CH$_3$OH  4$_{2,3}$--5$_{1,4}$  &0.180  &58.59\\
&235.151720  &SO$_2$ 4$_{22}$-3$_{13}$  &0.180  &58.59\\
&236.452293  &SO 1$_2$--2$_1$  &0.179  &58.59\\
&237.983380  &$^{13}$CH$_3$OH  5$_{1,4}$--4$_{1,3}$  &0.089  &58.59\\
&239.137925  &CH$_3$CN  13$_0$--12$_0$  &0.088  &58.59\\
\hline
\enddata
\end{deluxetable*}

\begin{deluxetable*}{lllllllll}
\tabletypesize{\scriptsize}
\tablecaption{ALMA Observational log for AGE-PRO Ophiuchus sample \label{Table:Obs_log}}
\tablewidth{0pt}
\tablehead{
\colhead{Setup} & \colhead{Source Name} & \colhead{UTC Date} & \colhead{Config}  & \colhead{Baselines} & 
\colhead{N$_{\rm ant}$} & \colhead{Elev} & \colhead{PWV} & \colhead{Calibrators} 
\\
\colhead{} & \colhead{} &  \colhead{} & \colhead{} & \colhead{(m)} & 
\colhead{} & \colhead{(deg)} & \colhead{(mm)} & \colhead{}
}
\startdata
  Oph CO (SS1)        &  Oph 1-6, 8-10            &  2022-03-29 06:06:   &  C43-2  &  14.6 - 313.7   &     45&   65.3&    2.2&  J1517-2422, J1625-2527   \\
   &  Oph 1-6, 8-10            &  2022-03-30 06:34:   &  C43-2  &  14.6 - 313.7   &     45&   72.5&    1.1&  J1517-2422, J1625-2527   \\
   &  Oph 1-6, 8-10            &  2022-04-08 05:35:   &  C43-2  &  14.6 - 313.7   &     45&   67.2&    0.5&  J1517-2422, J1625-2527   \\
   &  Oph 1-6, 8-10            &  2022-07-03 03:10:   &  C43-5  &  15.3 - 1301.6  &     42&   67.4&    0.5&  J1517-2422, J1700-2610   \\
   &  Oph 1-6, 8-10            &  2022-07-05 02:17:   &  C43-5  &  15.1 - 1996.7  &     41&   76.6&    1.5&  J1517-2422, J1700-2610   \\
   &  Oph 1-6, 8-10            &  2022-07-05 03:50:   &  C43-5  &  15.1 - 1996.7  &     41&   55.8&    1.5&  J1517-2422, J1700-2610   \\
   &  Oph 1-6, 8-10            &  2022-07-16 21:35:   &  C43-6  &  15.1 - 2617.4  &     42&   49.2&    0.7&  J1517-2422, J1700-2610   \\
   &  Oph 1-6, 8-10            &  2022-07-16 22:59:   &  C43-6  &  15.1 - 2617.4  &     41&   68.3&    0.6&  J1517-2422, J1700-2610   \\
   &  Oph 1-6, 8-10            &  2022-07-18 01:32:   &  C43-6  &  15.3 - 2617.4  &     44&   75.4&    0.5&  J1517-2422, J1700-2610   \\
   &  Oph 1-6, 8-10            &  2022-07-18 23:10:   &  C43-6  &  15.3 - 2617.4  &     43&   71.8&    1.4&  J1517-2422, J1700-2610   \\
   Oph CO (FAUST) &  Oph 7                    &  2019-01-21 10:46:   &  C43-2  &  15.0 - 313.7  &     48&   61.7&    2.1&  J1625-2527 , J1700-2610  \\ 
   &  Oph 7                    &  2018-12-02 17:09:   &  C43-4  &  15.1 - 783.5  &     44&   70.6&    0.7&  J1625-2527, J1700-2610, J1626-2951   \\ 
   &  Oph 7                    &  2019-05-02 04:53:   &  C43-4  &  15.1 - 783.5  &     43&   76.5&    1.0&  J1625-2527, J1700-2610, J1626-2951   \\
   &  Oph 7                    &  2020-03-17 09:46:   &  C43-5  &  15.2 - 1231.4  &     -&   77.2&    2.7&  J1625-2527, J1700-2610, J1626-2951   \\
   Oph CO (eDisk) &  Oph 7                    &  2022-06-14 04:13:   &  C43-5  &  15.1 - 1301.6  &     42&   70.3&    0.3&  J1650-2943, J1700-2610 \\ 
   &  Oph 7                    &  2022-06-15 00:45:   &  C43-5  &  15.1 - 1301.6  &     43&   62.8&    0.5&  J1650-2943, J1700-2610   \\
   \hline
  Oph C$^{17}$O (SS2) &  Oph 1-10                 &  2022-04-03 11:09:   &  C43-2  &  14.6 - 360.6   &     44&   40.8&    1.3&  J1924-2914, J1700-2610   \\
   &  Oph 1-10                 &  2022-04-09 05:31:   &  C43-2  &  15.0 - 313.7   &     41&   67.0&    0.9&  J1517-2422, J1625-2527   \\
   &  Oph 1-10                 &  2022-04-09 08:03:   &  C43-2  &  15.1 - 313.7   &     41&   77.9&    0.8&  J1517-2422, J1625-2527   \\
   &  Oph 1-10                 &  2022-07-05 05:14:   &  C43-5  &  15.1 - 1996.7  &     41&   35.7&    1.5&  J1924-2914, J1700-2610   \\
   &  Oph 1-10                 &  2022-07-19 02:05:   &  C43-6  &  15.3 - 2617.4  &     44&   66.3&    0.9&  J1517-2422, J1700-2610   \\
   &  Oph 1-10                 &  2022-07-22 22:40:   &  C43-6  &  15.3 - 2617.4  &     42&   70.4&    1.3&  J1517-2422, J1700-2610   \\
  &  Oph 1-10                 &  2022-07-29 22:18:   &  C43-6  &  15.1 - 2517.3  &     38&   71.5&    1.2&  J1517-2422, J1700-2610   \\
   &  Oph 1-10                 &  2022-07-30 01:32:   &  C43-6  &  15.1 - 2517.3  &     38&   66.2&    1.1&  J1517-2422, J1700-2610   \\
   &  Oph 1-10                 &  2022-07-31 01:36:   &  C43-5  &  15.1 - 1996.7  &     44&   62.1&    1.0&  J1517-2422, J1700-2610   \\
   &  Oph 1-10                 &  2022-08-01 21:37:   &  C43-5  &  15.1 - 1301.6  &     42&   65.0&    1.1&  J1517-2422, J1700-2610   \\
   &  Oph 1-10                 &  2022-08-02 00:52:   &  C43-5  &  15.1 - 1301.6  &     42&   70.1&    0.6&  J1517-2422, J1700-2610   \\
   &  Oph 1-10                 &  2022-08-03 01:26:   &  C43-5  &  15.1 - 1301.6  &     43&   61.5&    0.9&  J1517-2422, J1700-2610   \\
\hline
\enddata
\end{deluxetable*}

\section{234 GHz continuum and Line Observations}
\label{App:Cont_Mom05}

\begin{figure*}
\centering
\gridline{\fig{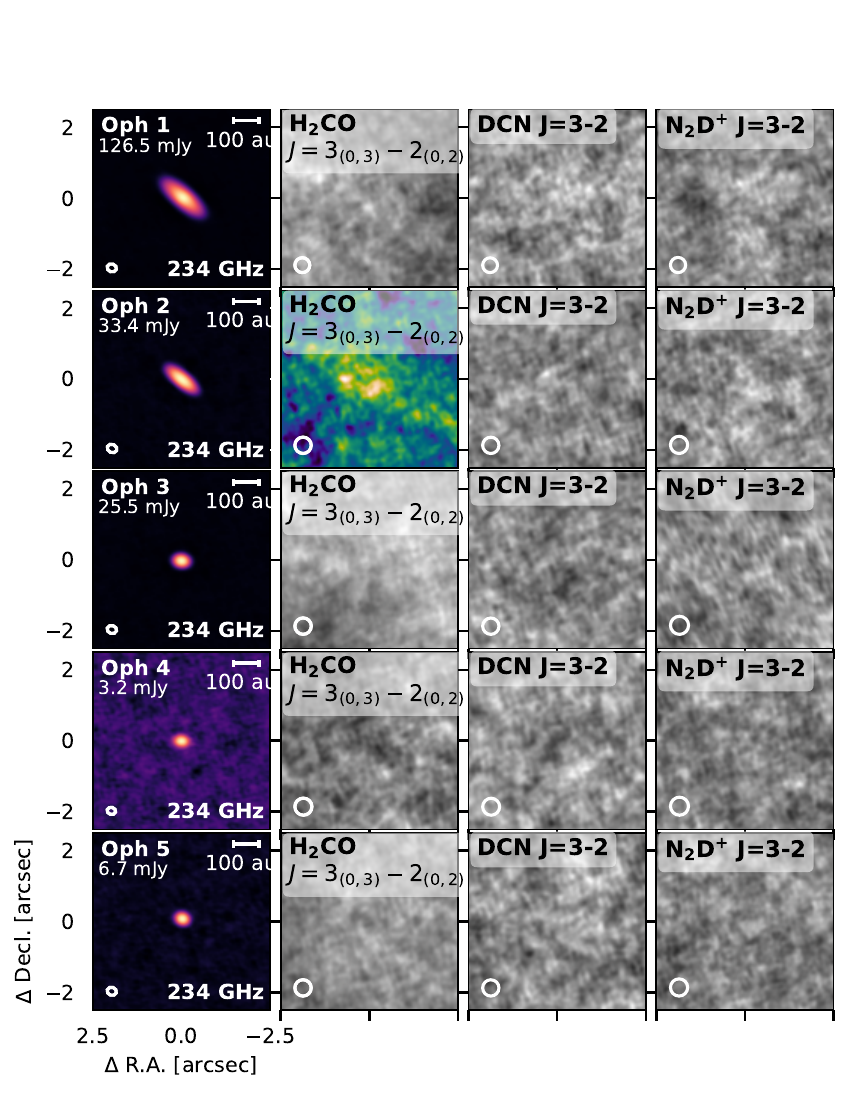}{0.5\textwidth}{}
   \fig{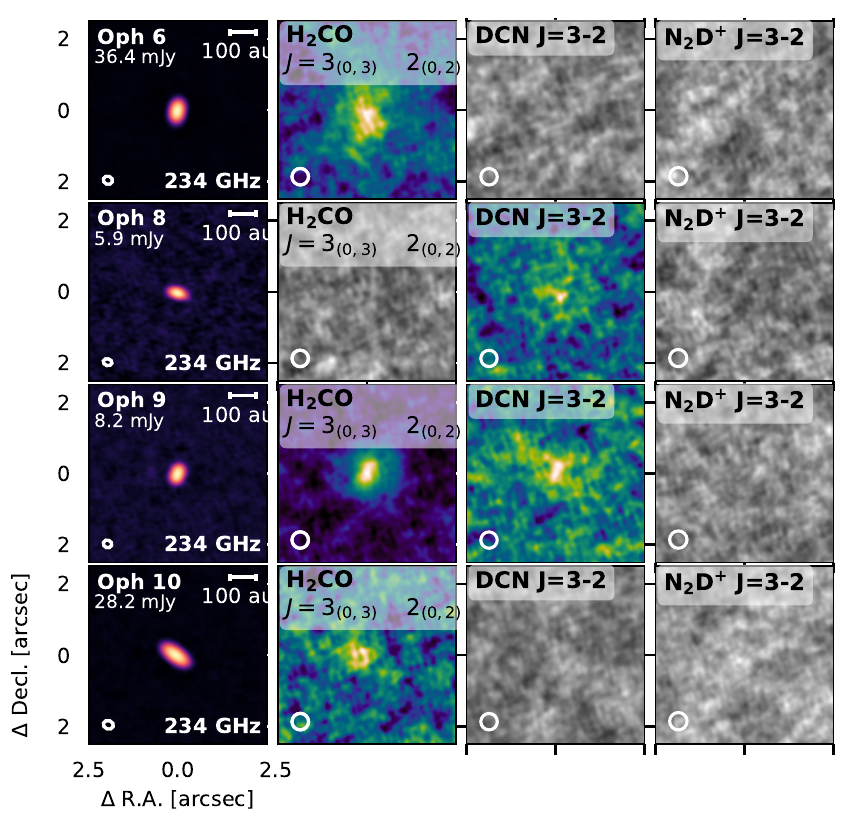}{0.5\textwidth}{}}
\figcaption{234 GHz continuum images and H$_{2}$CO, DCN v=0 $J$=3$-$2, and N$_{2}$D$^{+}$ $J$=3$-$2 Moment-0 maps of the AGE-PRO Ophiuchus targets. These ALMA products are obtained with a robust parameter of 1 in the velocity ranges displayed in Table \ref{Table:Mom_parameters} and generate an average beam size of 0.5 $^{"}$. Flux peak values are added at the left upper side. The resulting beam sizes are represented with the white circles at the lower left corner. Colored and black-and-white images indicate detections and non-detections in our sample, respectively.
\label{Fig:Mom0SS1}
}
\end{figure*}

\begin{figure*}
\centering
\gridline{\fig{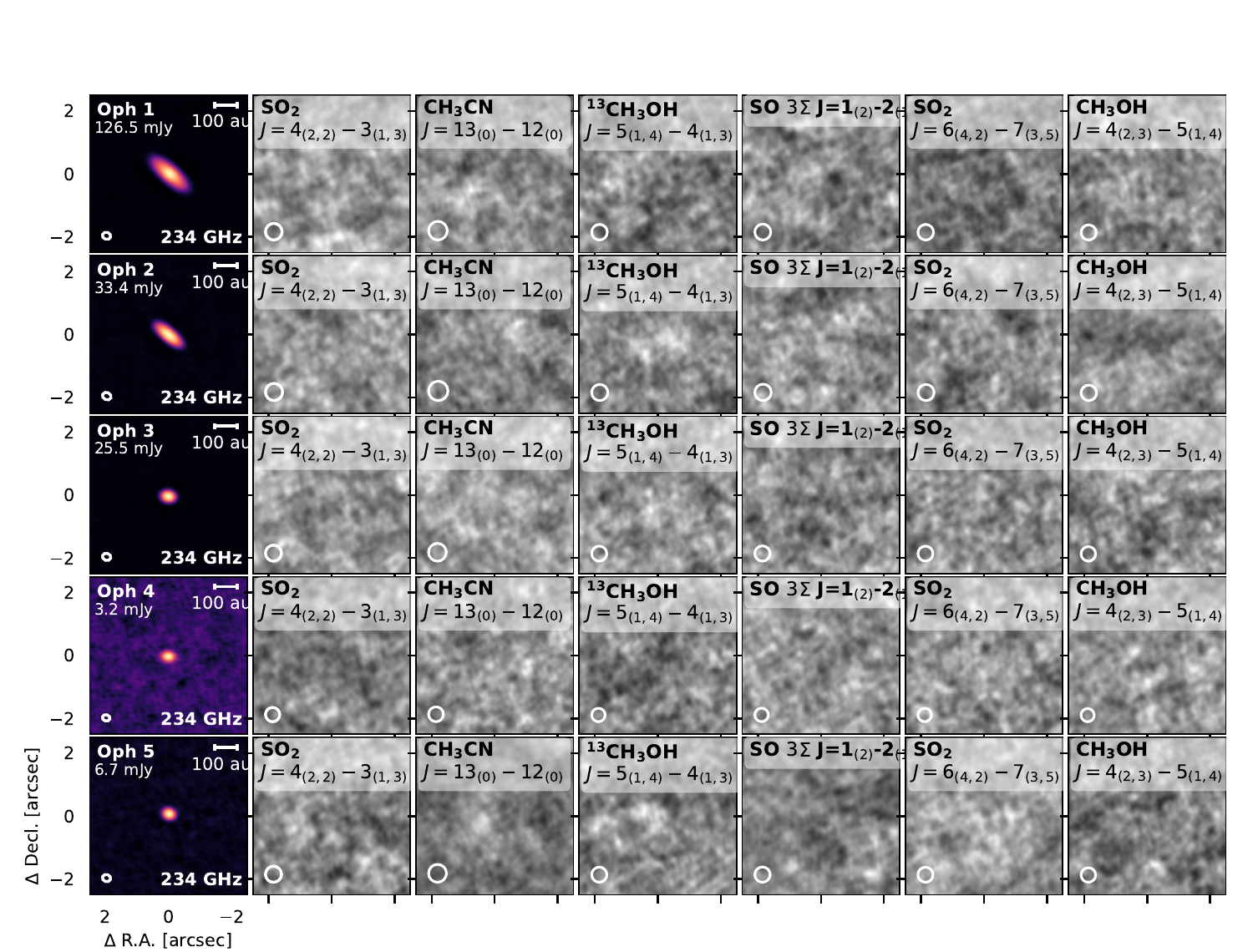}{0.5\textwidth}{}
   \fig{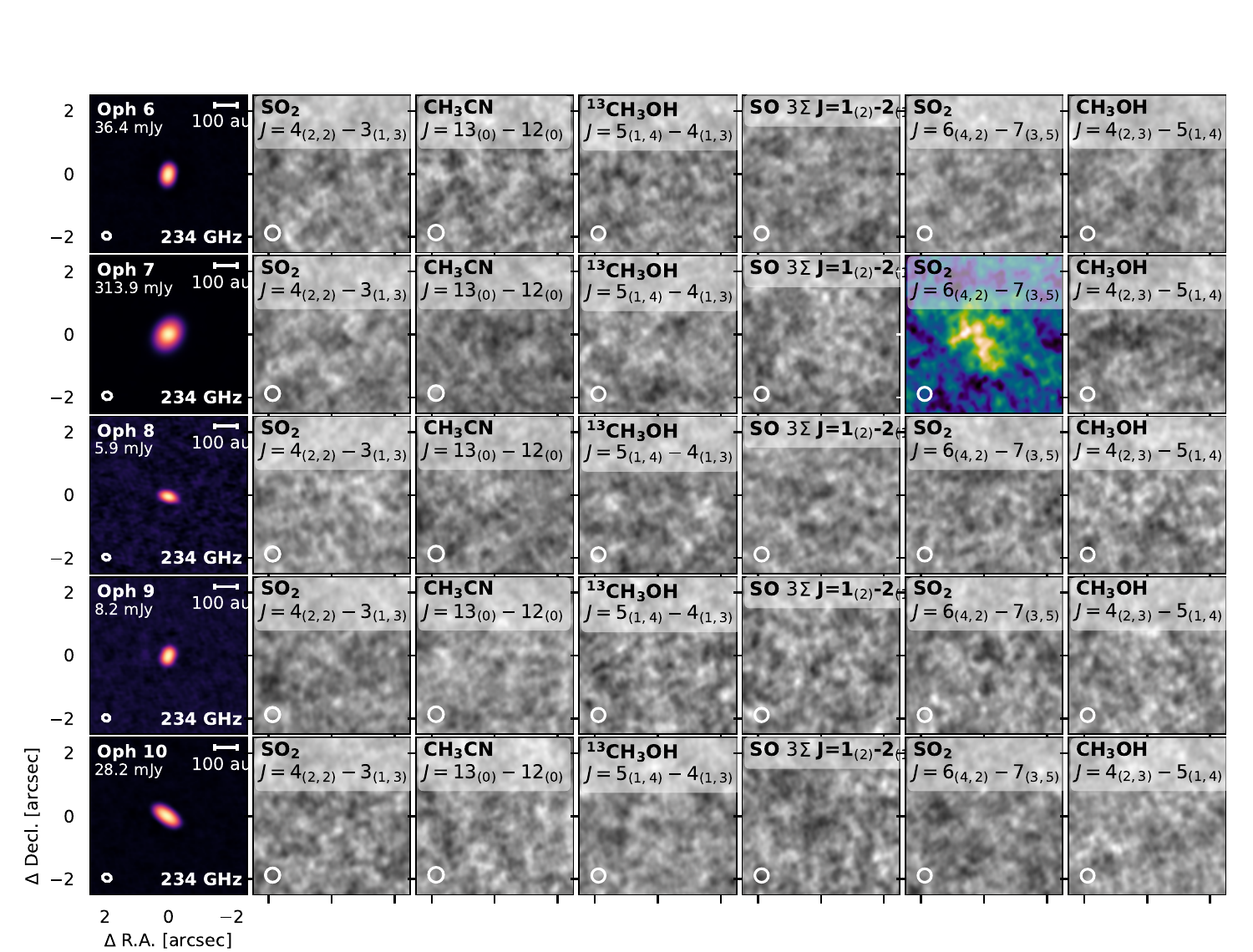}{0.5\textwidth}{}}
\figcaption{234 GHz continuum images and O$_{2}$ v=0, CH$_{3}$CN v=0, $^{13}$CH$_{3}$OH v=0, SO 3$\Sigma$, SO$_{2}$ v=0, and CH$_{3}$OH v=0 Moment-0 maps of the AGE-PRO Ophiuchus targets. These ALMA products are obtained with a robust parameter of 1 in the velocity ranges displayed in Table \ref{Table:Mom_parameters} and generate an average beam size of 0.5 $^{"}$. Flux peak values are added at the left upper side. The resulting beam sizes are represented with the white circles at the lower left corner. Colored and black-and-white images indicate detections and non-detections in our sample, respectively.
\label{Fig:Mom0SS2}
}
\end{figure*}

As presented in Section \ref{sec:Reduction}, we extracted 234 GHz and line data cubes for SS1 and SS2. Unfortunately, very few lines were detected toward most of the targets. The lack of detections may be due to the highly embedded nature or the high inclinations of these sources. In addition, we generated continuum emission maps imaged with robust parameters of 0.5 and 1. Throughout the main text, results and analysis are based on products imaged with a robust parameter of 1. From continuum products imaged with a Briggs of 0.5, we also estimated radial extensions at 68$\%$ and 90$\%$ of the total flux following the approach shown in Section \ref{Sec:Flux_Radii}. Table \ref{Tab:Continuum05} shows radial estimates for 220 and 234 GHZ continuum emissions with a robust parameter of 0.5. Figures \ref{Fig:Mom0SS1} and \ref{Fig:Mom0SS2} show 234 GHz continuum imaged with a robust parameter of 0.5 and line emission lines for SS1 and SS2 imaged with a robust parameter of 1, respectively. 

\begin{deluxetable*}{l c  c c c c c c c}
\tabletypesize{\scriptsize}
\label{Tab:Continuum05}
\tablecaption{Disk Parameters from  234 and 220 GHz Continuum emission with Briggs of 0.5}
\tablehead{\colhead{Source} & \colhead{R 90$\%$} & \colhead{R 68$\%$} & \colhead{Flux at 234 GHz} & \colhead{R 90$\%$} &  \colhead{R 68$\%$}\tablenotemark{b} & \colhead{Flux at 220 GHz}\tablenotemark{b}} 
\startdata
	   Oph 1\tablenotemark{a} &  1	& 0.72	& 125.7	& 0.997	& 0.714	& 104\\
     \midrule
     Oph 2   & 0.894	& 0.627	& 33.3	& 0.944	& 0.656	& 27.3\\
     \midrule
     Oph 3\   & 0.498	& 0.347	& 23.7	& 0.515	& 0.37	& 20.4\\
     \midrule
     Oph 4\ & 0.428	& 0.31	& 2 &	0.407	& 0.305	& 1.7\\
     \midrule
     Oph 5\  & 0.399	& 0.295	& 4.9	& 0.533	& 0.34	& 4.3 \\
     \midrule
     Oph 6\ & 0.612	& 0.434	& 36.1	& 0.668	& 0.458	& 34.5 \\
     \midrule
     Oph 7\  & 	--&-- &-- & 0.695	& 0.483	& 305.7\\
     \midrule
     Oph 8\ & 0.85	& 0.542	& 4.8	& 0.99	& 0.604	& 4.2\\
     \midrule
     Oph 9\ & 0.747	& 0.503 & 	8.4 &	0.85	&0.545	& 8 \\
     \midrule
     Oph 10\ & 0.677	& 0.495	& 27.4	& 0.777	& 0.542	& 26.5  \\
     \enddata
\end{deluxetable*}



\section{CO Isotopologue channel maps}
\label{App:Channel}

\begin{figure*}[h!]
\includegraphics[width=\linewidth]{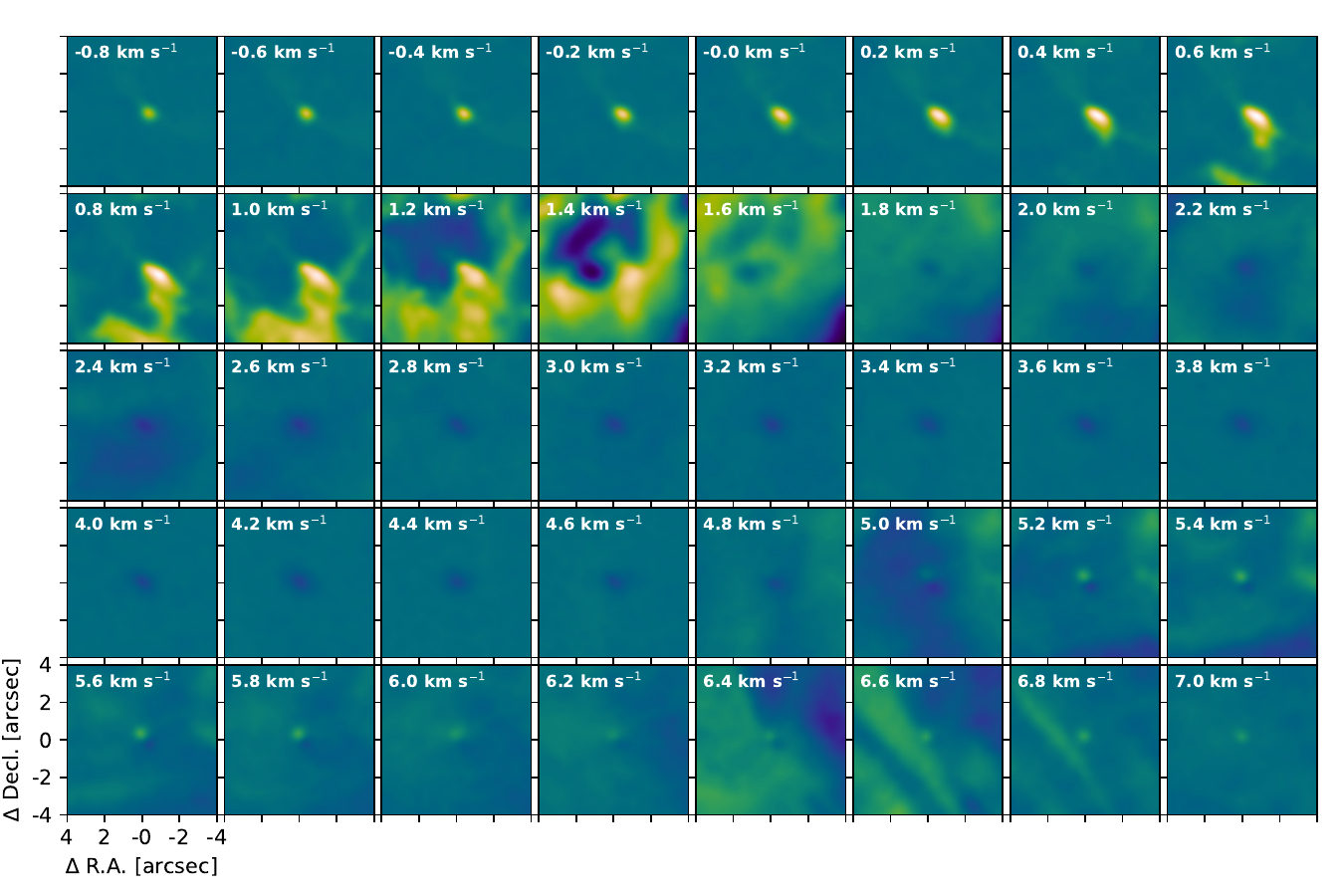}
\figcaption{Channel maps of the $^{12}$CO line emission for \textit{Oph 1}.
\label{Fig:Oph112CO}
}
\end{figure*}

\begin{figure*}[h!]
\includegraphics[width=\linewidth]{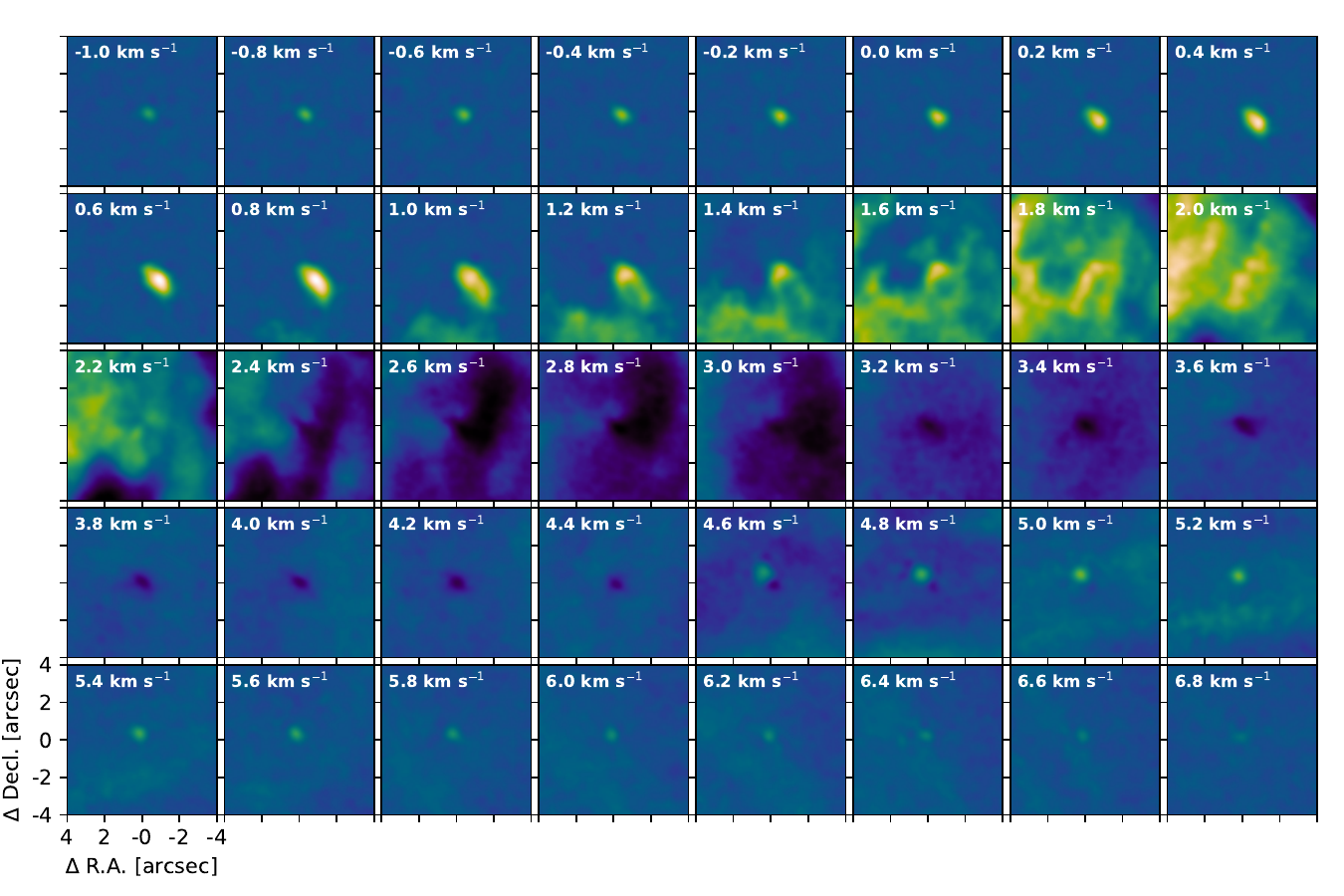}
\figcaption{Channel maps of the $^{13}$CO line emission for \textit{Oph 1}.  
\label{Fig:Oph113CO}
}
\end{figure*}

\begin{figure*}[h!]
\includegraphics[width=\linewidth]{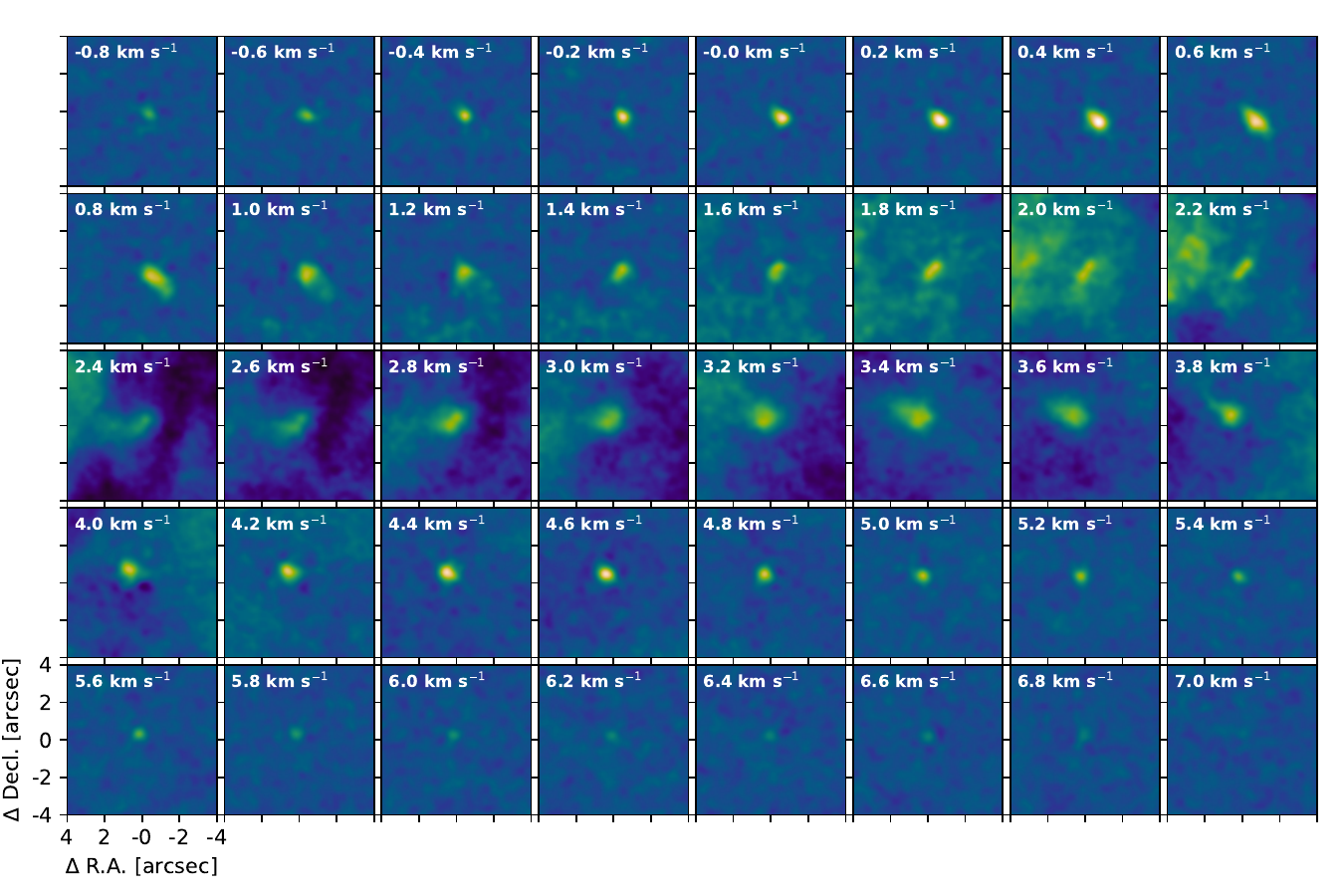}
\figcaption{Channel maps of the C$^{18}$O line emission for \textit{Oph 1}.  
\label{Fig:Oph1C18O}
}
\end{figure*}

\begin{figure*}[h!]
\includegraphics[width=\linewidth]{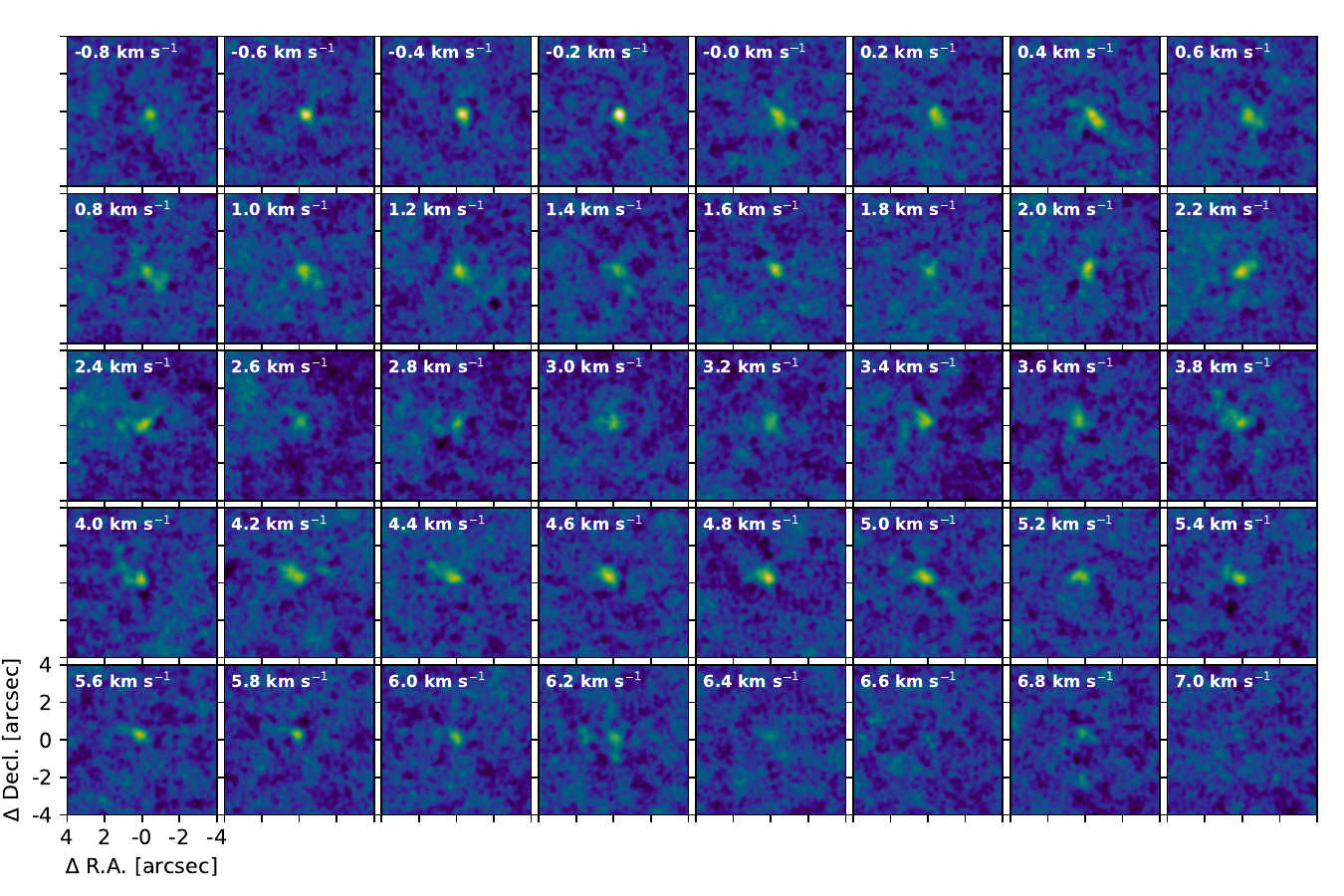}
\figcaption{Channel maps of the C$^{17}$O line emission for \textit{Oph 1}.  
\label{Fig:Oph1C17O}
}
\end{figure*}

\begin{figure*}[h!]
\includegraphics[width=\linewidth]{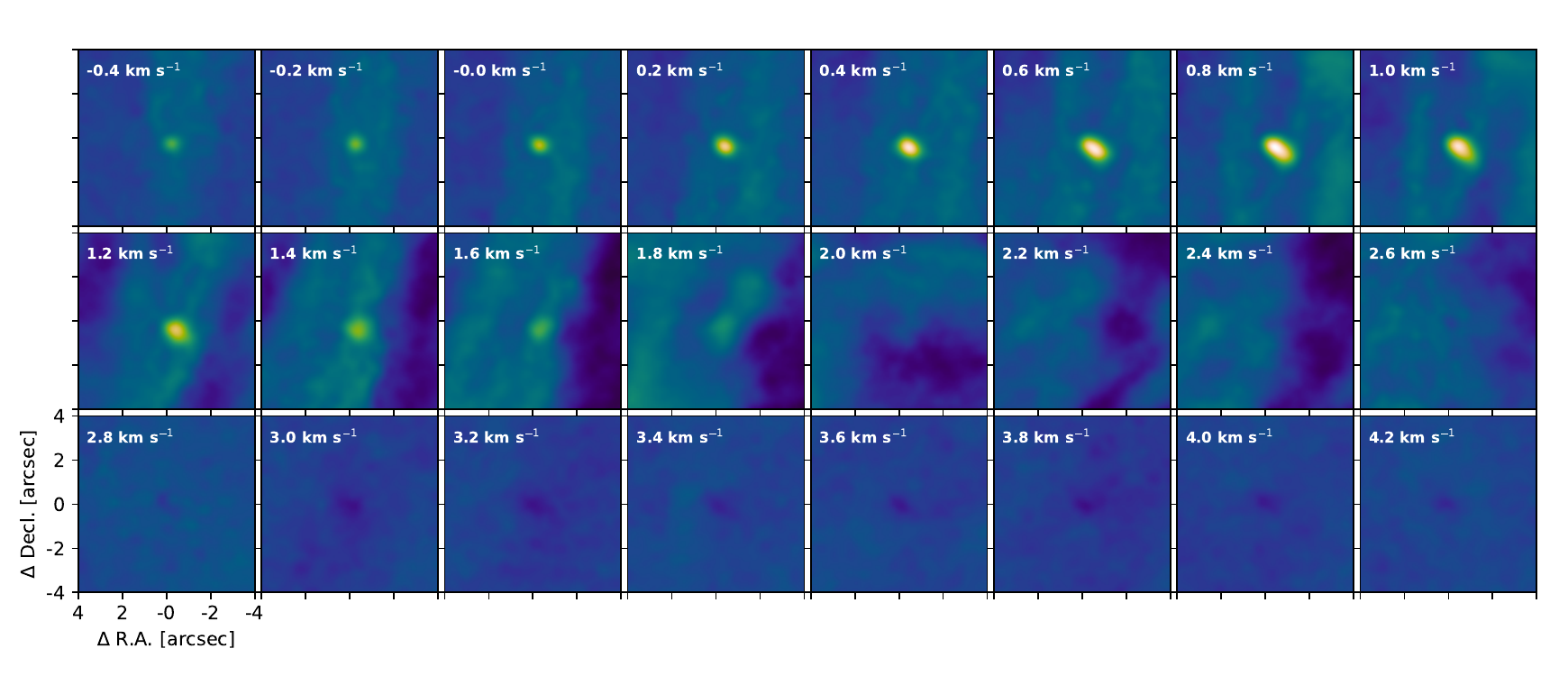}
\figcaption{Channel maps of the $^{12}$CO line emission for \textit{Oph 2}. 
\label{Fig:Oph212CO}
}
\end{figure*}

\begin{figure*}[h!]
\includegraphics[width=\linewidth]{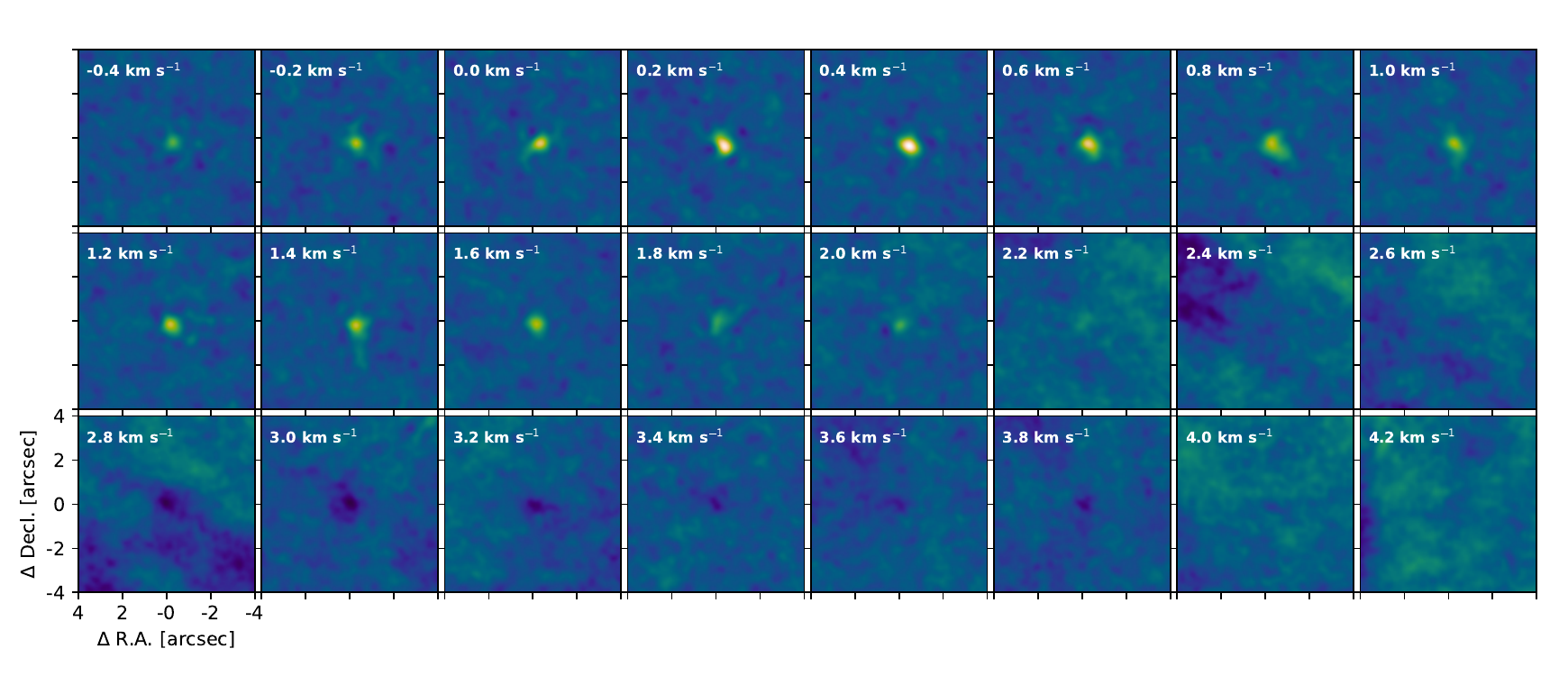}
\figcaption{Channel maps of the $^{13}$CO line emission for \textit{Oph 2}. 
\label{Fig:Oph213CO}
}
\end{figure*}

\begin{figure*}[h!]
\includegraphics[width=\linewidth]{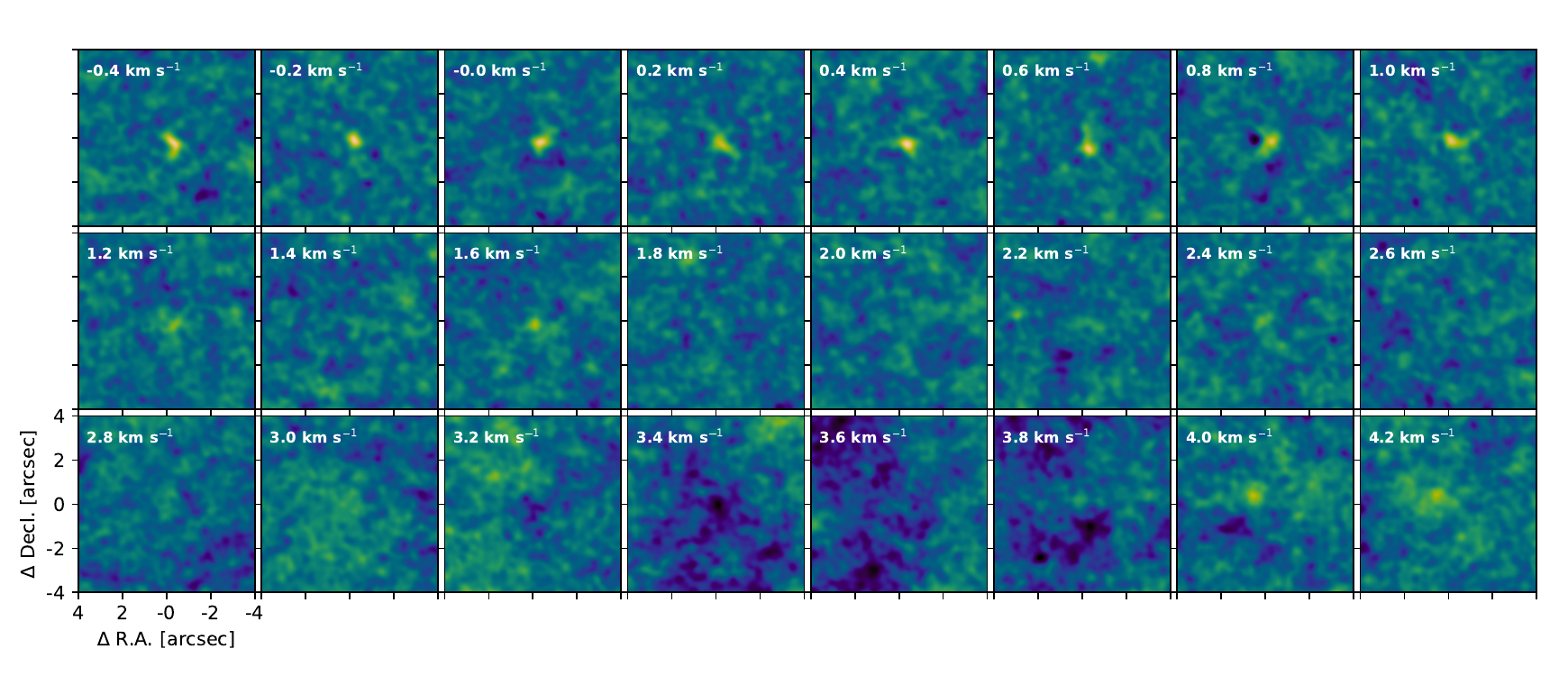}
\figcaption{Channel maps of the C$^{18}$O line emission for \textit{Oph 2}. 
\label{Fig:Oph2C18O}
}
\end{figure*}

\begin{figure*}[h!]
\includegraphics[width=\linewidth]{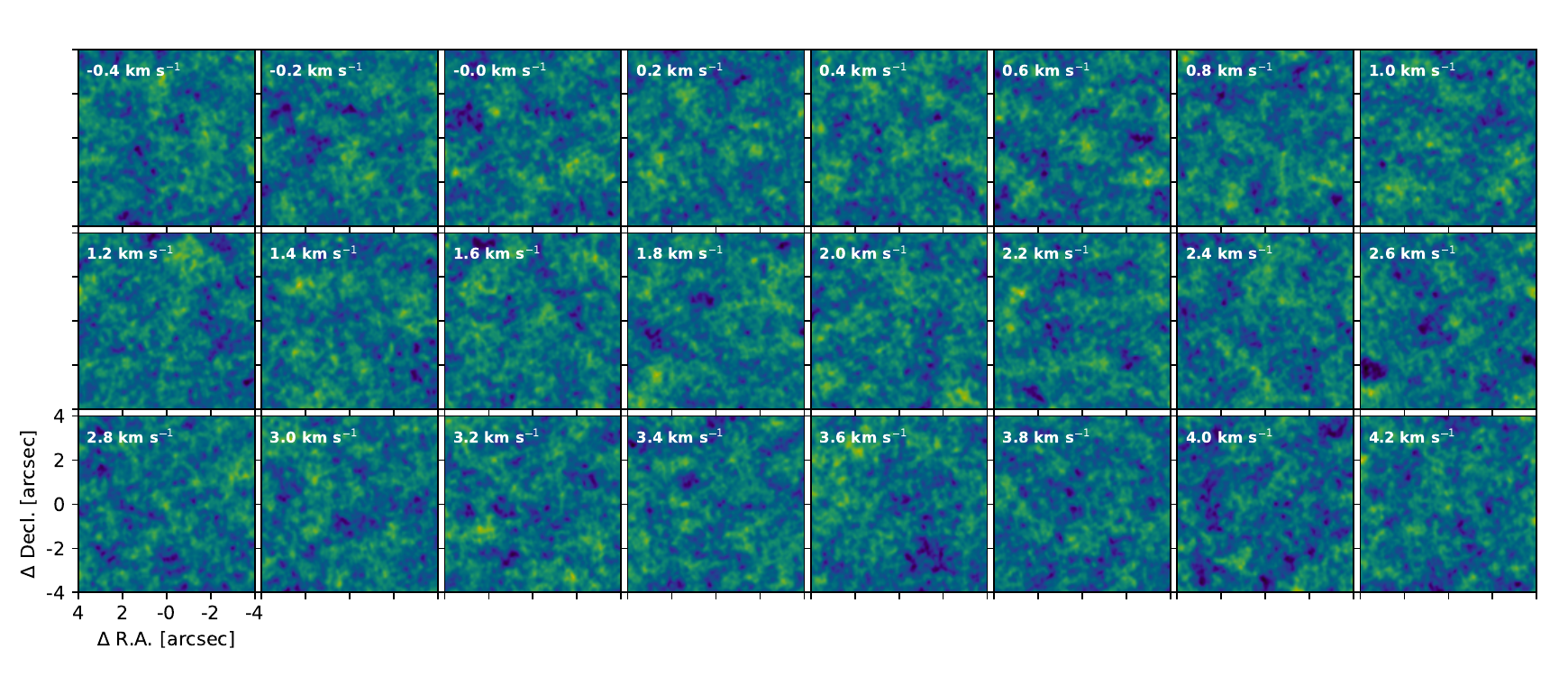}
\figcaption{Channel maps of the C$^{17}$O line emission for \textit{Oph 2}. 
\label{Fig:Oph2C17O}
}
\end{figure*}

\begin{figure*}[h!]
\includegraphics[width=\linewidth]{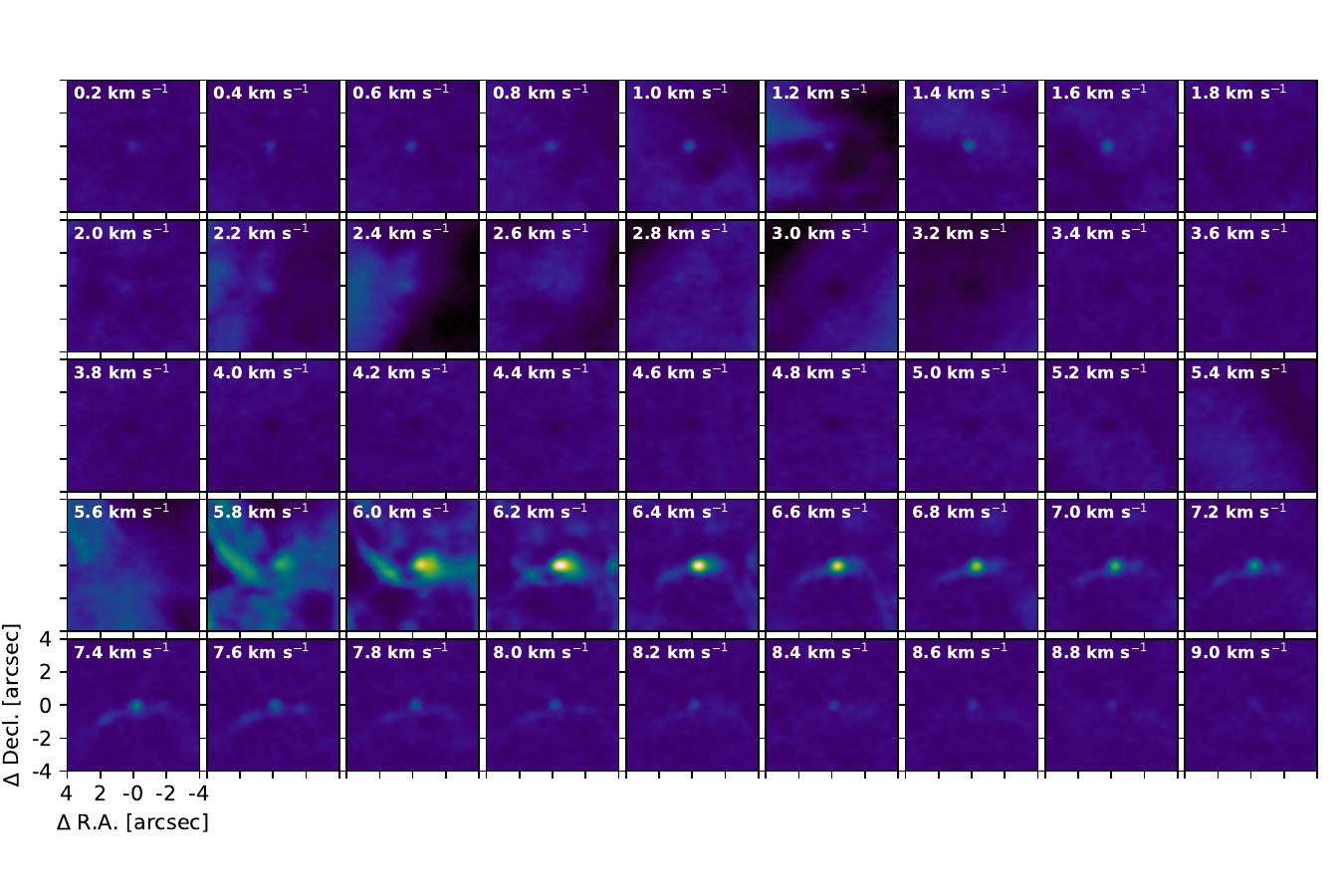}
\figcaption{Channel maps of the $^{12}$CO line emission for \textit{Oph 3}. 
\label{Fig:Oph312CO}
}
\end{figure*}

\begin{figure*}[h!]
\includegraphics[width=\linewidth]{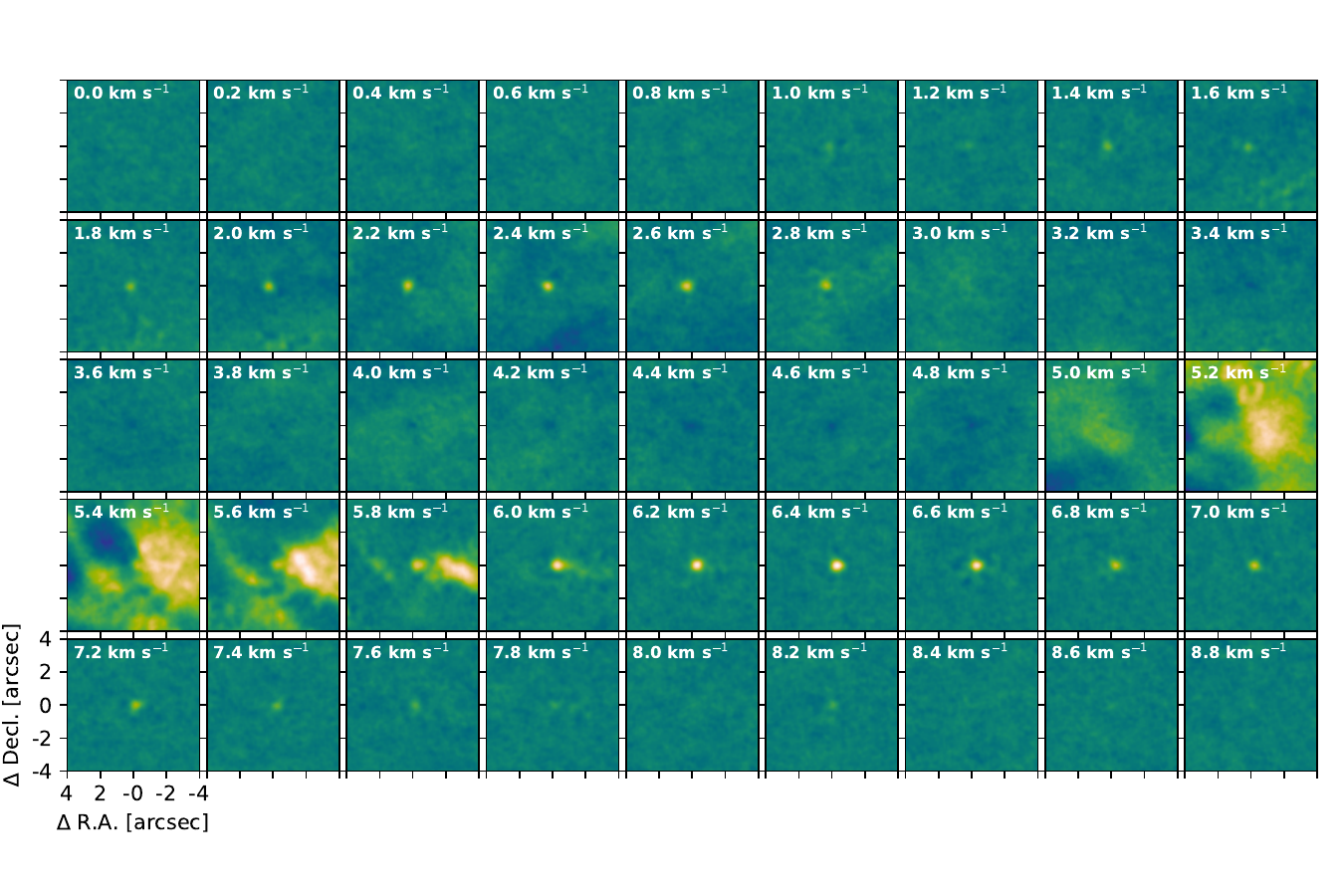}
\figcaption{Channel maps of the $^{13}$CO line emission for \textit{Oph 3}.  
\label{Fig:Oph313CO}
}
\end{figure*}

\begin{figure*}[h!]
\includegraphics[width=\linewidth]{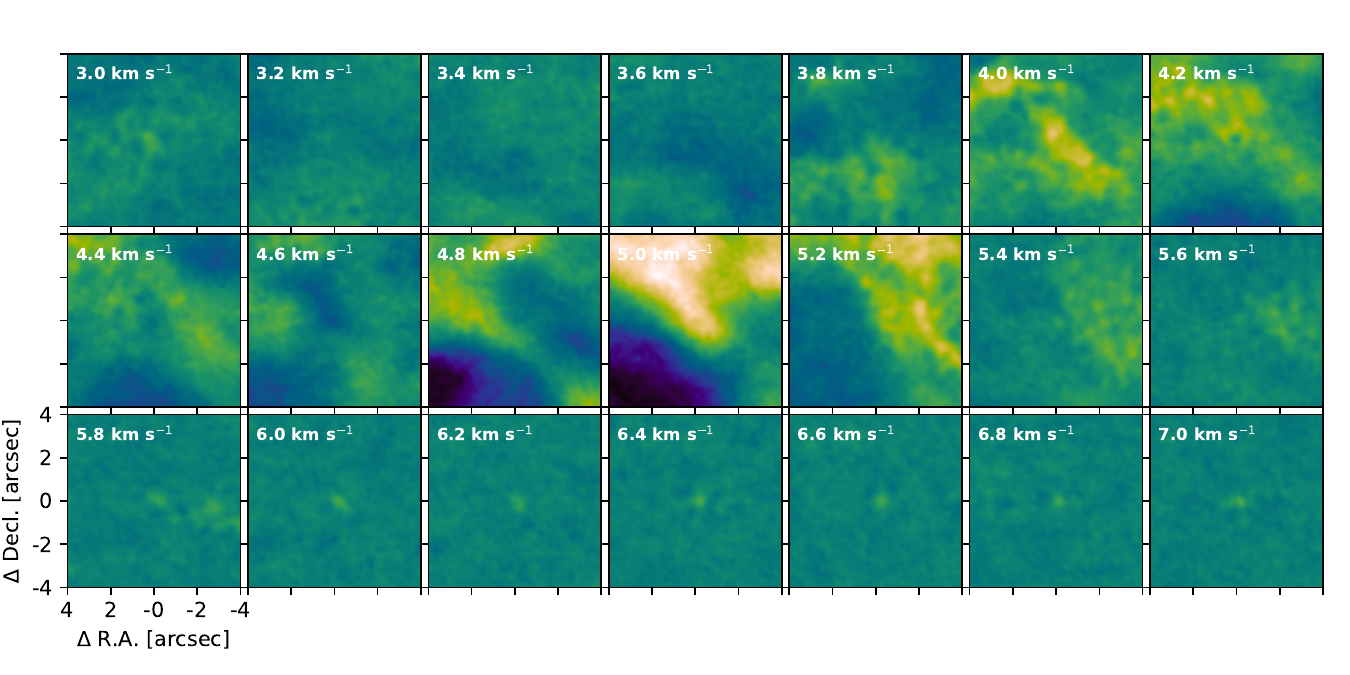}
\figcaption{Channel maps of the C$^{18}$CO line emission for \textit{Oph 3}.  
\label{Fig:Oph3C18O}
}
\end{figure*}

\begin{figure*}[h!]
\includegraphics[width=\linewidth]{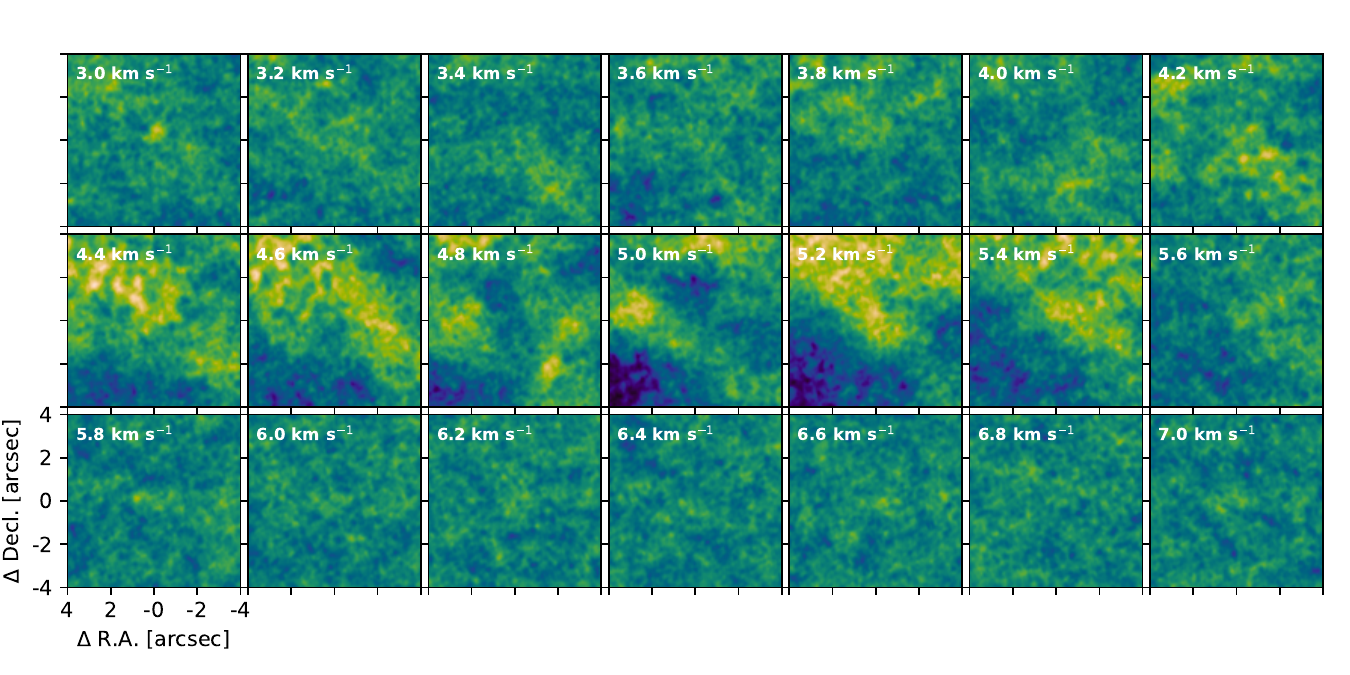}
\figcaption{Channel maps of the C$^{17}$O line emission for \textit{Oph 3}. 
\label{Fig:Oph3C17O}
}
\end{figure*}

\begin{figure*}[h!]
\includegraphics[width=\linewidth]{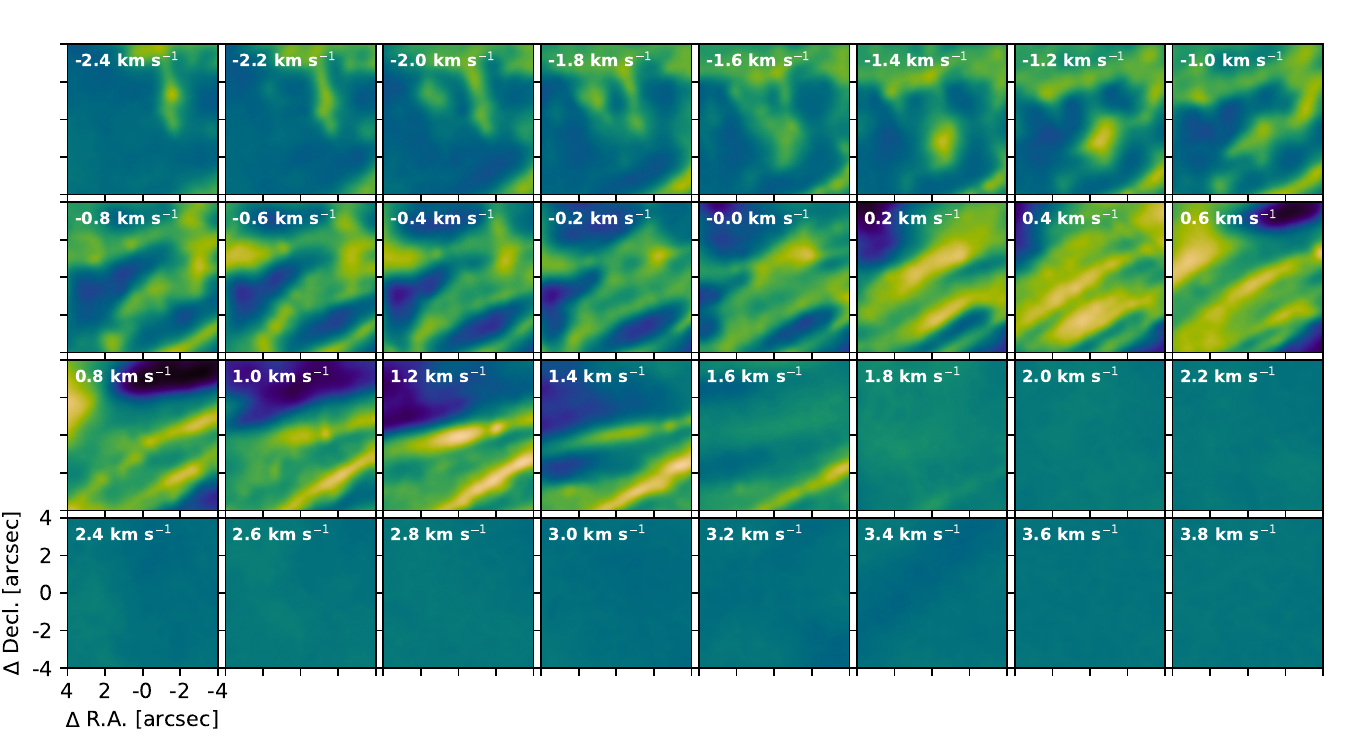}
\figcaption{Channel maps of the $^{12}$CO line emission for \textit{Oph 4}.  
\label{Fig:Oph412CO}
}
\end{figure*}

\begin{figure*}[h!]
\includegraphics[width=\linewidth]{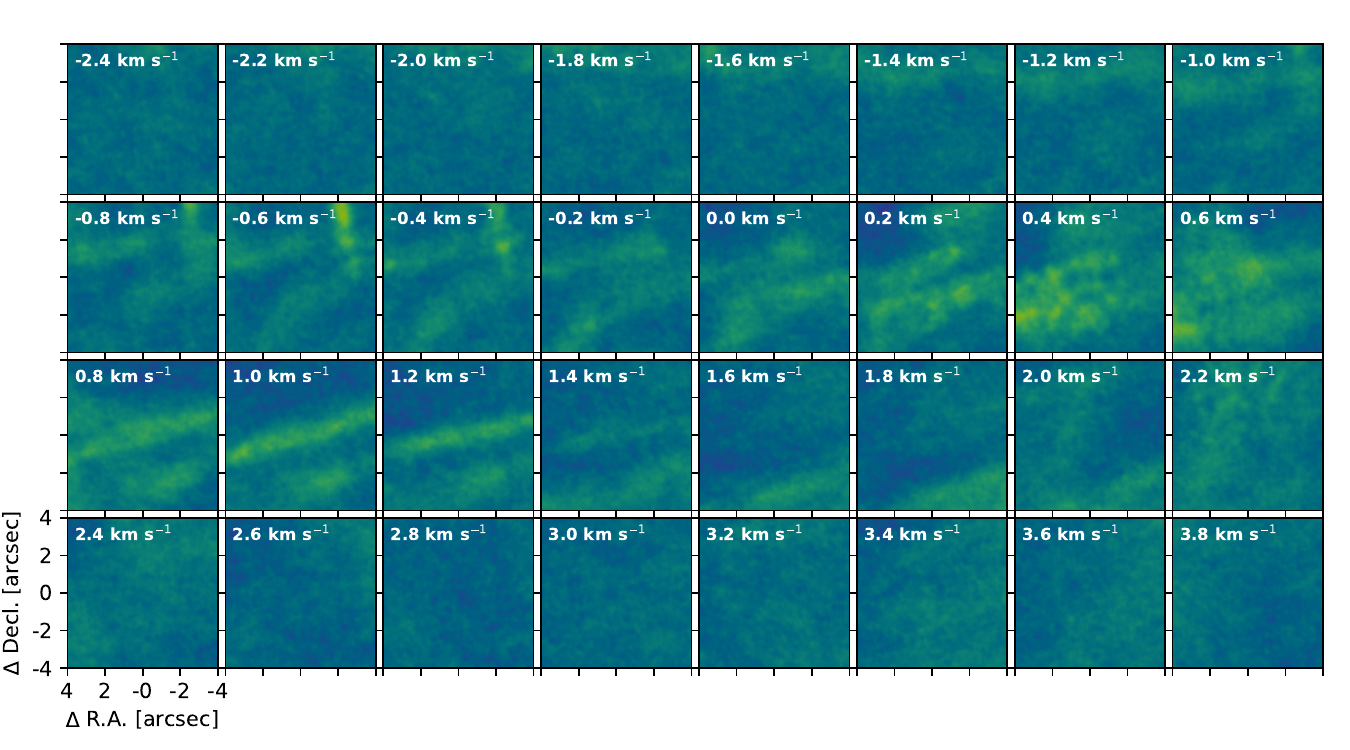}
\figcaption{Channel maps of the $^{13}$CO line emission for \textit{Oph 4}.  
\label{Fig:Oph413CO}
}
\end{figure*}

\begin{figure*}[h!]
\includegraphics[width=\linewidth]{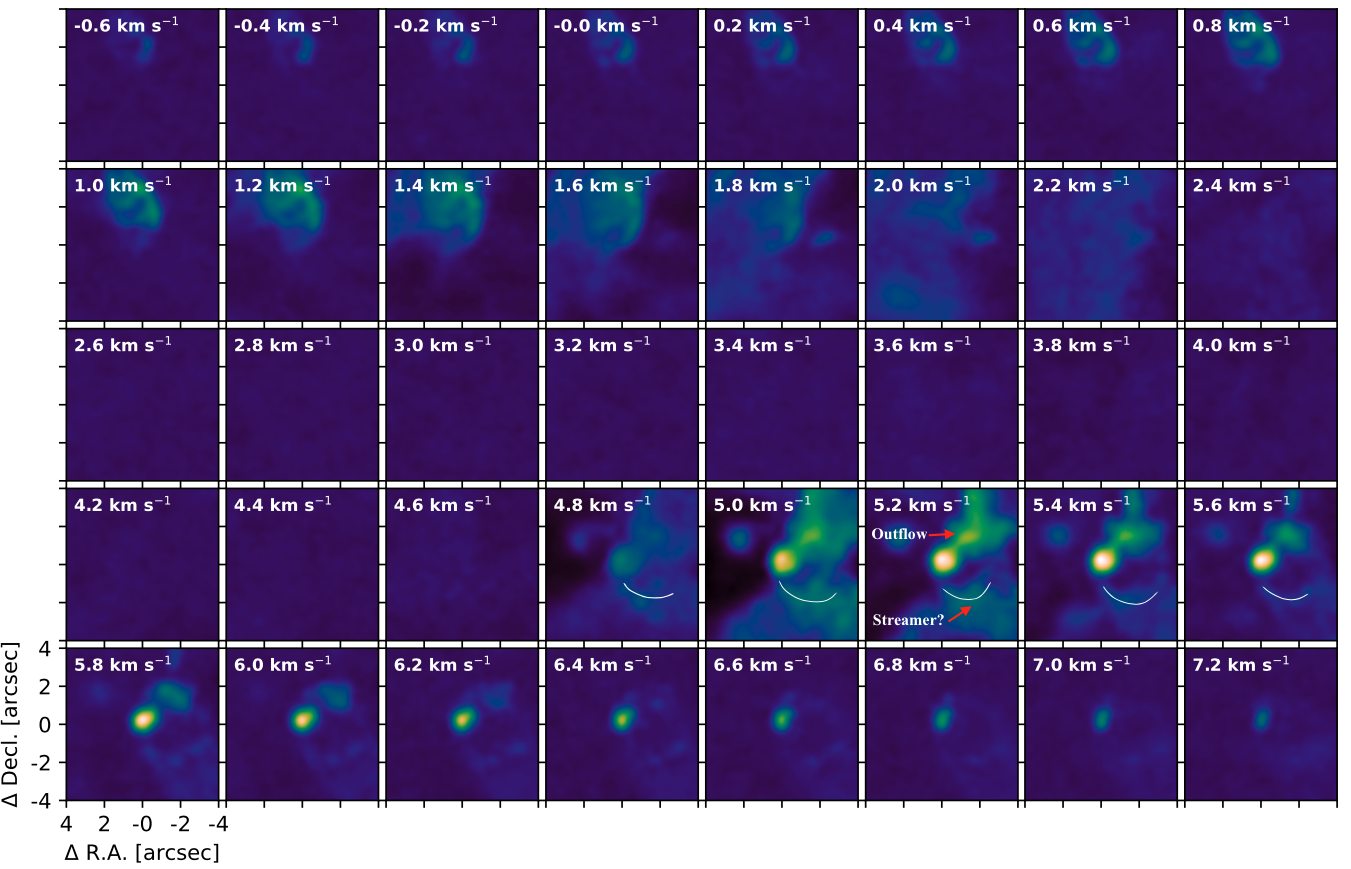}
\figcaption{Channel maps of the $^{12}$CO line emission for \textit{Oph 5}. White lines present an arc feature possibly due to an accretion streamer feedind the central system.
\label{Fig:Oph512CO}
}
\end{figure*}

\begin{figure*}[h!]
\includegraphics[width=\linewidth]{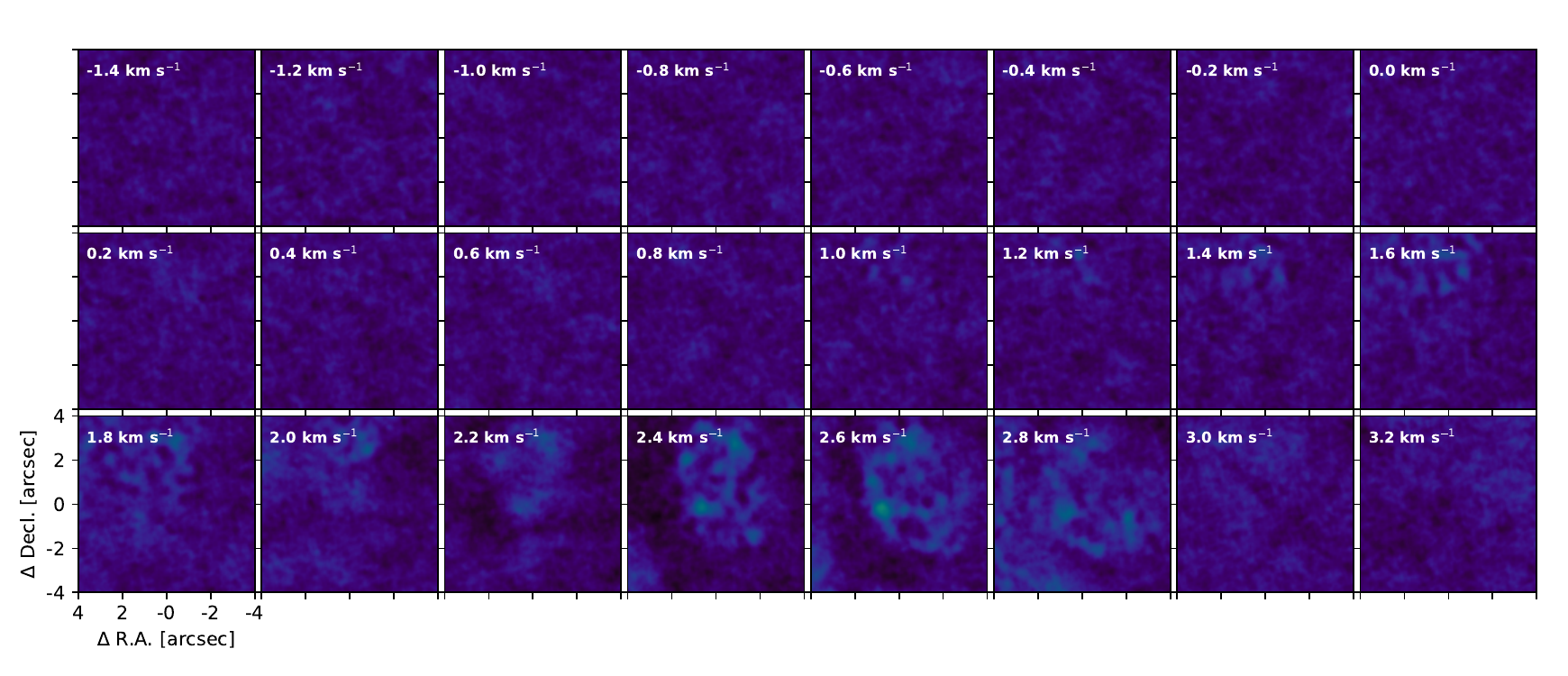}
\figcaption{Channel maps of the $^{13}$CO line emission for \textit{Oph 5}.  
\label{Fig:Oph513CO}
}
\end{figure*}

\begin{figure*}[h!]
\includegraphics[width=\linewidth]{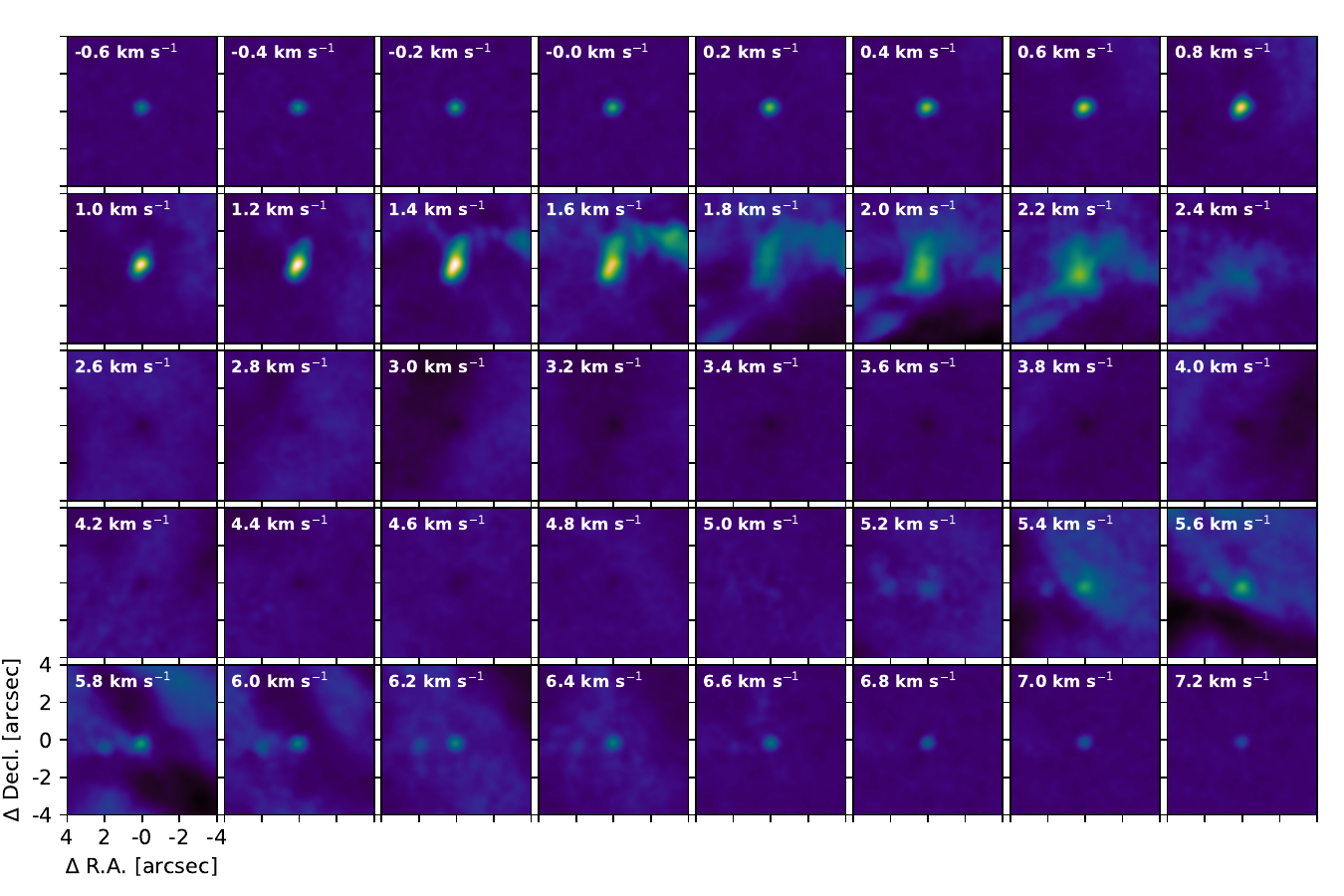}
\figcaption{Channel maps of the $^{12}$CO line emission for \textit{Oph 6}.  
\label{Fig:Oph612CO}
}
\end{figure*}

\begin{figure*}[h!]
\includegraphics[width=\linewidth]{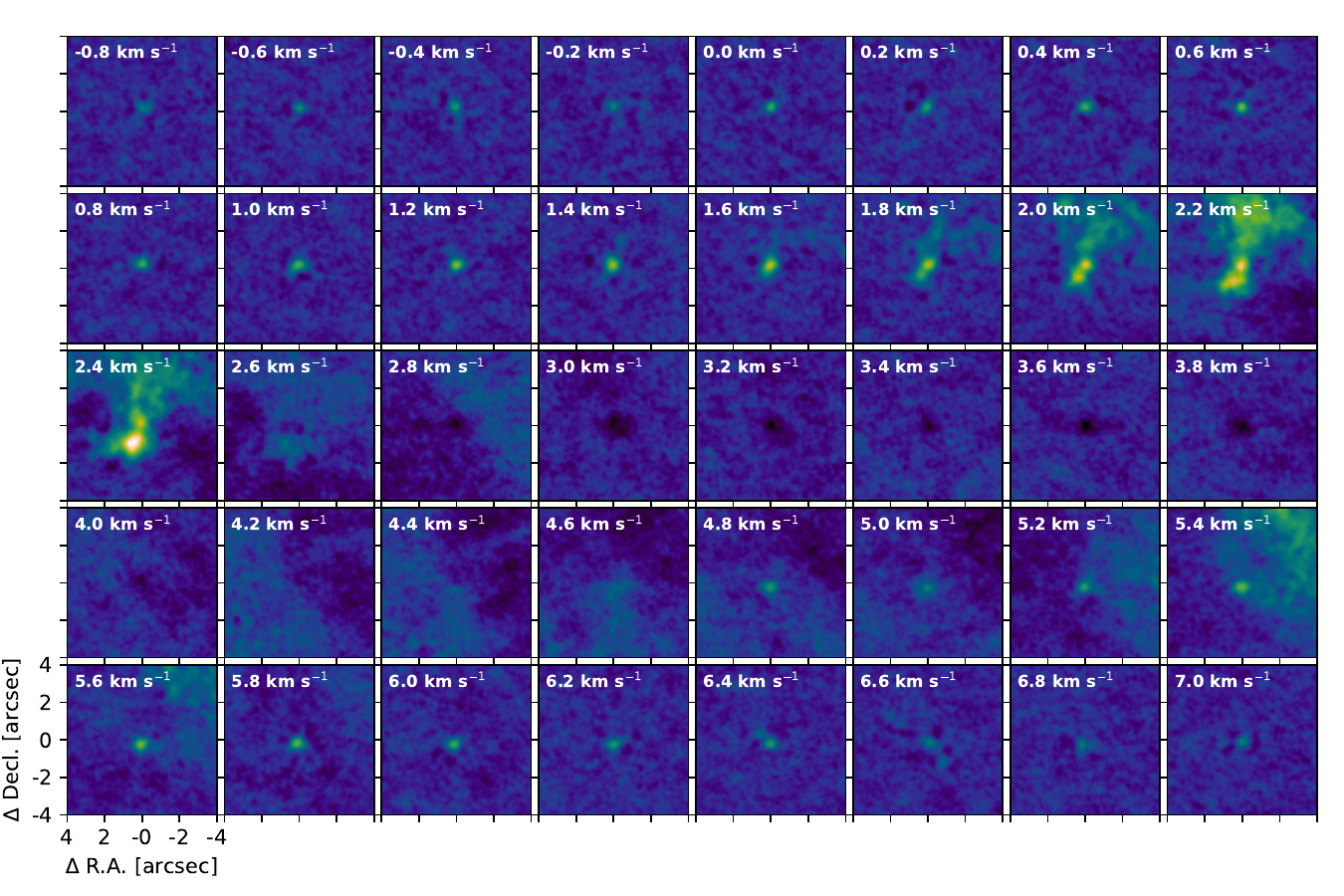}
\figcaption{Channel maps of the $^{13}$CO line emission for \textit{Oph 6}.  
\label{Fig:Oph613CO}
}
\end{figure*}

\begin{figure*}[h!]
\includegraphics[width=\linewidth]{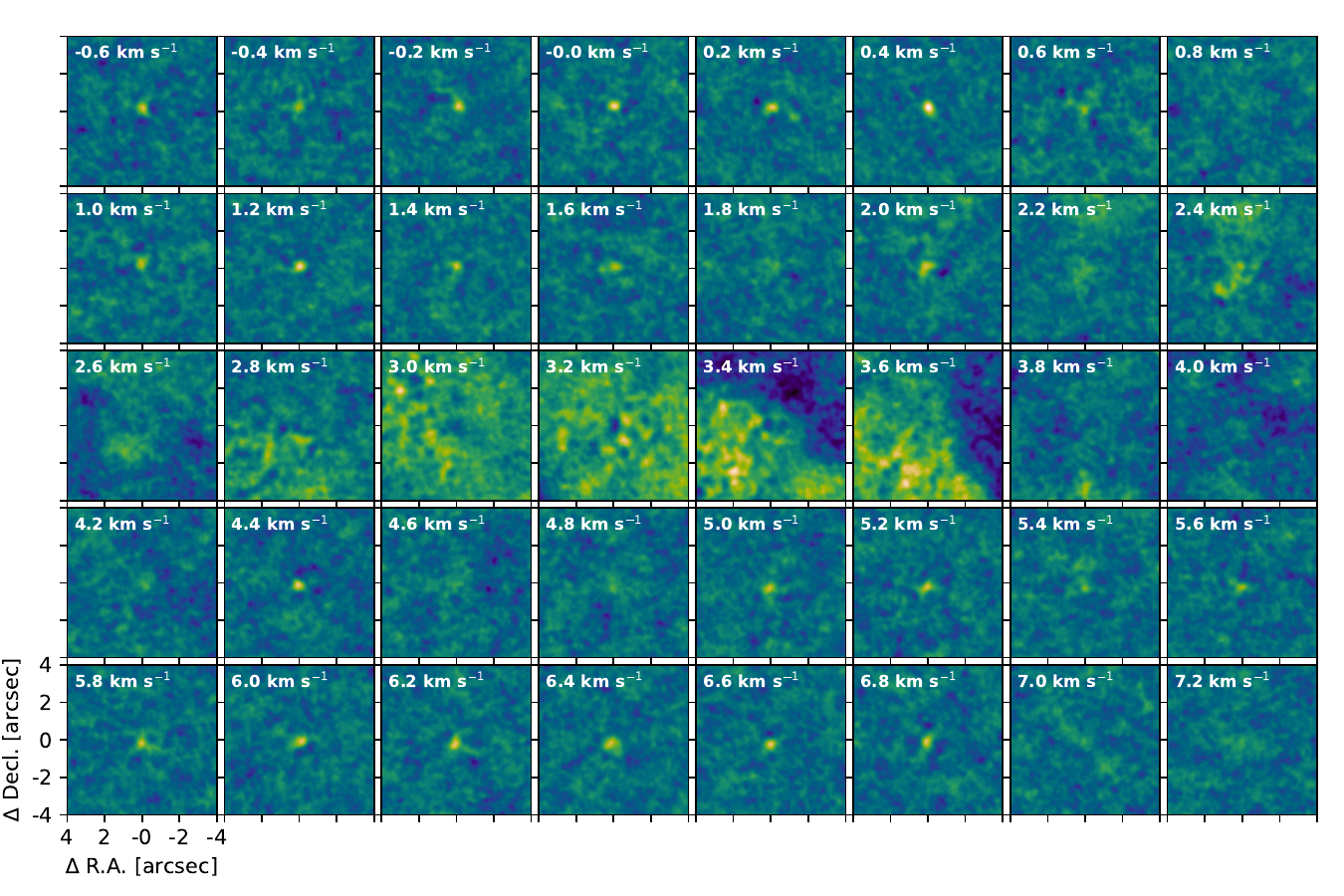}
\figcaption{Channel maps of the C$^{18}$O line emission for \textit{Oph 6}.  
\label{Fig:Oph6C18O}
}
\end{figure*}

\begin{figure*}[h!]
\includegraphics[width=\linewidth]{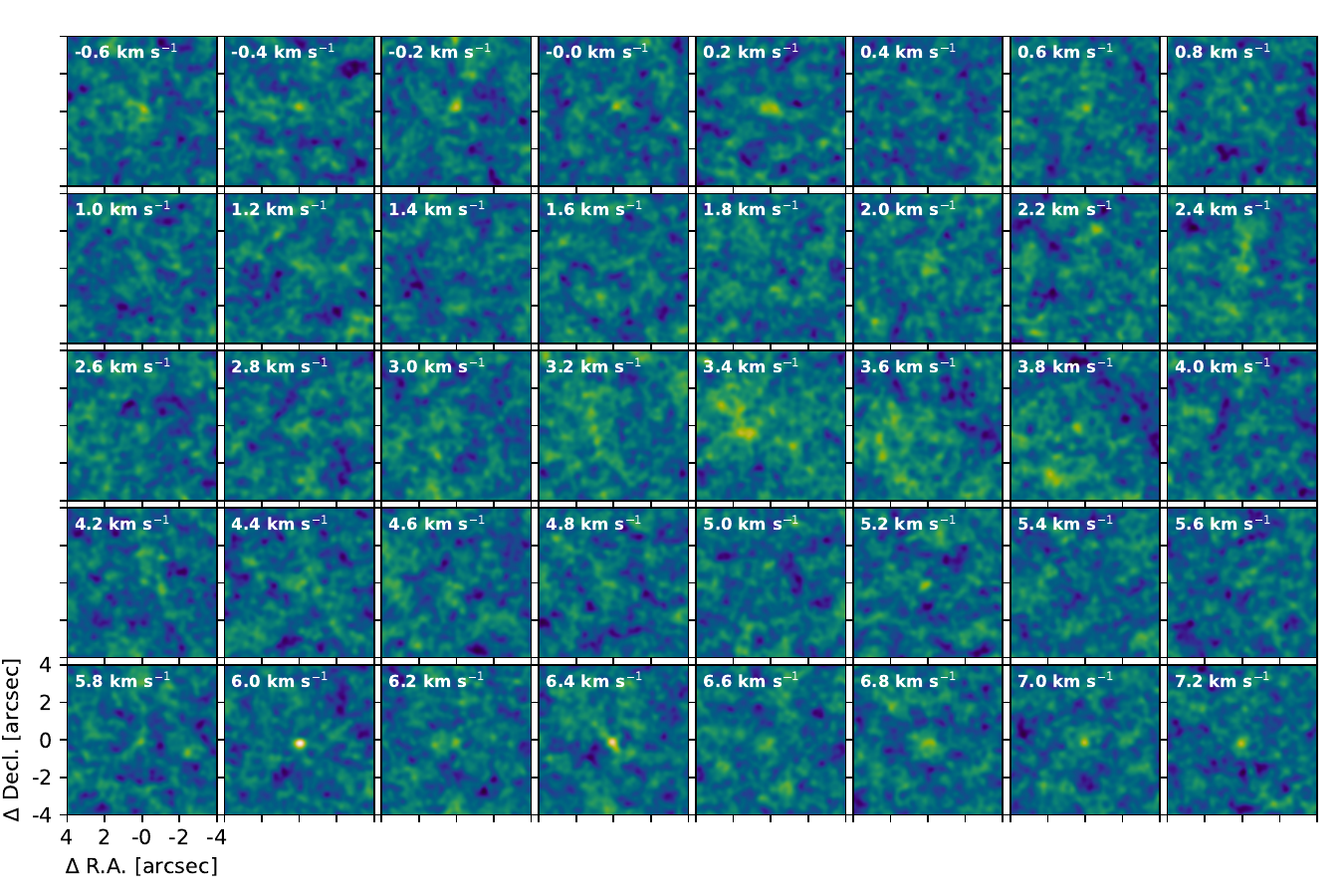}
\figcaption{Channel maps of the C$^{17}$O line emission for \textit{Oph 6}.  
\label{Fig:Oph6C17O}
}
\end{figure*}

\begin{figure*}[h!]
\includegraphics[width=\linewidth]{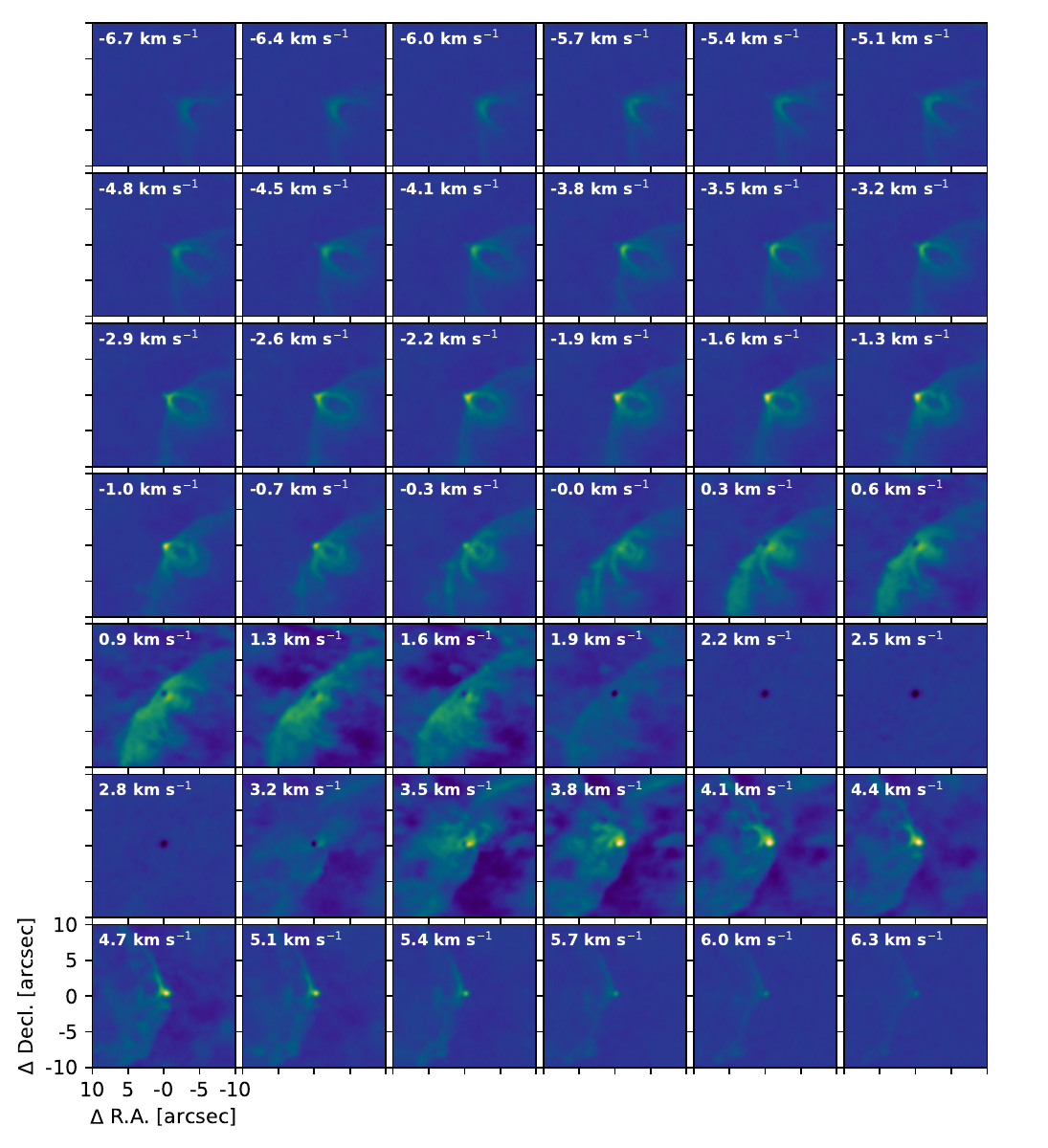}
\figcaption{Channel maps of the $^{12}$CO line emission for \textit{Oph 7}.  
\label{Fig:Oph712CO}
}
\end{figure*}

\begin{figure*}[h!]
\includegraphics[width=\linewidth]{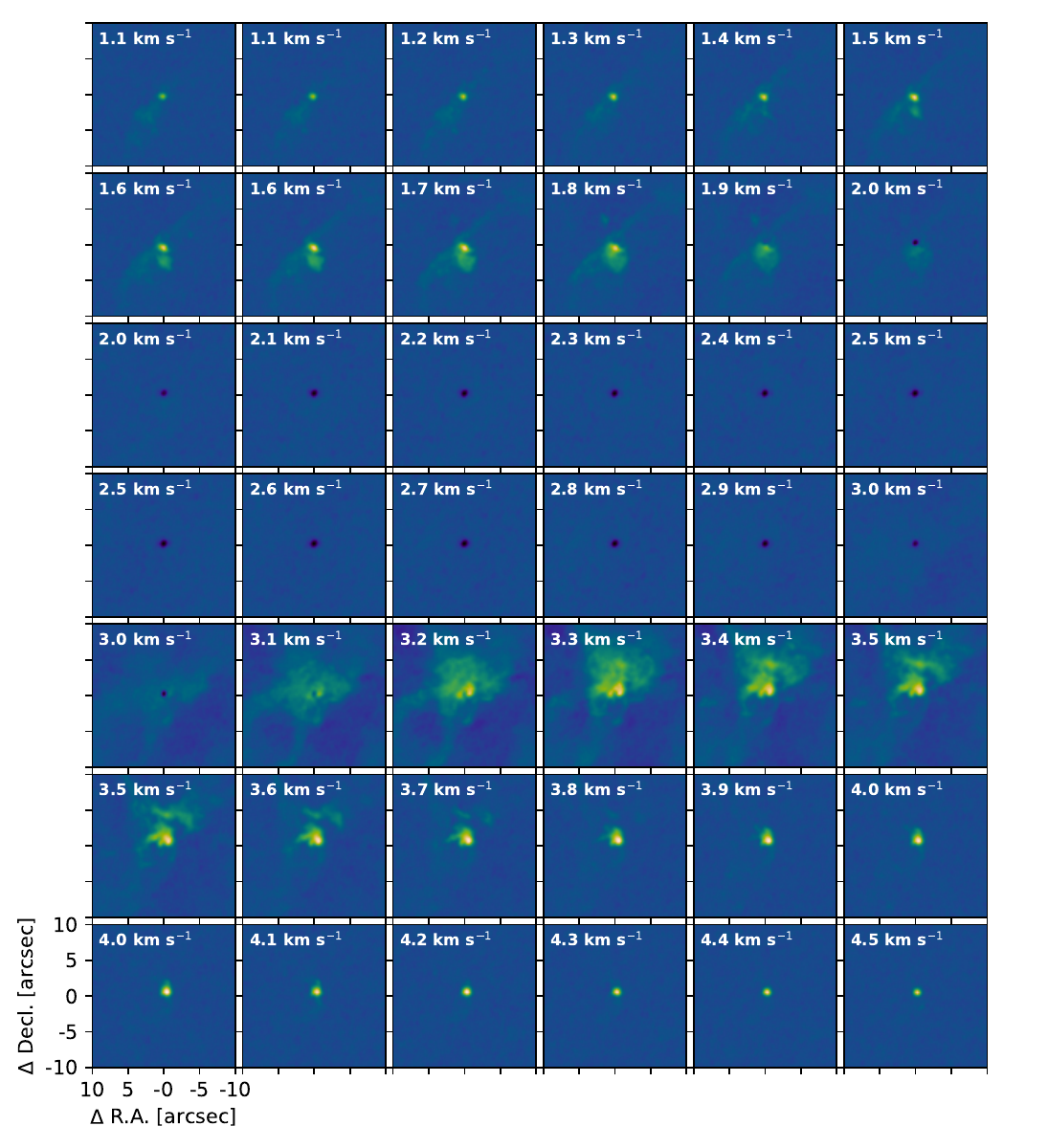}
\figcaption{Channel maps of the $^{13}$CO line emission for \textit{Oph 7}.  
\label{Fig:Oph713CO}
}
\end{figure*}

\begin{figure*}[h!]
\includegraphics[width=\linewidth]{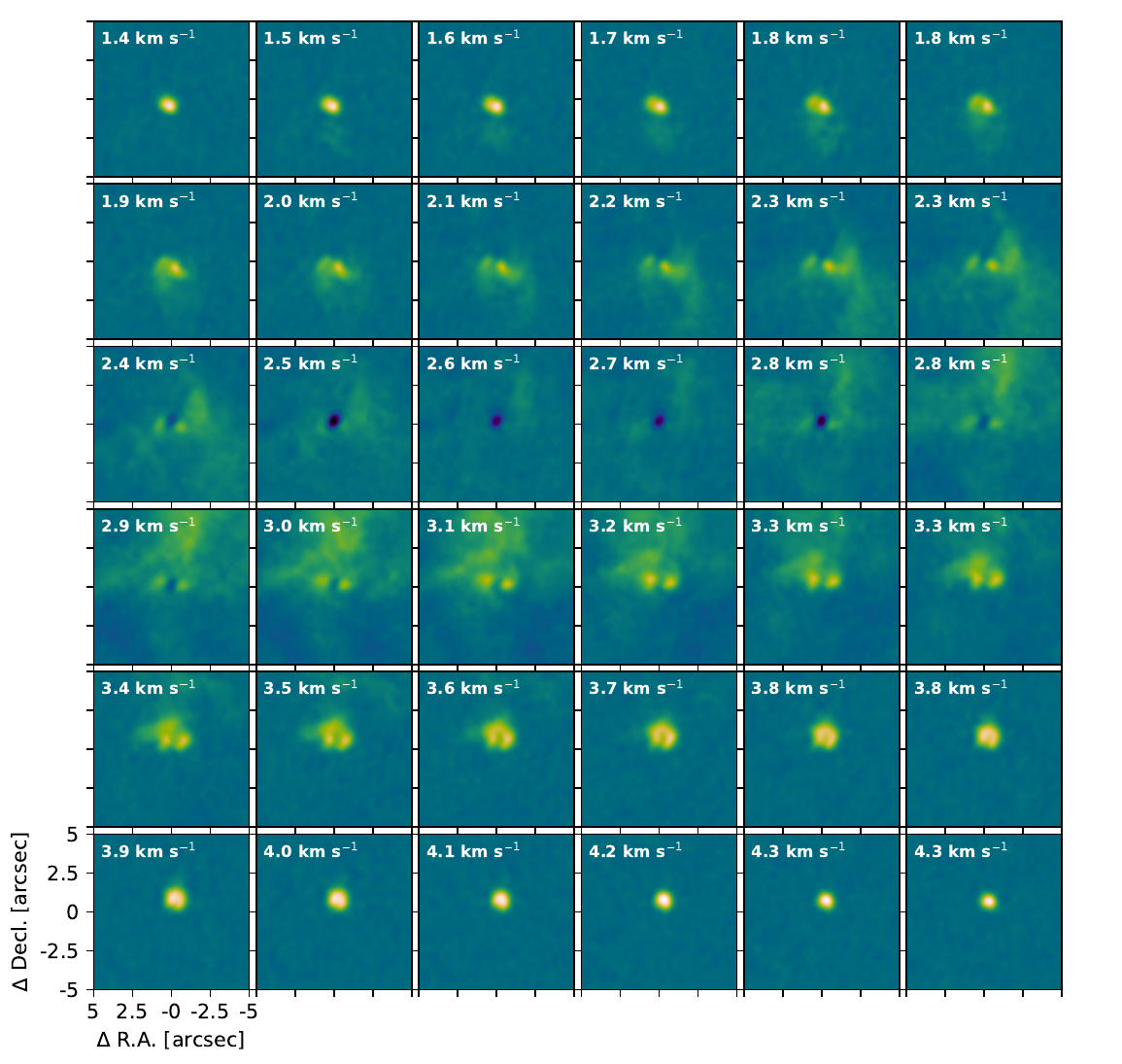}
\figcaption{Channel maps of the C$^{18}$O line emission for \textit{Oph 7}.  
\label{Fig:Oph7C18O}
}
\end{figure*}

\begin{figure*}[h!]
\includegraphics[width=\linewidth]{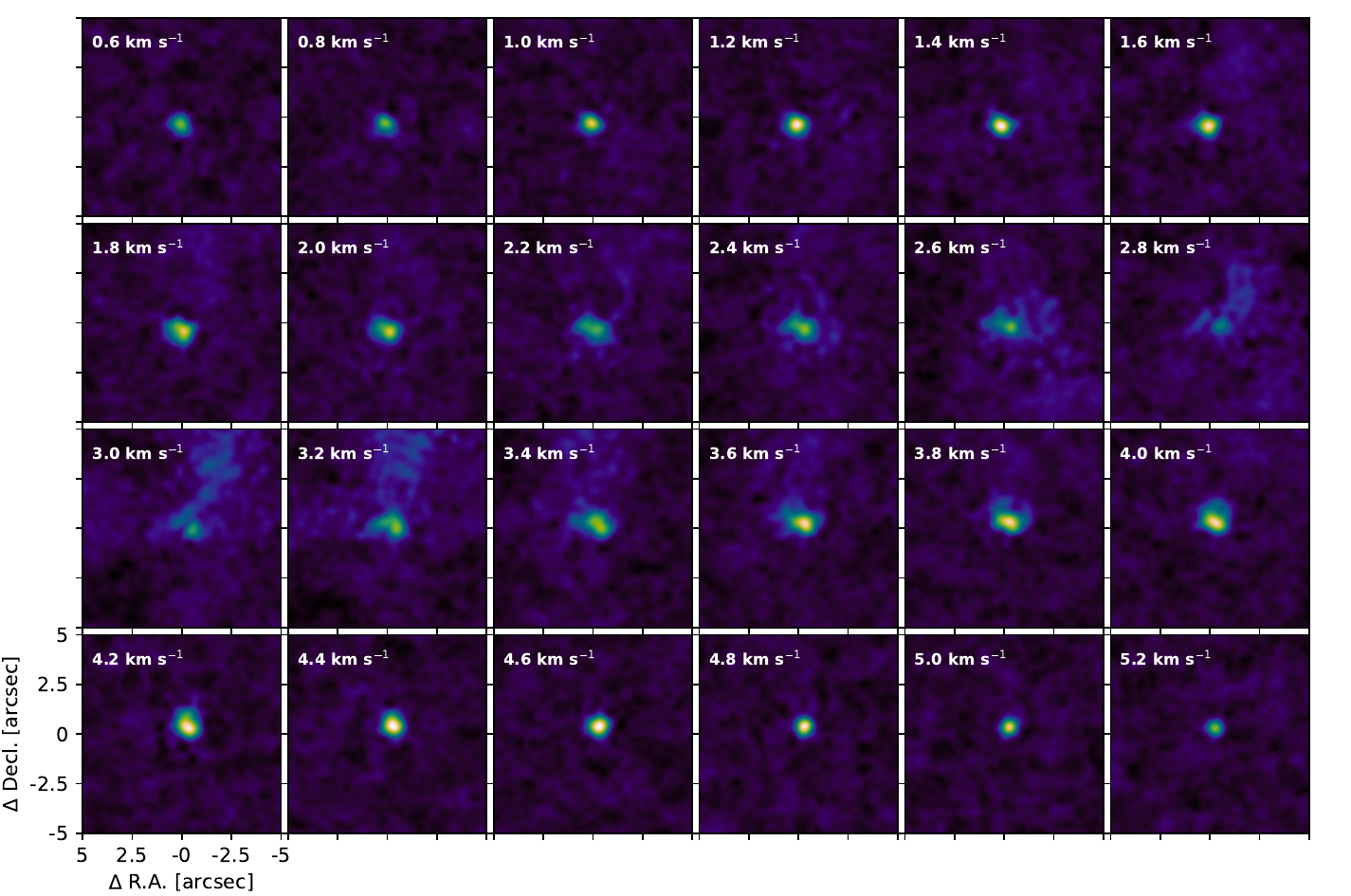}
\figcaption{Channel maps of the C$^{17}$O line emission for \textit{Oph 7}.  
\label{Fig:Oph7C17O}
}
\end{figure*}

\begin{figure*}[h!]
\includegraphics[width=\linewidth]{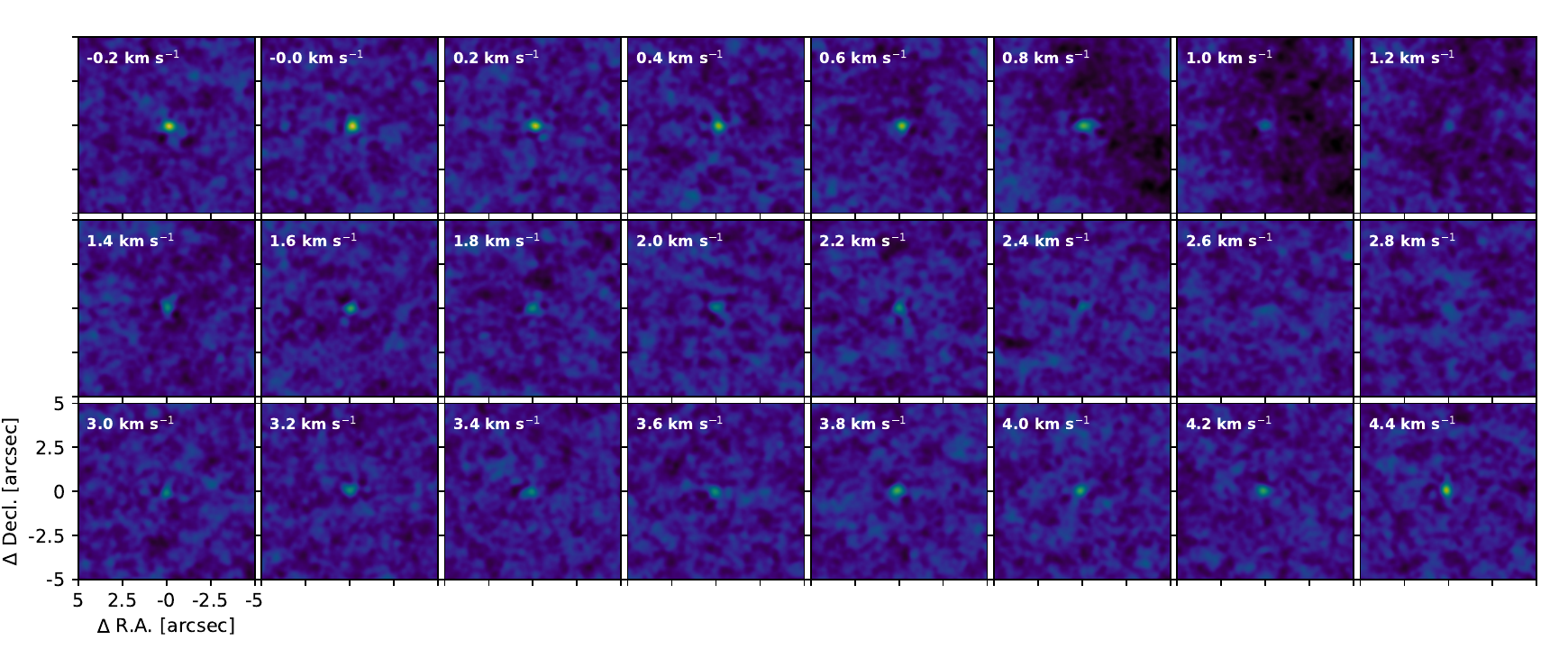}
\figcaption{Channel maps of the $^{12}$CO line emission for \textit{Oph 8}.  
\label{Fig:Oph812CO}
}
\end{figure*}

\begin{figure*}[h!]
\includegraphics[width=\linewidth]{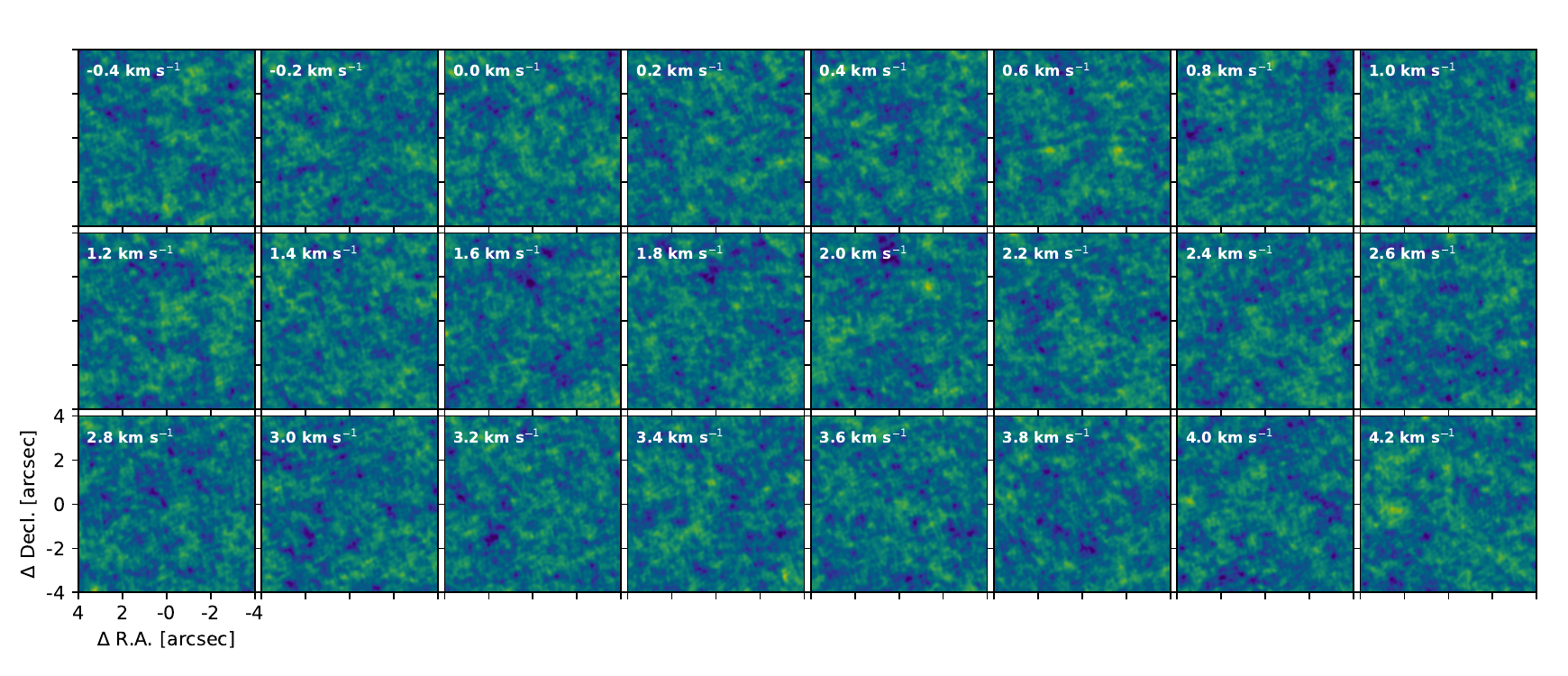}
\figcaption{Channel maps of the $^{13}$CO line emission for \textit{Oph 8}.  
\label{Fig:Oph813CO}
}
\end{figure*}

\begin{figure*}[h!]
\includegraphics[width=\linewidth]{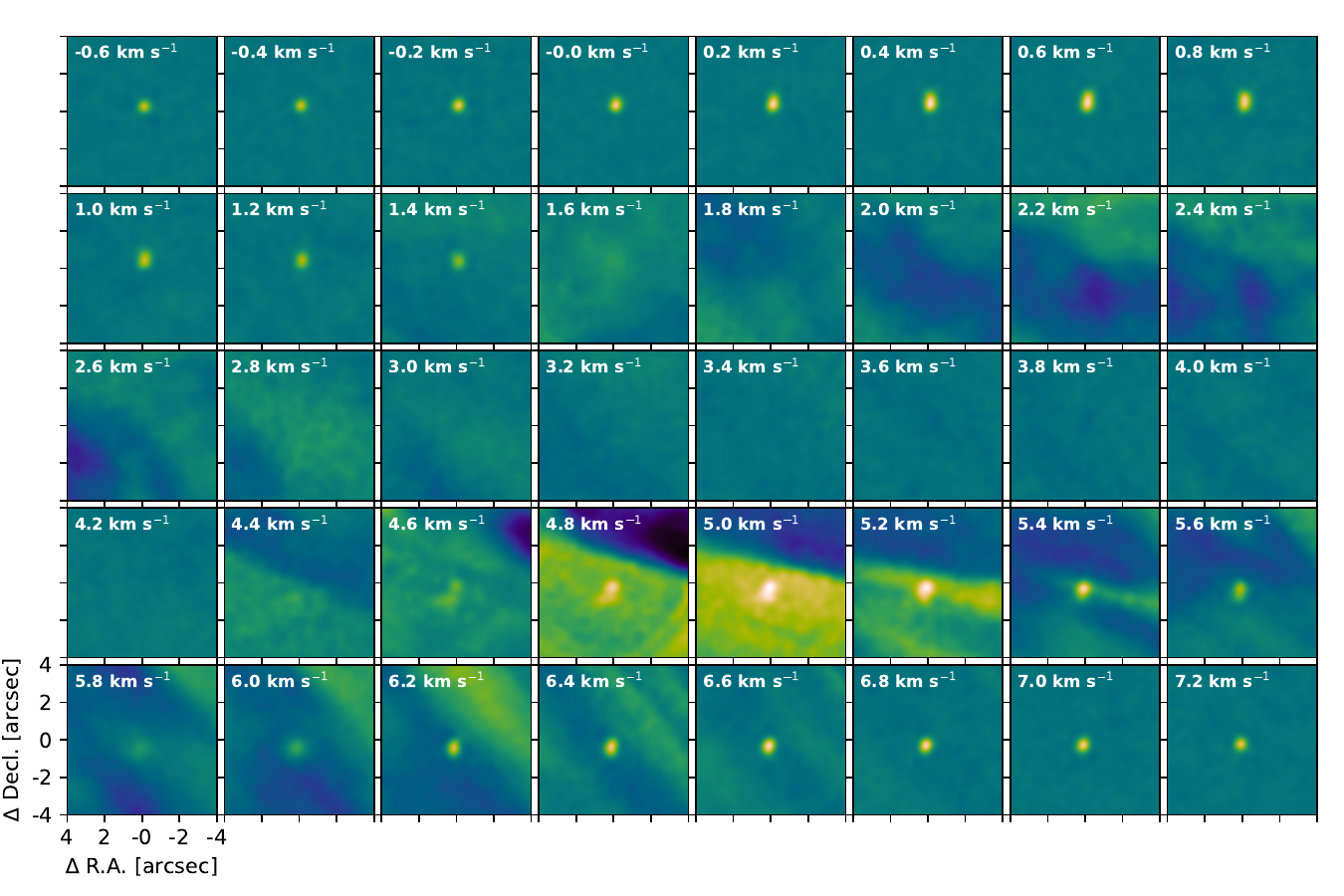}
\figcaption{Channel maps of the $^{12}$CO line emission for \textit{Oph 9}.  
\label{Fig:Oph912CO}
}
\end{figure*}

\begin{figure*}[h!]
\includegraphics[width=\linewidth]{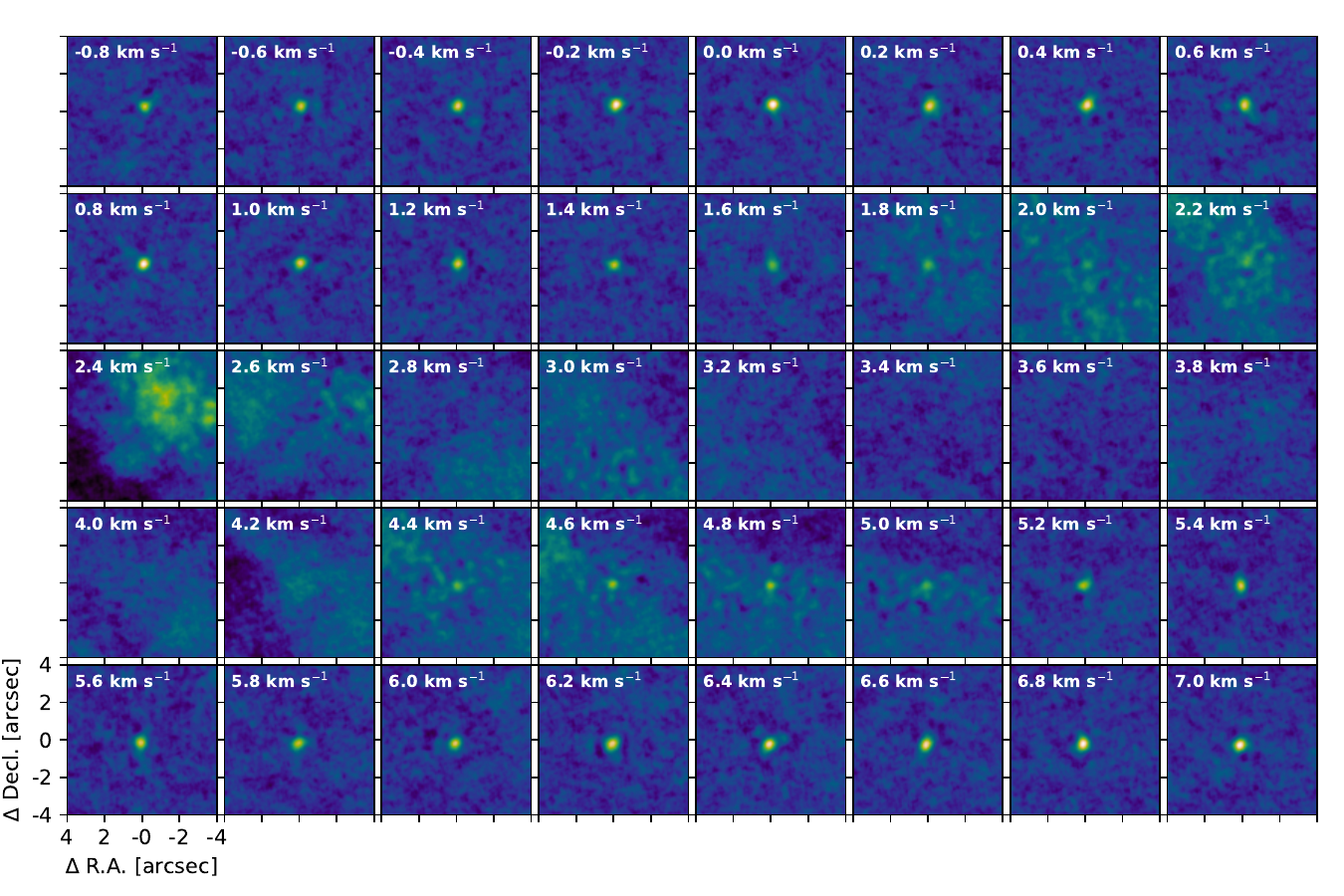}
\figcaption{Channel maps of the $^{13}$CO line emission for \textit{Oph 9}.  
\label{Fig:Oph913CO}
}
\end{figure*}

\begin{figure*}[h!]
\includegraphics[width=\linewidth]{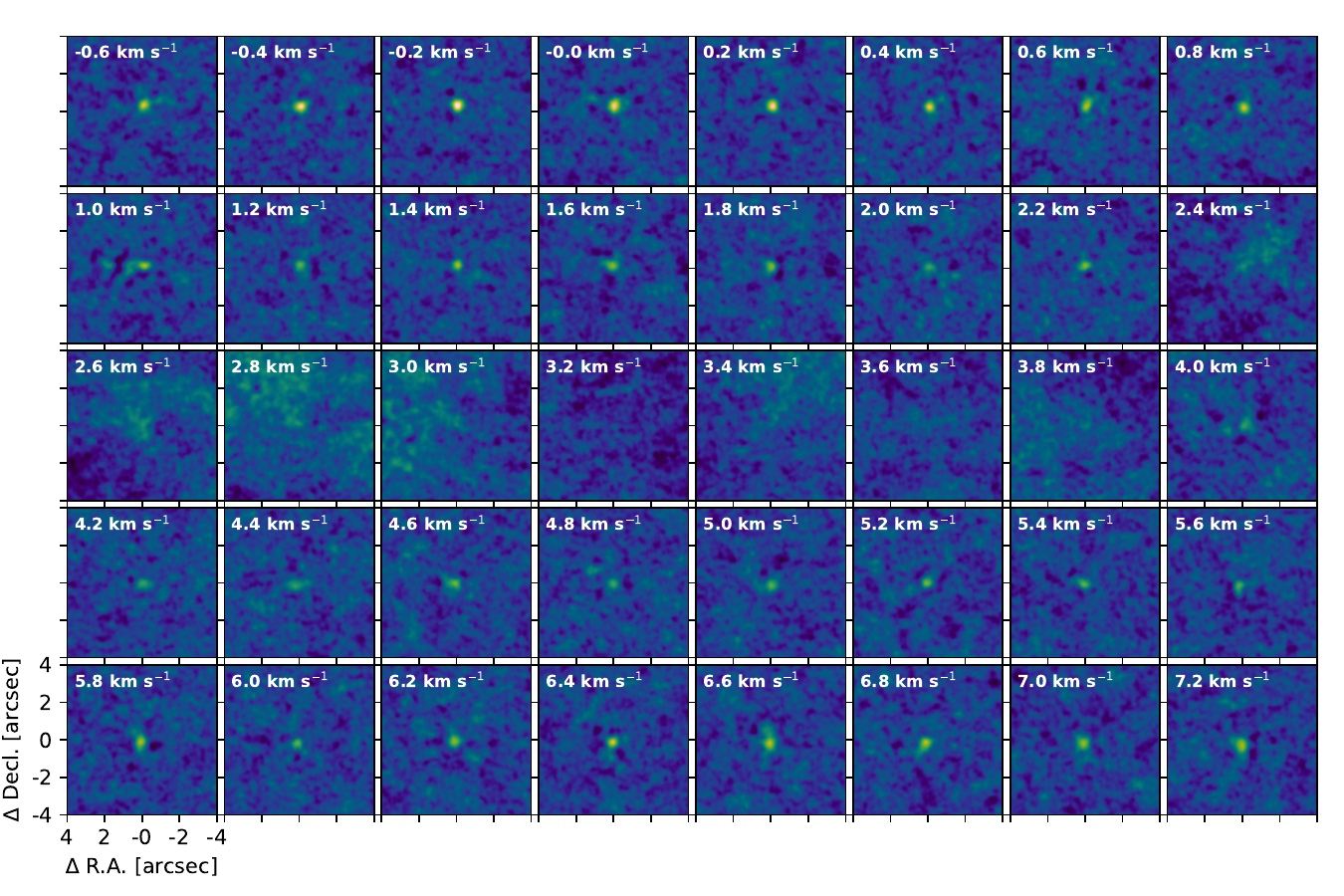}
\figcaption{Channel maps of the C$^{18}$O line emission for \textit{Oph 9}. 
\label{Fig:Oph9C18O}
}
\end{figure*}

\begin{figure*}[h!]
\includegraphics[width=\linewidth]{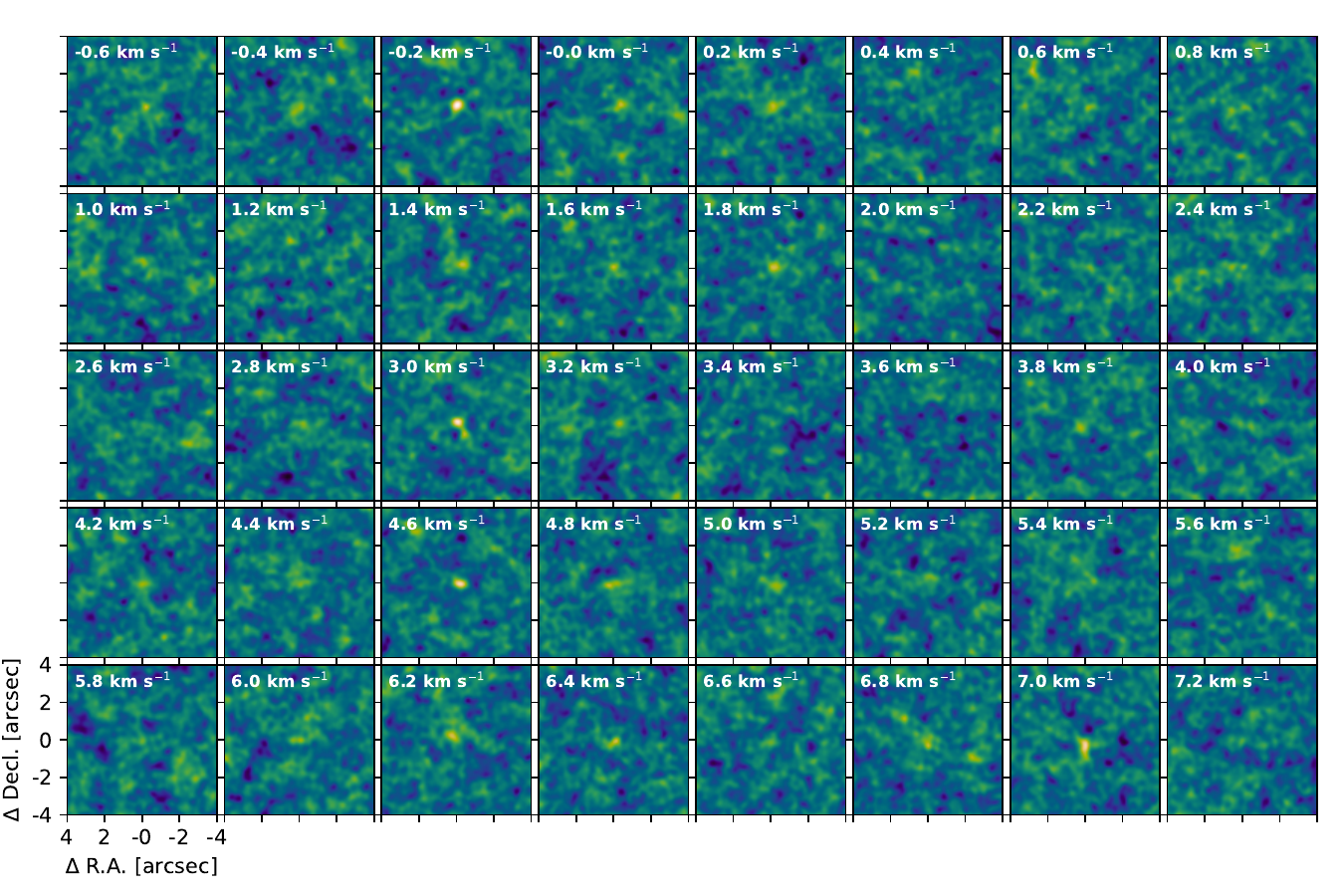}
\figcaption{Channel maps of the C$^{17}$O line emission for \textit{Oph 9}.  
\label{Fig:Oph9C17O}
}
\end{figure*}

\begin{figure*}[h!]
\includegraphics[width=\linewidth]{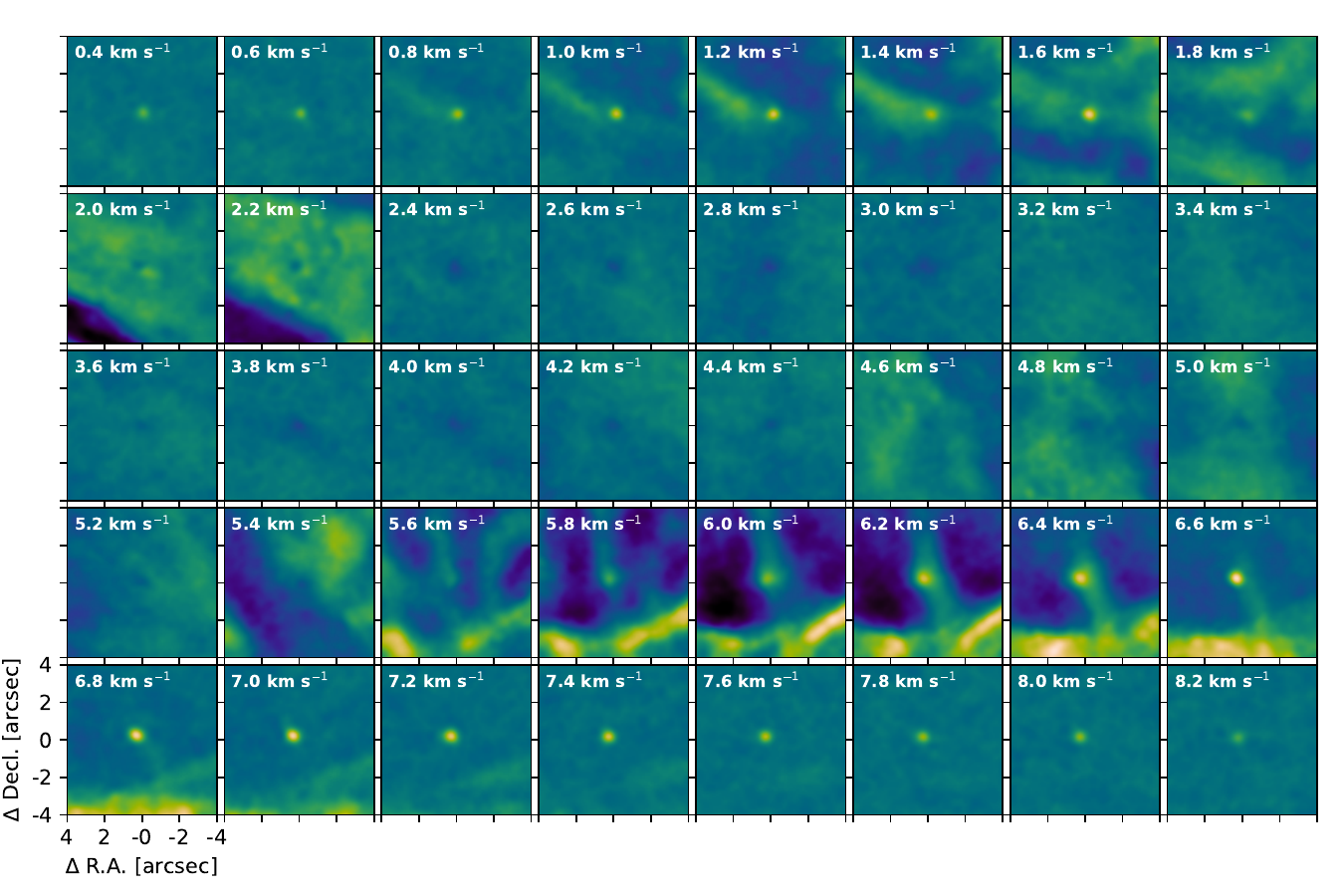}
\figcaption{Channel maps of the $^{12}$CO line emission for \textit{Oph 10}.  
\label{Fig:Oph1012CO}
}
\end{figure*}

\begin{figure*}[h!]
\includegraphics[width=\linewidth]{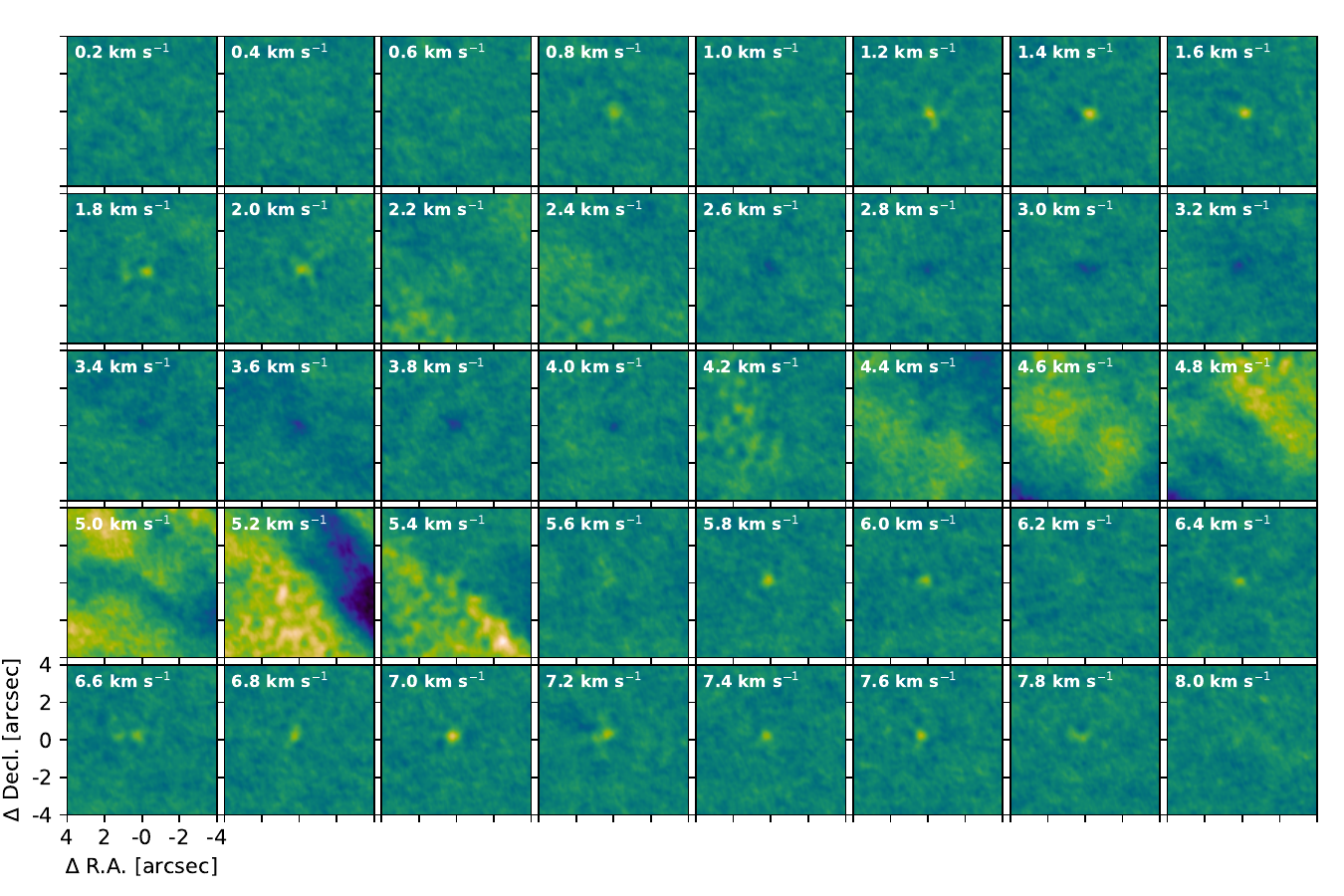}
\figcaption{Channel maps of the $^{13}$CO line emission for \textit{Oph 10}.  
\label{Fig:Oph1013CO}
}
\end{figure*}

\begin{figure*}[h!]
\includegraphics[width=\linewidth]{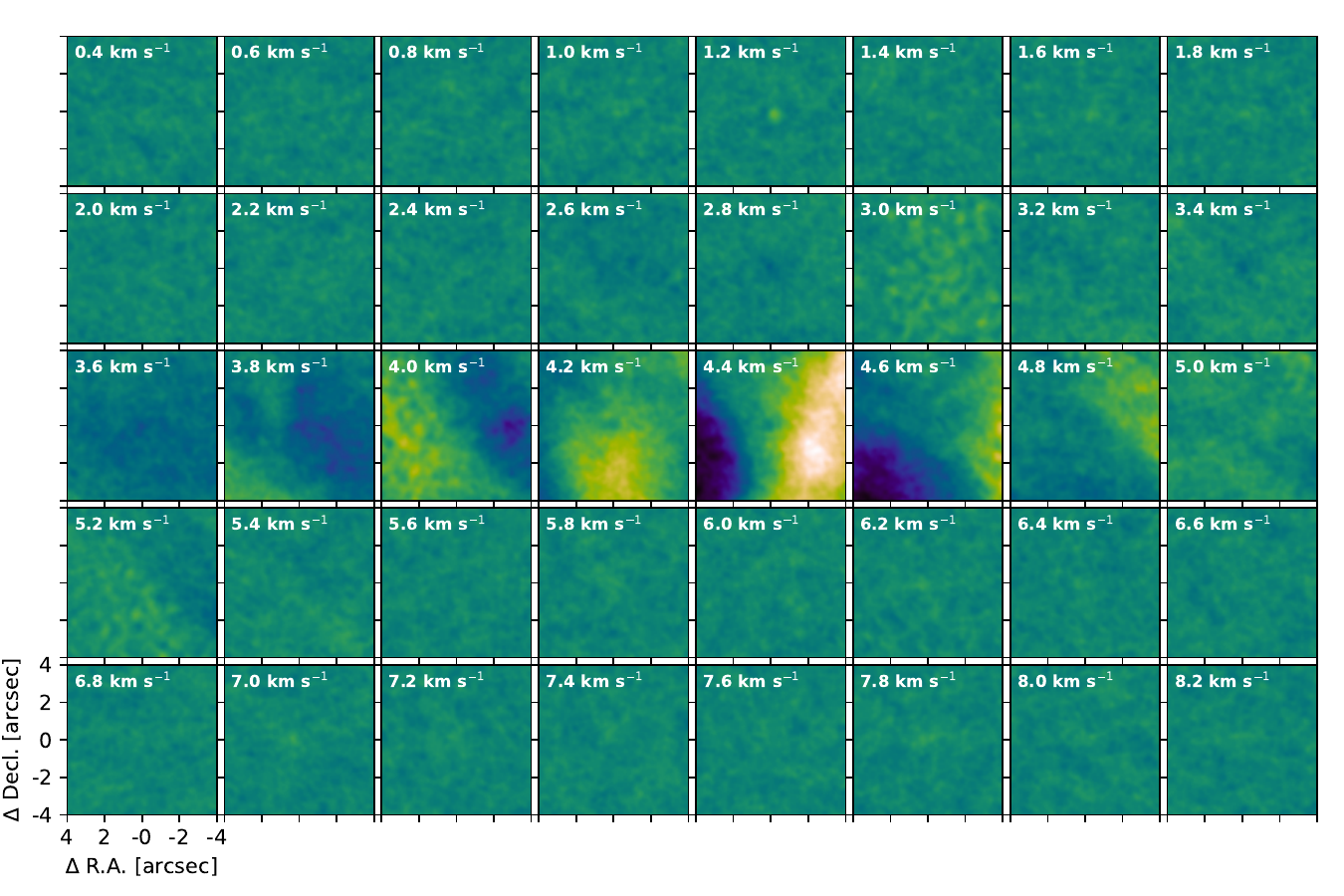}
\figcaption{Channel maps of the C$^{18}$O line emission for \textit{Oph 10}.  
\label{Fig:Oph10C18O}
}
\end{figure*}

\begin{figure*}[h!]
\includegraphics[width=\linewidth]{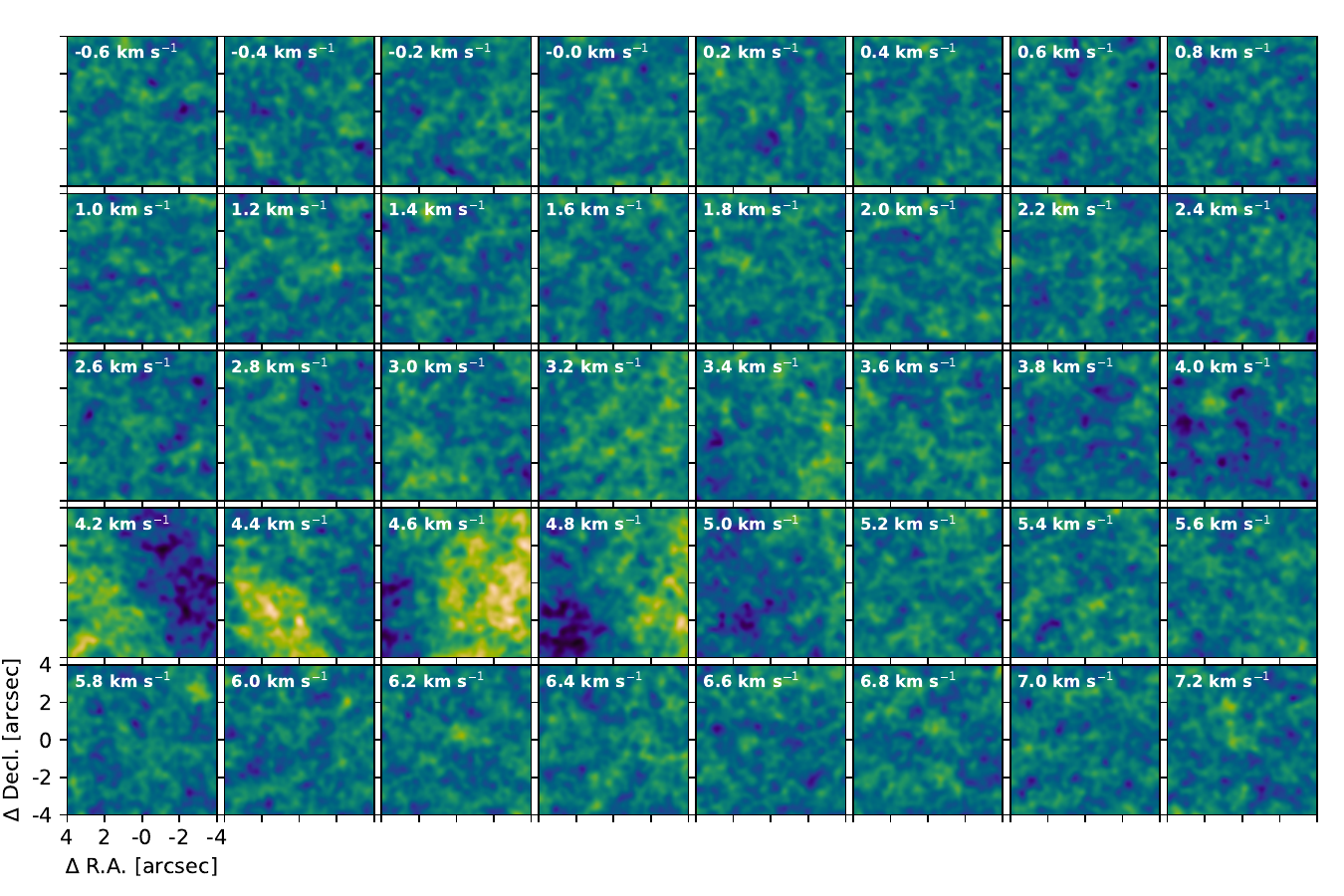}
\figcaption{Channel maps of the C$^{17}$O line emission for \textit{Oph 10}.  
\label{Fig:Oph10C17O}
}
\end{figure*}

The individual $^{12}$CO, $^{13}$CO, C$^{18}$O, and C$^{17}$O channel maps used to generate moment maps (Section \ref{Sec:Moments}) are shown in Figures \ref{Fig:Oph112CO} - \ref{Fig:Oph10C17O}.

\section{Velocity Range to Generate AGE-PRO Products}
\label{App:VelRange}

\begin{figure*}[h!]
\gridline{\fig{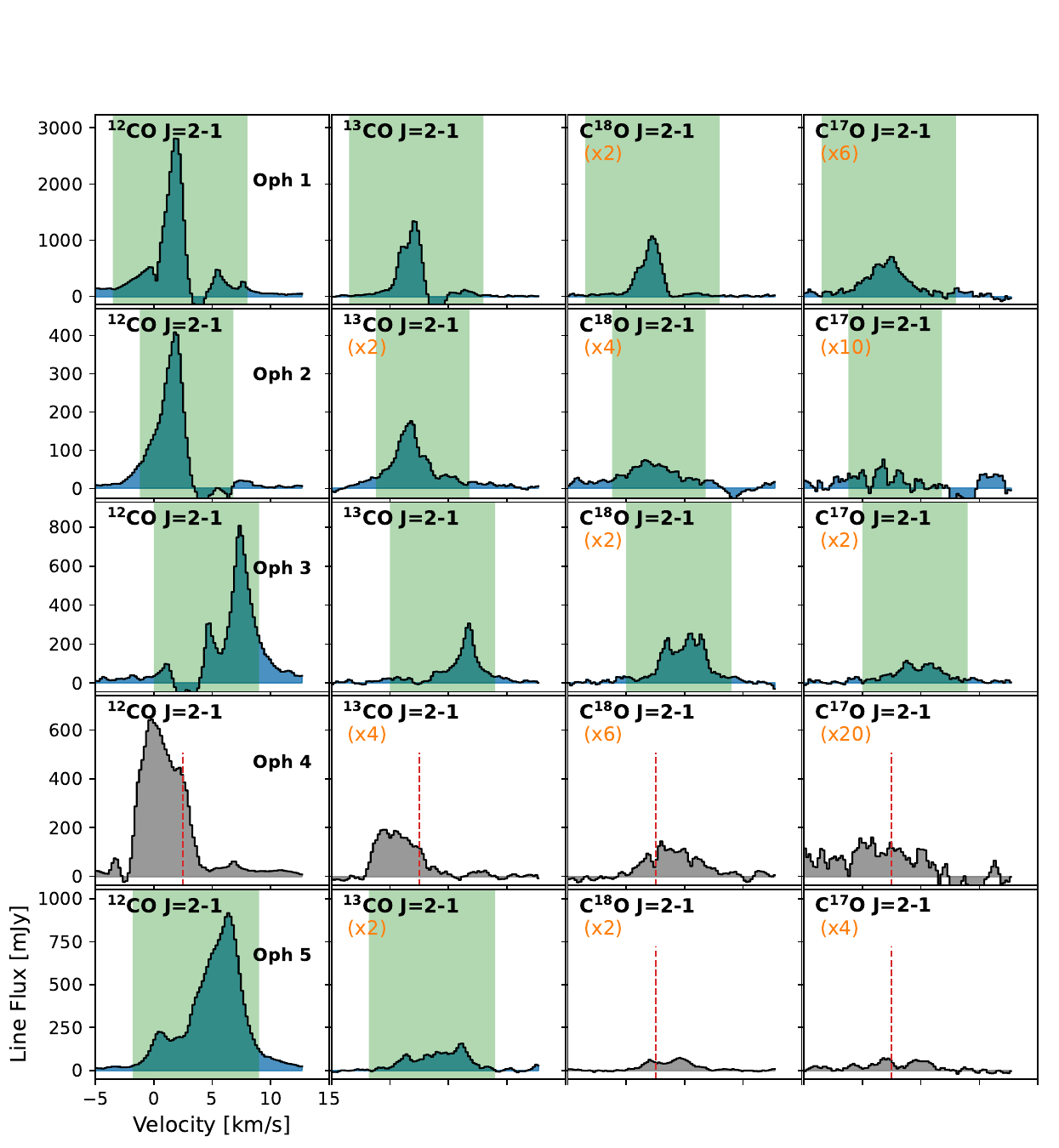}{0.5\textwidth}{}
   \fig{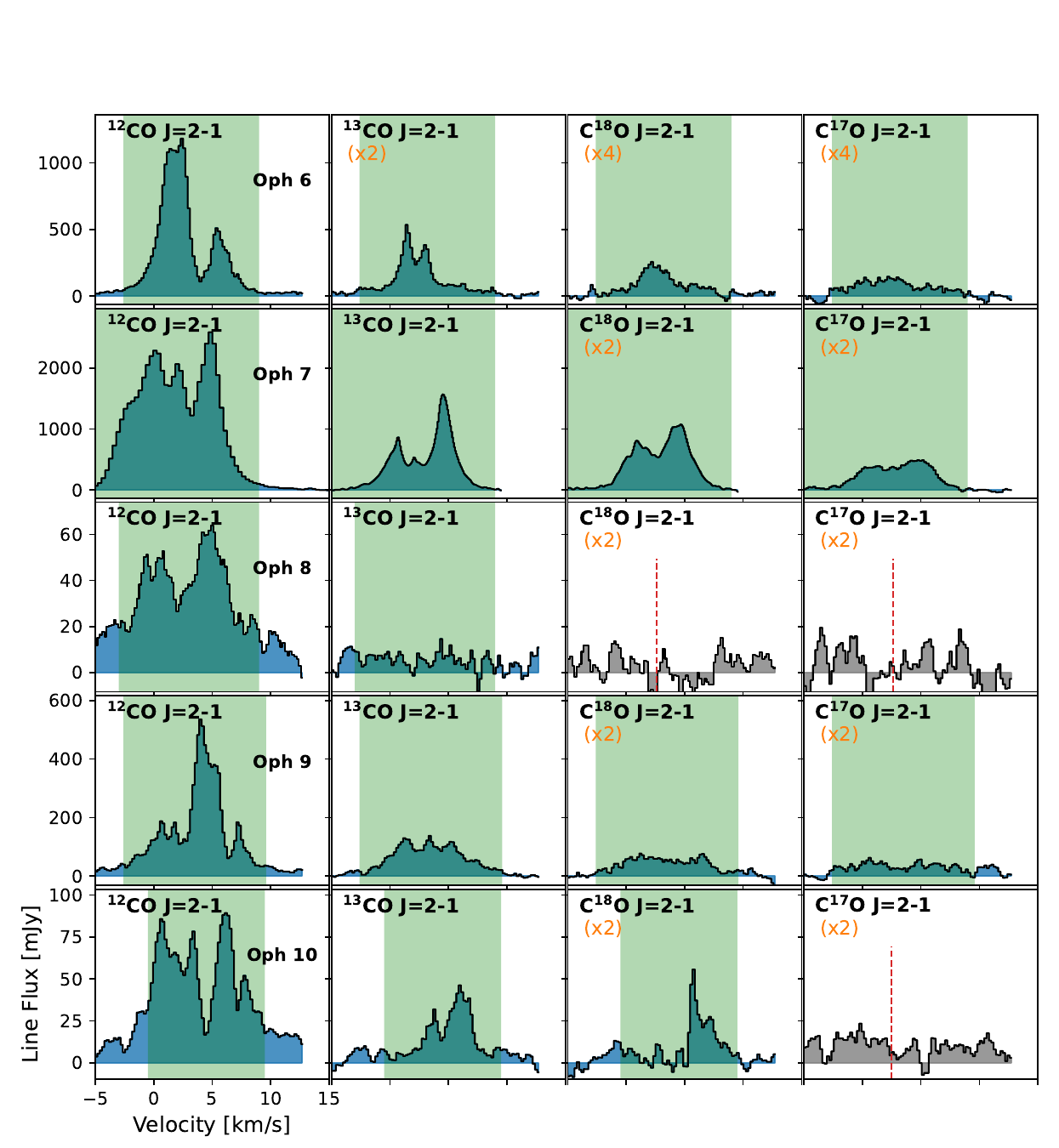}{0.5\textwidth}{}
            }
\figcaption{CO spectra extracted from an elliptical aperture of 2$^{\prime \prime}$
diameter and centered at the continuum peak positions of the Ophiuchus sample. In each panel, we report the factor used to scale emission at higher levels to compare with other molecules visually. Green-shaded areas represent the velocity range used to generate CO moment maps, see Section \ref{Sec:Moments}.
\label{Fig:Spectra}
}
\end{figure*}

Figure \ref{Fig:Spectra} displays the extracted CO velocity-stacked spectra to define velocity ranges for emission integration, and Table \ref{Table:Mom_parameters} presents an overview of the parameters applied to generate Moment Maps of the CO isotopologues, see Section \ref{Sec:Moments}.

\begin{table*}[ht]
\caption{Values used to generate moment maps} 
\centering 
\begin{tabular}{l c c c c} 
\hline\hline
	   Source  & V$_{\rm LRSK}$ & Velocity Range & $\sigma$ & Radius \\
      & km s$^{-1}$ & km s$^{-1}$ & mJy km s$^{-1}$ &arcsec\\ 
	 \midrule
	   Oph 1 &  2.4 & -3.5 $-$ 8.0 & 3.3 & 3.0  \\
     
     \midrule
     Oph 2 & 2.4 & -1.2 $-$ 6.8 & 2.9 & 1.5 \\
    
     \midrule
     Oph 3\ & 2.5 & 0.0 $-$ 9.0 & 3.0 & 2.0 \\
     
     \midrule
     Oph 4\  & 2.5 & -5.0 $-$ 8.5 & 2.0 & 1.0 \\
     
     \midrule
     Oph 5\ & 2.5 & -1.8 $-$ 9.0 & 2.0 & 1.2 \\
     
     \midrule
     Oph 6\ & 3.0 & -2.6 $-$ 9.0 & 3.5  & 2.0 \\
    
     \midrule
     Oph 7\ & 2.9 & -7.0 $-$ 9.0 & 7.5 & 2.0 \\
     
     \midrule
     Oph 8\ & 2.6 & -3.0 $-$ 9.0 & 3.0 & 1.5 \\
     
     \midrule
     Oph 9\ & 3.0 & -2.6 $-$ 9.6 & 3.5 & 2.0 \\
     
     \midrule
     Oph 10\ & 2.5 & -0.5 $-$ 9.5 & 7.0 & 1.5 \\
     
     \midrule
	\bottomrule
    \end{tabular}
\label{Table:Mom_parameters}
\end{table*}


\section{Outflows Detected in AGE-PRO: Ophiuchus}
\label{App:Outflows}

\begin{figure*}
\includegraphics[width=\linewidth]{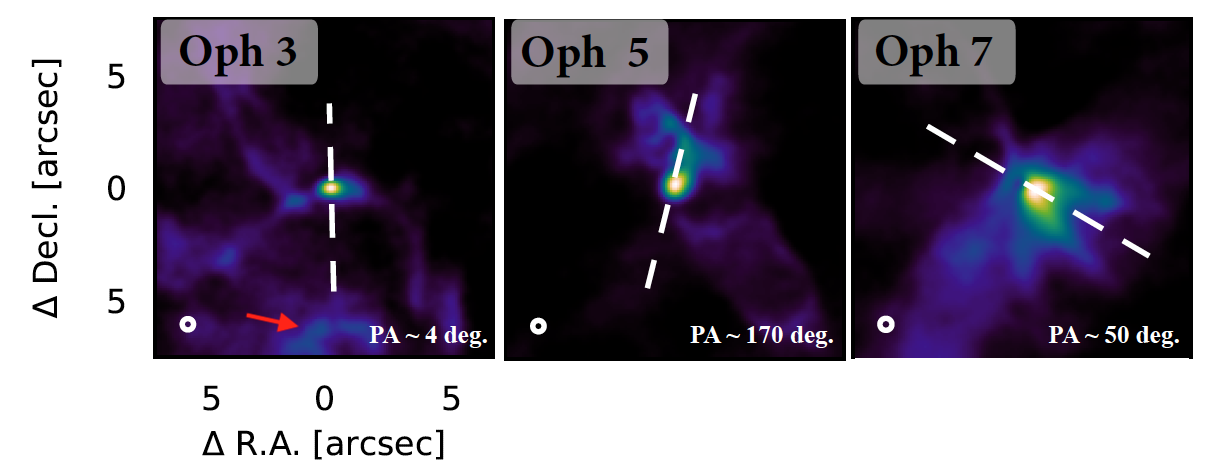}
\figcaption{Moment-0 of $^{12}$CO lines of the Ophiuchus targets associated with outflows. These maps were created by integrating emission over the same velocity range as the moment 0 maps shown in Figure \ref{Fig:Mom0}. North is up, east is left.
\label{Fig:Mom0357}
}
\end{figure*}

\begin{figure*}
\includegraphics[width=\linewidth]{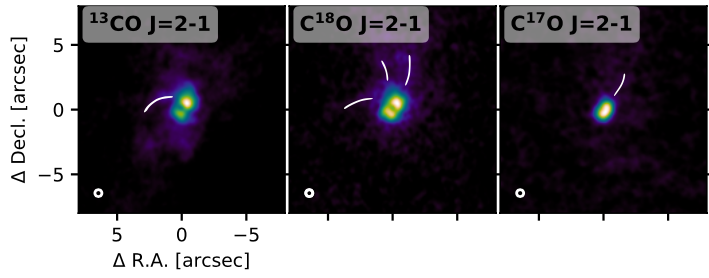}
\figcaption{Moment-0 maps of $^{12}$CO, C$^{18}$O, and C$^{17}$O lines of \textit{Oph 7}. These maps were created by integrating emission over the same velocity range as the moment 0 maps shown in Figure \ref{Fig:Mom0}. North is up, east is left. White lines display potential accretion streamers in \textit{Oph 7}, and were studied recently by \citet{Flores2023}.
\label{Fig:Mom0_streamer}
}
\end{figure*}

Figure \ref{Fig:Mom0357} shows outflows/streamers associated with \textit{Oph 3}, \textit{Oph 5}, and \textit{Oph 7} and detected in $^{12}$CO, while \ref{Fig:Mom0_streamer} shows the $^{12}$CO, C$^{18}$O, and C$^{17}$O mom-0 maps of \textit{Oph 7} where it is marked the potential location of streamers associated with its disk.

\section{Total Flux and Outer Disk Radii Estimates using a Keplerian Mask}
\label{App:Radii_without_mask}

\begin{figure*}[h!]
\gridline{\fig{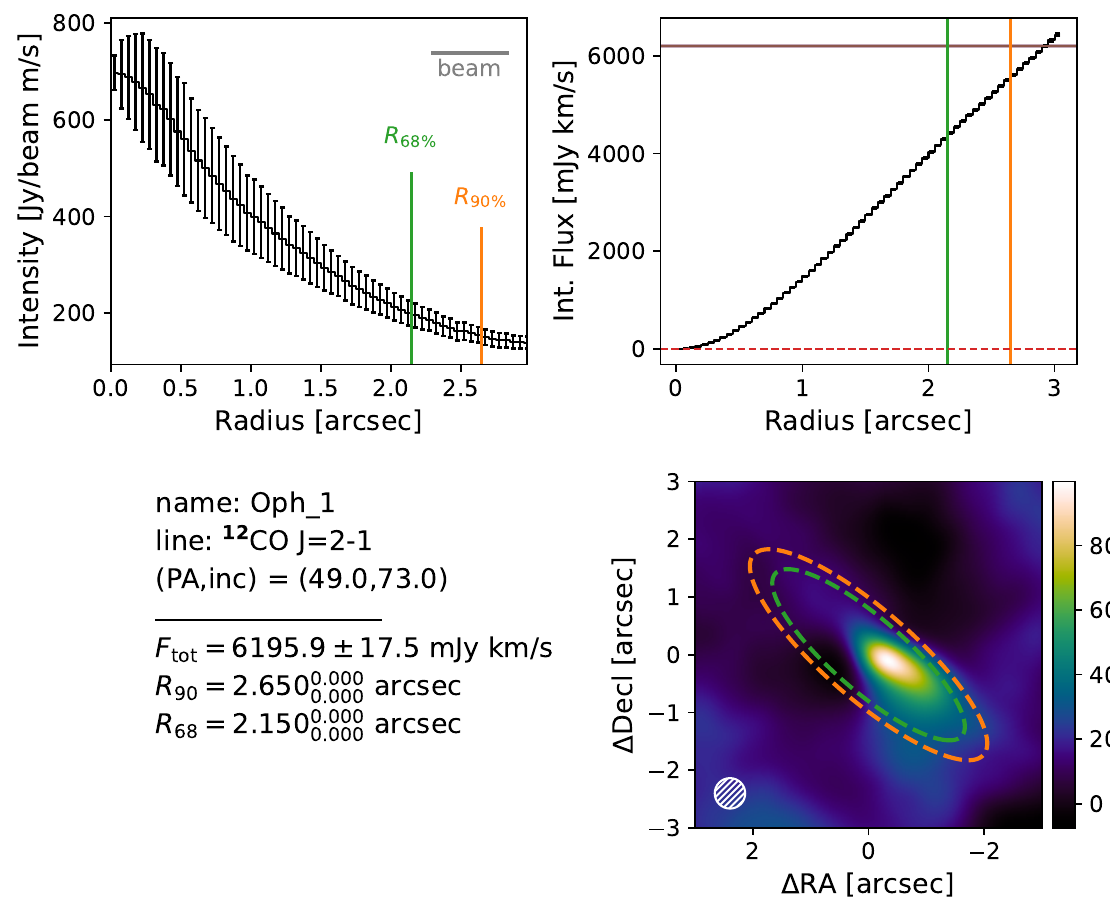}{0.3\textwidth}{}
   \fig{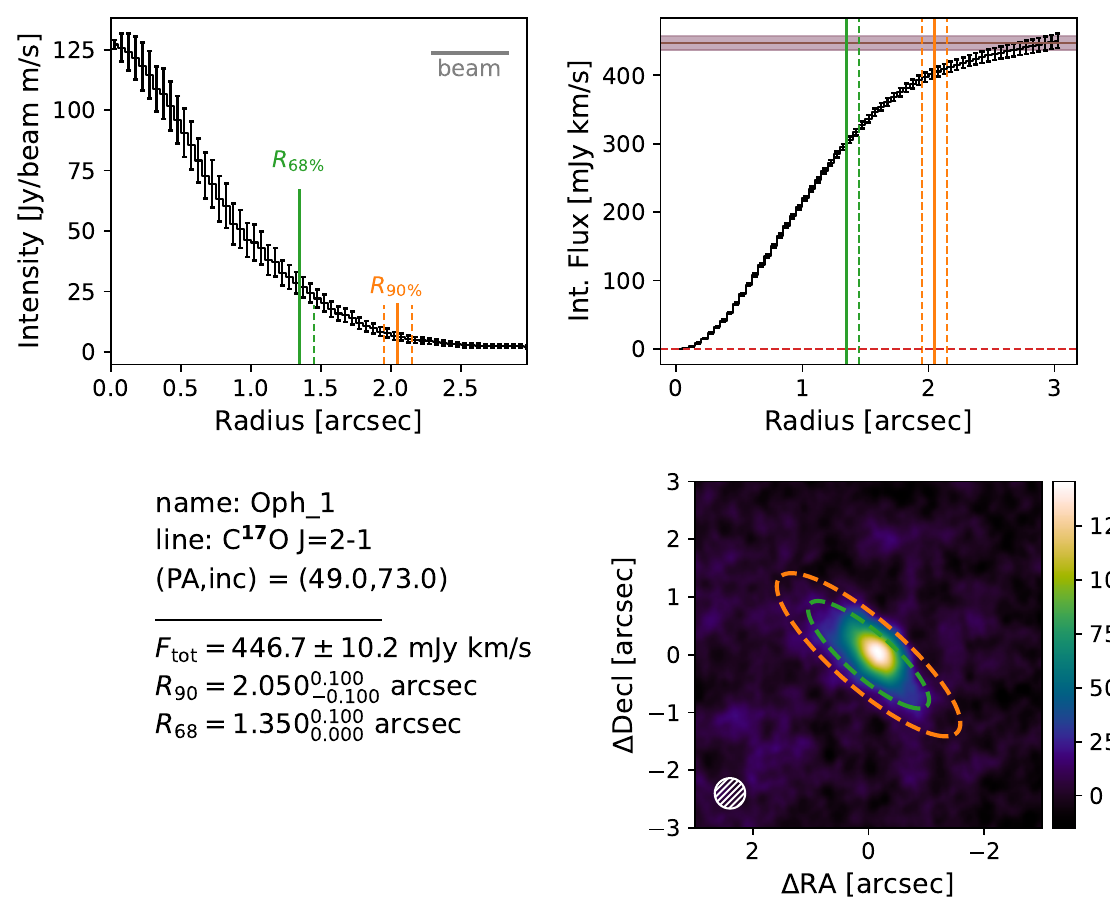}{0.3\textwidth}{}
   \fig{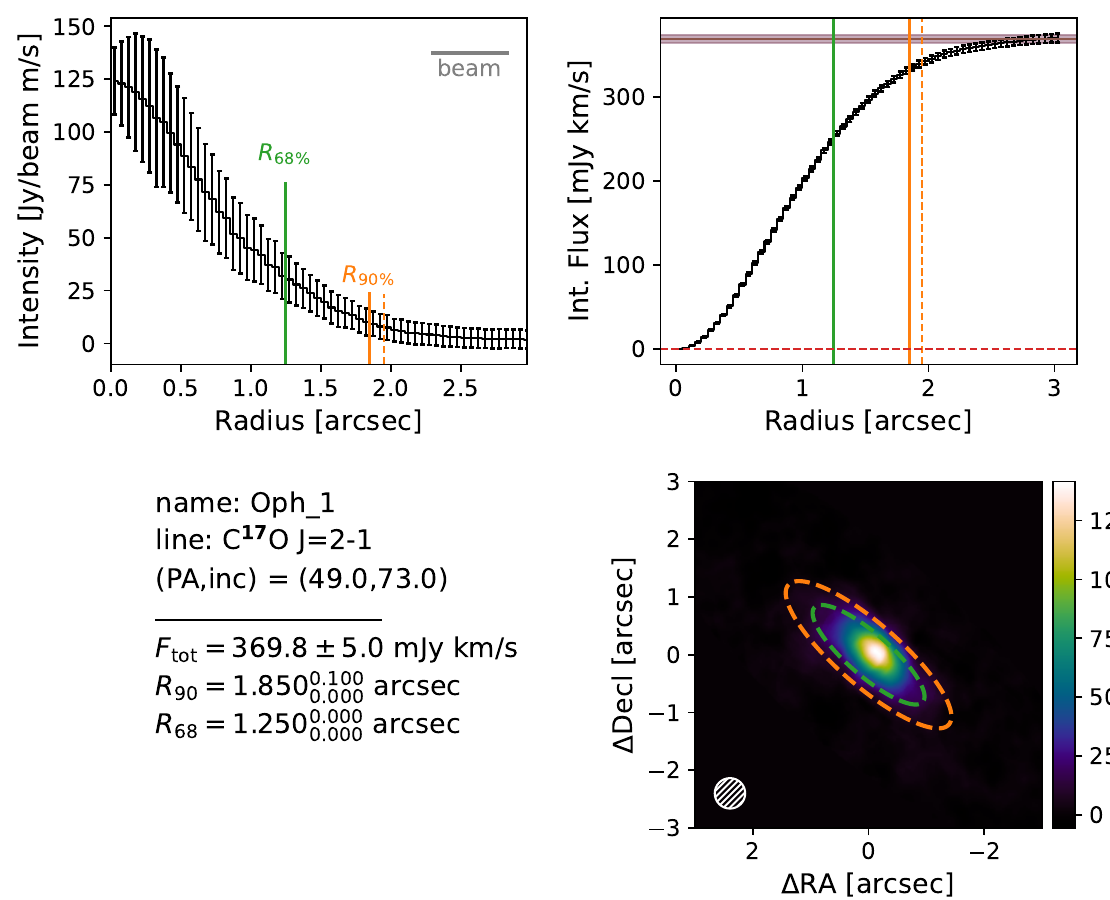}{0.3\textwidth}{}
            }
\figcaption{Curve of growth analysis of the $^{12}CO$ emission for the Oph 1 target
\label{Fig:Curve}
}
\end{figure*}

Overall, the AGE-PRO program has used Keplerian masks for extracting line data products and estimating integrated fluxes and radii of the line and continuum disk emission. However, the masking method is not suitable for observed line emission from a source with a non-Keplerian velocity pattern, and on top of that severe self-absorption, cloud emission, and outflows. This is the case for most of the data products from the Ophiuchus sources where high levels of contamination prevented the efficient alignment and stacking of projected Keplerian velocities at the moment to build a mask and generate moment-0 and -1 maps. This results in parameters that do not necessarily reflect the true features of the disk. For instance, the highly embedded Oph 1 suffers from self-absorption and molecular cloud emission making it difficult to trace a rotation pattern that can be applied in the data analysis. Figure \ref{Fig:Curve} shows that the curve of growth analysis of the $^{12}$CO emission for the Oph 1 target does not converge into a maximum value due to the large emission extension.
On the contrary, a similar analysis applied to the compacted and thin tracer C$^{17}$O data results in more accurate measurements of the gas's outer radius and integrated disk flux. However, we noticed that applying a Keplerian mask for collapsing the spectral cube along the spectral axis leads to lower flux values. This is expected as one of the main functions of the mask is to disentangle various components in the observations, in this case, extra emission from outflows and circumstellar material. Table \ref{Table:CO_mask} shows flux and radius values extracted from C$^{18}$O and C$^{17}$O by using a Keplerian mask following the approach presented in Section \ref{Sec:Flux_Radii}.

\begin{table*}
\centering

    \caption{Disk Parameters C$^{18}$O and C$^{17}$O lines using keplerian mask.}
    \begin{tabular}{c c c c c}\toprule
	   Source & Line &Radius\tablenotemark{a} & Radius\tablenotemark{a} & Integrated Flux   \\
            &        & 68$\%$ & 90$\%$ & [mJy km s$^{-1}$] \\
	 \midrule
      Oph& C$^{18}$O  & 1.65 $_{0.007}^{0.007}$  & 2.35 $_{0.017}^{0.017}$  &   1084 $\pm$ 9  \\
     1&C$^{17}$O & 1.25 $_{0.018}^{0.018}$ & 1.85 $_{0.055}^{0.048}$ &  370 $\pm$ 5  \\
     \midrule

      Oph& C$^{18}$O  & 1.14 $_{0.041}^{0.039}$  & 1.63 $_{0.137}^{0.105}$  &   910 $\pm$ 3  \\
     2&C$^{17}$O & 1.10 $_{0.128}^{0.120}$ &  1.40 $_{0.270}^{0.192}$  &  21.0 $\pm$ 3  \\
     \midrule

      Oph& C$^{18}$O & 1.05  $_{0.029}^{0.029}$ & 1.45 $_{0.119}^{0.105}$  & 203 $\pm$ 5  \\
     3&C$^{17}$O & 1.05 $_{0.094}^{0.083}$ & 1.55 $_{0.269}^{0.191}$ &  97.0 $\pm$ 6 \\
     \midrule


     Oph& C$^{18}$O & --  &  -- & --   \\
     5&C$^{17}$O & -- &  -- &  --      \\
     \midrule

      Oph& C$^{18}$O  & 1.05 $_{0.047}^{0.044}$  &  1.75 $_{0.064}^{0.063}$  & 161 $\pm$ 4  \\
     6&C$^{17}$O & 0.95 $_{0.064}^{0.061}$ &  1.45 $_{0.191}^{0.138}$ &  117.5 $\pm$ 11  \\
     \midrule
     
      Oph& C$^{18}$O  & 1.15$_{0.010}^{0.010}$ & 1.65$_{0.016}^{0.016}$ &  960 $\pm$ 5 \\
     7&C$^{17}$O & 0.85$_{0.016}^{0.016}$ & 1.45$_{0.060}^{0.060}$ &  829 $\pm$ 5   \\
     \midrule
 
     Oph& C$^{18}$O  & \textcolor{red}{0.55$_{0.200}^{0.400}$}  &  \textcolor{red}{0.85$_{0.400}^{0.200}$} &  \textcolor{red}{7.0 $\pm$ 3.1}  \\
     8&C$^{17}$O & \textcolor{red}{0.65$_{0.100}^{0.200}$}  &  \textcolor{red}{0.95$_{0.300}^{0.200}$} &  \textcolor{red}{17.0 $\pm$ 4.1}  \\
     \midrule
      Oph& C$^{18}$O  & 0.85 $_{0.020}^{0.023}$  &  1.25 $_{0.100}^{0.131}$ &  191 $\pm$ 4  \\
     9&C$^{17}$O & 0.85 $_{0.042}^{0.042}$  &  1.35 $_{0.142}^{0.117}$ &  107 $\pm$ 4   \\
       \midrule
      Oph& C$^{18}$O  & \textcolor{red}{0.95$_{0.200}^{0.400}$}  &  \textcolor{red}{1.25$_{0.300}^{0.200}$} &  \textcolor{red}{10.2 $\pm$ 3.0}  \\
     10&C$^{17}$O & \textcolor{red}{1.05$_{0.100}^{0.200}$}  &  \textcolor{red}{1.35$_{0.300}^{0.100}$} &  \textcolor{red}{15.0 $\pm$ 3.6}  \\
     \midrule
     
     \midrule
	\bottomrule
    \end{tabular}
\label{Table:CO_mask}
\tablenotetext{a}{Units are arcseconds.}
\end{table*}

\section{COMPARISON OF CONTINUUM FLUXES: AGE-PRO VS. ODISEA}
\label{App:Masses}

\begin{figure*}[h!]
\centering
\fig{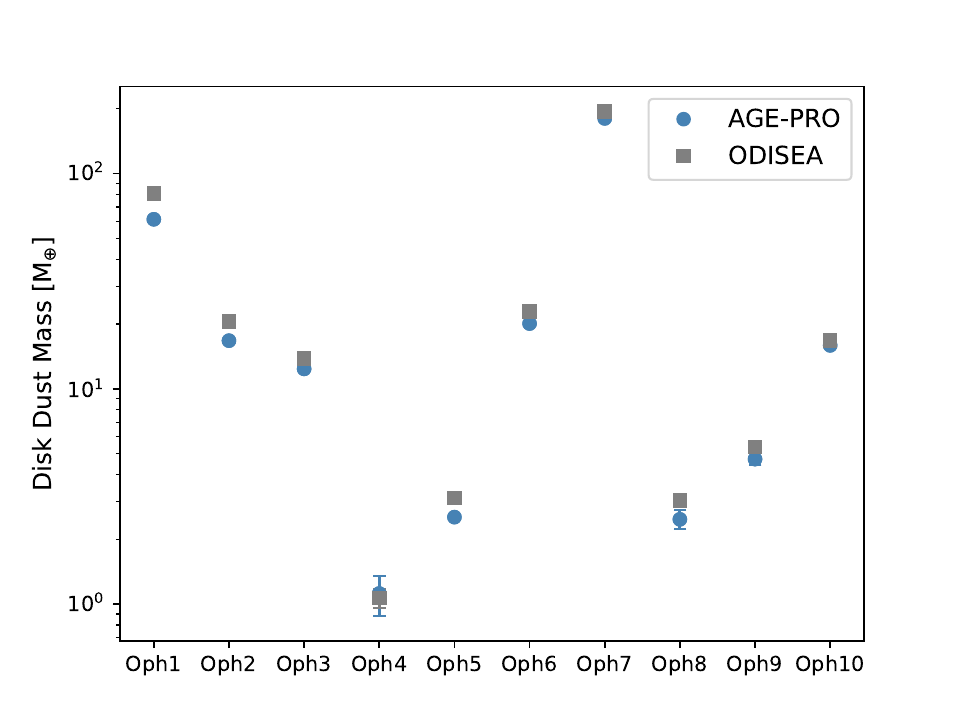}{0.5\textwidth}{}
\figcaption{AGE-PRO disk dust masses estimated from the 220 GHz continuum fluxes (blue dots) compared with the ODISEA disk masses estimated from 230 GHz continuum fluxes (gray squares) \citep{Cieza2019}. 
\label{Fig:AGE_ODISEA_Mass}
}
\end{figure*}

Figure \ref{Fig:AGE_ODISEA_Mass} displays a direct comparison between the AGE-PRO and ODISEA disk dust masses estimated using a $\rm T_{Dust}$ of 20 K. The ALMA 232 GHz continuum fluxes are taken from \citep{Cieza2019}. Overall, the AGE-PRO and ODISEA Ophiuchus masses are in good agreement, where the latter masses tend to be slightly higher than our measurements. This small discrepancy could be due to the methodology used to estimate these values.

\clearpage

\bibliography{Sample631.bib}{}
\bibliographystyle{aasjournal}
\end{document}